\documentclass[onecolumn]{article}

\usepackage{standalone}

\usepackage[a4paper,portrait]{geometry}
\usepackage{subcaption}

\usepackage{pdflscape}

\usepackage[utf8]{inputenc}
\usepackage[T1]{fontenc}

\usepackage{xparse}

\usepackage[sorting=nyt, 
style=alphabetic, 
backref, 
maxbibnames=100, 
doi=false,
url=false]{biblatex}

\usepackage{graphicx}
\graphicspath{{figs/}}

\usepackage{zref-clever}
\zcsetup{cap=true,comp=true,sort=true,nameinlink=true}
\usepackage{amsmath,amssymb,mathtools}
\usepackage{tcolorbox}
\usepackage{keytheorems}
\usepackage{hyperref}
\usepackage{stmaryrd}
\usepackage{comment}

\usepackage{scalerel}

\usepackage[dvipsnames]{xcolor}
\usepackage[]{authblk}

\usepackage{algorithm}
\usepackage{algpseudocodex}

\usepackage{tikz}
\usetikzlibrary{positioning, fit, arrows.meta, backgrounds, calc}

\usepackage{soul} %

\newkeytheorem{definition}[parent=section, tcolorbox-no-titlebar={colback=red!10}]
\newkeytheorem{example,lemma,conjecture,claim,theorem,proposition,corollary,fact,remark,assumption,construction,ingredient}[sibling=definition, tcolorbox-no-titlebar={colback=red!10}]

\DeclareMathOperator{\tr}{tr}
\DeclareMathOperator{\supp}{supp}
\DeclareMathOperator{\diag}{diag}

\NewDocumentCommand{\badfaults}{}{\ensuremath{\mathcal{F}}}

\DeclareMathOperator*{\wtsum}{\boxplus}
\DeclareMathOperator*{\wtprod}{\circledast}
\newcommand{\bigwtprod}{\mathop{\scalerel*{\wtprod}{\bigcup}}\displaylimits}
\newcommand{\bigwtsum}{\mathop{\scalerel*{\wtsum}{\bigcup}}\displaylimits}

\NewDocumentCommand\compCodeType{o}{\ensuremath{\IfNoValueTF{#1}{\mathsf{CompCode}}{\mathsf{CompCode}(#1)}}}
\NewDocumentCommand\gadget{}{\ensuremath{\mathsf{Gad}}}
\NewDocumentCommand\qOp{}{\ensuremath{\mathsf{Op}}}
\NewDocumentCommand\qGate{}{\ensuremath{\mathsf{g}}}

\NewDocumentCommand{\gateAdv}{}{\ensuremath{\mathsf{ADV}}}
\NewDocumentCommand{\gateH}{}{\ensuremath{\mathsf{H}}}
\NewDocumentCommand{\gateS}{}{\ensuremath{\mathsf{S}}}
\NewDocumentCommand{\gateTOFF}{}{\ensuremath{\mathsf{TOFF}}}

\NewDocumentCommand{\gateCNOT}{}{\ensuremath{\mathsf{CNOT}}}
\NewDocumentCommand{\gateCS}{}{\ensuremath{\mathsf{CS}}}
\NewDocumentCommand{\gateI}{}{\ensuremath{\mathsf{Id}}}
\NewDocumentCommand{\gateX}{}{\ensuremath{\mathsf{X}}}
\NewDocumentCommand{\gateCtoQ}{}{\ensuremath{\mathsf{CtoQ}}}
\NewDocumentCommand{\gateQtoC}{}{\ensuremath{\mathsf{QtoC}}}
\NewDocumentCommand{\gatePREP}{}{\ensuremath{\mathsf{PREP}}}

\NewDocumentCommand{\gateSWAP}{}{\ensuremath{\mathsf{SWAP}}}

\NewDocumentCommand{\Nprog}{}{\ensuremath{{N_{\mathrm{prog}}}}}

\NewDocumentCommand{\Nprogi}{}{\ensuremath{{N^{(i)}_{\mathrm{prog}}}}}
\NewDocumentCommand{\Kprogi}{}{\ensuremath{{K^{(i)}_{\mathrm{prog}}}}}

\NewDocumentCommand{\gateInject}{}{\ensuremath{\mathsf{INJECT}}}
\NewDocumentCommand{\gateEject}{}{\ensuremath{\mathsf{EJECT}}}

\NewDocumentCommand{\poly}{}{\ensuremath{\mathrm{poly}}}
\NewDocumentCommand{\polylog}{}{\ensuremath{\mathrm{polylog}}}

\NewDocumentCommand{\maxwidth}{}{\ensuremath{W_{\mathrm{max}}}}

\NewDocumentCommand{\gatelayer}{O{C} r() d()}{\ensuremath{\IfNoValueTF{#3}{{#1}\langle#2\rangle}{{#1}\langle#2,#3\rangle}}}
\NewDocumentCommand{\locations}{o}{\ensuremath{\mathsf{LOC}\IfNoValueF{#1}{\left(#1\right)}}}
\NewDocumentCommand{\circuitMap}{m}{\ensuremath{#1}}
\NewDocumentCommand{\weightenum}{mm}{\ensuremath{\mathcal{W}{\left(#1;\,#2\right)}}}
\NewDocumentCommand{\fault}{o}{%
  \IfNoValueTF{#1}
   {\ensuremath{\mathbf{f}}}
   {\ensuremath{\mathbf{#1}}}%
}

\NewDocumentCommand{\swap}{o}{\ensuremath{\mathsf{SWAP}\IfNoValueF{#1}{(#1)}}}
\NewDocumentCommand{\indirectswap}{o}{\ensuremath{\mathsf{INDSWAP}\IfNoValueF{#1}{(#1)}}}
\NewDocumentCommand{\xswap}{o}{\ensuremath{\mathsf{XSWAP}\IfNoValueF{#1}{(#1)}}}
\NewDocumentCommand{\perm}{o}{\ensuremath{\mathsf{PERM}\IfNoValueF{#1}{(#1)}}}
\NewDocumentCommand{\xchg}{o}{\ensuremath{\mathsf{XCHG}\IfNoValueF{#1}{(#1)}}}

\NewDocumentCommand{\ket}{m}{\ensuremath{|#1\rangle}}

\NewDocumentCommand{\braket}{mo}{\ensuremath{\IfNoValueTF{#2}{\langle #1 | #1 \rangle}{\langle #1 | #2 \rangle}}}
\NewDocumentCommand{\ketbra}{mo}{\ensuremath{\IfNoValueTF{#2}{\rangle #1 | #1 \langle}{\rangle #1 | #2 \langle}}}

\NewDocumentCommand\encoder{o}{\IfNoValueTF{#1}{\ensuremath{\mathsf{enc}}}{\ensuremath{\mathsf{enc}_{#1}}}}
\NewDocumentCommand\decoder{o}{\IfNoValueTF{#1}{\ensuremath{\mathsf{dec}}}{\ensuremath{\mathsf{dec}_{#1}}}}

\newcommand{\BB}{{\mathbb{B}}}
\newcommand{\calF}{{\mathcal{F}}}
\newcommand{\calS}{{\mathcal{S}}}
\DeclarePairedDelimiter{\abs}{|}{|}

\begin{document}

\title{Fault-tolerant quantum processor \\with minimal external control}

\author[1]{Christopher A. Pattison}
\author[2,3]{Jeongwan Haah}
\author[4,5]{John Preskill}
\affil[1]{{\small \it Simons Institute for the Theory of Computing, University of California, Berkeley}}
\affil[2]{{\small \it Leinweber Institute for Theoretical Physics, Stanford University}}
\affil[3]{{\small \it Google Quantum AI}}
\affil[4]{{\small \it Institute for Quantum Information and Matter, California Institute of Technology}}
\affil[5]{{\small \it Oratomic, Inc.}}

\maketitle

\begin{abstract}
Contemporary classical information processors integrate tens of billions of transistors printed on a chip, yet are fully controlled through comparatively few external wires. This is possible because data flows through hard-coded operations in step with a global clock; the controller only needs to steer the data along a desired path through the chip.
Today's solid-state quantum processors are different -- every qubit receives instructions from outside the chip via its own dedicated control line, and error correction adds a further torrent of syndrome data that is typically decoded off-chip. 

Here we show that the external control of a quantum chip can be greatly simplified. 
We propose a fault-tolerant architecture for universal quantum computation, inspired by modern classical processors, in which quantum information flows through a fixed, clocked device, propelled by geometrically local gates in a planar array.
Our quantum processor receives an input stream of classical instructions via a small, constant number of external wires, each attached to a single qubit at a fixed location; this number is independent of the size and target accuracy of the computation.
Except for those few qubits coupled to external wires, all qubits evolve under a depth-3 circuit of geometrically local gates, which is applied periodically and fixed at the time the device is fabricated.
Just as for a classical CPU, the layout and internal dynamics of the quantum chip are independent of the program that it runs -- it is manufactured once and then told what to compute. 

Assuming a noiseless global clock (so that every operation is applied on schedule, though imperfectly),
we establish an accuracy threshold theorem:
For physical noise below a universal positive threshold -- afflicting every gate, measurement, and wire -- a chip with $n$ physical qubits can run any quantum algorithm that uses~$n / \polylog(n/\epsilon)$ logical qubits, where $\epsilon$ is the error rate per logical gate.
The instructions are protected by a classical code that the noisy chip decodes fault-tolerantly, and the results of the computation are returned as an output stream protected in the same way, decoded by the user. 

Each logical gate takes time~$n / \polylog(n/\epsilon)$;
an alternative design, in which quantum operations remain strictly local and noisy,
but on-chip wires broadcast noiseless classical signals instantaneously, 
reduces the time per logical gate to $\polylog(n/\epsilon)$. 
\end{abstract}

\section{Introduction}

In modern classical information processors,
tens of billions ($10^{10}$) of transistors are integrated 
on a lithographically manufactured chip, yet are controlled externally 
by only hundreds ($10^2$) to thousands ($10^3$) of wires.
This is possible because the layout of wires and transistors, which is fixed when the chip is manufactured, already determines most of what the processor does. 
Information is represented by electric voltages on input and output wires of each logic gate, so the flow of computation can be traced along the spatial layout of the chip.
A global clock, ticking a few billion ($10^9$) times per second, drives the data forward. On each tick, memory elements in the device record their inputs and hold them, and the transistor circuitry settles to a computed function of the recorded inputs before the next tick. Universal computation is achieved by routing: instructions steer data through hard-coded operations, which together form a universal set. The vast majority of the chip need not be told what to do in each step; it is a fixed clocked machine through which data flows.

In contrast, typical proposals for quantum information processors
primarily use in-place operations.
This is fundamentally because quantum information cannot be copied; therefore an input qubit for a quantum unitary gate
must be replaced by an output qubit.
This no-cloning principle has led to architectures
in which qubits that hold logical information are conceptually static
and every logical operation is performed by external control.\footnote{Measurement-based quantum computation~\cite{Raussendorf_2007} has no static logical qubits,
but it remains true that every qubit must be externally controlled.}

The complexity of such a quantum architecture is
exacerbated by the need for quantum error correction.
Most of the time, a quantum processor measures
Pauli stabilizers of an underlying quantum error-correcting code,
and the associated decoder -- often centralized -- 
processes syndrome bits collected from most parts of the system.
Occasionally, externally controlled operations
deviate from the periodic Pauli stabilizer measurement schedule
to implement logical operations.
As quantum processors become more relevant to real-world problems at scale,
it is important to ask how much of this external control is actually necessary.

In this paper we answer: very little. We propose a fault-tolerant quantum architecture, inspired by modern classical information processors, in which quantum information flows through a fixed, clocked machine. Our quantum processor design has the following features.

\begin{itemize}
    \item \textbf{Minimal external control.}
    There are only a few ($\le 20$) external wires,
    attached to fixed locations on the chip, each controlling exactly one qubit.\footnote{An \emph{external wire} is a line crossing the chip boundary that addresses a specific qubit with a time-varying, information-bearing signal. Additional static wiring connects components on the chip to a global clock just as in a classical processor.} Their number does not depend on the target logical error rate or the storage capacity of the processor.
    These wires stream classical bit strings into and out of the quantum chip.
    The input stream encodes the logical circuits that the quantum chip executes,
    and the output stream encodes the computation result.
    Only these few wires tell the chip which logical operations to perform.

    \item \textbf{Self-initialization}.
    The initial state of all qubits in our quantum chip may be arbitrary.
    When a user turns on our quantum chip,
    the external control line streams an initialization control sequence
    of length roughly proportional to the number of physical qubits in the chip,
    which is independent of the quantum algorithm that the user executes.
    
    \item {\bf Almost periodicity}.
    Except for those few operations involving qubits driven by external wires, all qubits on the chip undergo a depth-3 quantum circuit that is applied over and over again. 
    The first two layers of this depth-3 circuit consist entirely of unitary swap gates; the third layer 
    contains various unitary and nonunitary one- and two-qubit operations that are fixed when the chip is fabricated. Thus, the internal dynamics of the chip is independent of the program it runs. %
    
    \item {\bf 2d locality}.
    All elementary operations are geometrically local in a two-dimensional plane.
    Concretely, qubits occupy the sites of a square lattice with lattice spacing 1. All two-qubit gates couple qubits separated by Euclidean distance at most~$\sqrt{5}$.

    \item {\bf Fault tolerance}.
    Every operation on the chip, including the interface between
    qubits and external control wires, is noisy, with noise strength less than a universal positive constant (the accuracy threshold).
    We assume however that the global clock,
    according to which all operations are synchronized,
    is noiseless. 

    \item {\bf Efficiency}. \sloppy
    A chip with $n$ physical qubits can run any quantum algorithm
    that uses~$n / \polylog(n/\epsilon)$ logical qubits
    with one $1$d geometrically local logical gate at a time,
    where $\epsilon$ is the error rate per logical gate.
    The time overhead is at most a multiplicative factor of~$n \, \polylog(n/\epsilon)$.

\end{itemize}

Our main theorem establishes that a universal fault-tolerant quantum processor can be designed with all of the features listed above. Our overhead estimates are upper bounds that apply in the worst case under the above requirements, and can be improved appreciably if these requirements are relaxed. 

In particular, we also present a design variant that removes the factor of~$n$ in the time overhead by adding classical wires to the architecture. These classical wires convey classical bits across the chip instantaneously along a fixed path. 
This modification relaxes two-dimensional locality, but only for classical signals; all quantum operations remain strictly local, and only a constant number of classical wires cross any unit square on the chip. To motivate this model, we note that voltage signals can travel across today's classical chips in time scales on the order of nanoseconds, while time scales for quantum processing tend to be substantially longer.

Our processor design relies on the principle that increasing the physical qubit count can reduce the complexity of controlling the chip. Gadgets in one space and one time dimension may be unrolled into two-dimensional tiles, transforming a control \emph{sequence} into a fabricated \emph{layout} through which the data flows. This trade can be beneficial in a world where physical qubits are relatively cheap and local control systems are expensive in comparison.

\subsection{Realizing the model}
An assumption underlying our approach concerns the calibration of the processor. In today's solid-state quantum devices such as superconducting circuits and semiconductor quantum dots, an external control system not only executes gates and reports measurement outcomes, but also tunes the device, compensating for device-to-device variation and drift. In this respect, our model departs significantly from current practice. In a system with limited external control, where tuning individual qubits and couplers may be infeasible, either the tune-up should be multiplexed through a sparse interface, or the qubit performance should be robust against realistic manufacturing variations, or the fabrication should be considerably more precise than in current quantum hardware. Once high-quality solid-state qubits can be reliably fabricated, it may become advantageous to build quantum chips whose functionality is fixed at the time of manufacturing, just as for today's classical CPUs.
These qubits should also be small, so that many qubits can be placed on a chip.

To achieve higher thresholds and realize a model with quickly propagating classical signals, it will be convenient to co-locate classical computing elements in the qubit plane, allowing decoding to be performed locally in classical degrees of freedom.
While not explicit in our construction, our construction relies on the simplicity of decoding circuits for concatenated codes.
Depending on the available area, it may be possible to spend more classical resources on decoding as in~\cite{poulin2006optimal,goto2013fault}.
If qubits can also be transported or swapped forward with very high fidelity as in quantum dots~\cite{yoneda2021coherent,seidler2022conveyor,de2025high}, then the achievable threshold is likely to be improved.
Together, it may be possible to approach estimated threshold requirements in the concatenated code literature~\cite{knill2005quantum,goto2024many,yoshida2025concatenate} which are competitive with modern quantum low-density parity check (qLDPC) code threshold estimates.
It is also convenient that our scheme is naturally resilient to defective qubits.
The same fault tolerance that protects against transient errors also protects against errors that are localized to a point on the chip.\footnote{The \(\gateSWAP\) gates in our circuits also provide a natural location to reset qubit leakage.}

Perhaps the most natural fit to implement our scheme is a two-dimensional semiconductor spin qubit array such as proposed by Veldhorst \textit{et al.}~\cite{veldhorst2017silicon}.
Since the initial proposal, large crossbar arrays actuating circuit elements have been demonstrated~\cite{bavdaz2022quantum,borsoi2024shared,van2025cryo,tosato2026crossbar}.
Ref.~\cite{van2025cryo} additionally demonstrated a cyro-CMOS digital-to-analog converter capable of programming individual DC-biases.
While still insufficient~\cite{neyens2024probing,thomas2025rapid,loenders2026understanding,pellouxprayer2026highlyuniformfirstelectronposition}, it will also be extremely valuable to have very little manufacturing variation to reduce the amount and complexity of multiplexed control required for tune-up.
High fidelity operation of semiconductor spin qubits at high temperatures has also been demonstrated~\cite{petit2022design}.
High temperature enables substantially more cooling power as would be required for operating dissipative electronics such as digital CMOS logic.
Classical logic cointegrated in the qubit plane is likely to increase the threshold (using better decoders) and allows the classical wires of \zcref{sec:FastProcessorWithClassicalWiresFurtherDetail}.
Our work may also be interpreted as showing that the physical architecture proposed by Patom{\"a}ki \textit{et al.}~\cite{patomaki2024pipeline} is nearly compatible with fault tolerance.
However, techniques for local readout with conversion to a digital signal deep within a dense 2D array may need to be developed.
We refer interested readers to the review~\cite{dumoulin2026cmos} for progress on CMOS integration with semiconductor spin qubits.

Another potential avenue to realize our scheme is superconducting qubits.
Merged-element transmons (``mergemons''), in which the junction itself supplies most of the capacitance in place of a large shunt capacitor, may substantially reduce qubit footprints~\cite{zhao2020merged,mamin2021merged}. 
For classical control co-located with the qubits, a promising candidate is single-flux-quantum (SFQ) logic, built from Josephson junctions and operating at cryogenic temperatures with low power dissipation. An on-chip state-preparation and measurement scheme that avoids the large resonator used in dispersive readout has been demonstrated~\cite{abraham2026fast}, and a crossbar-like superconducting circuit has been proposed for programming large numbers of flux biases through a multiplexed interface~\cite{zhou2026superconducting}. Such flux biases could allow tune-up to be performed serially across a large array without control lines for individual qubits.

Protected qubits such as the \(0\)-\(\pi\) qubit~\cite{doucot2002pairing,ioffe2002possible,ioffe2002topologically,doucot2003topological,kitaev2006protected,brooks2013protected,gyenis2021experimental}
 Majorana qubits~\cite{kitaev2001unpaired,lutchyn2010majorana,oreg2010helical,aasen2025blueprint}, and GKP qubits~\cite{gottesman2000encoding,campagne2020quantum,sivak2023real,brock2025quantum} should in principle be robust to manufacturing and control signal variability.
 This benefit may be present even when not operating deep in the protected regime.
 However, if our proposed architecture is to benefit from such features, it will still be necessary for circuit elements (e.g. capacitors, cavities, etc) present in the protected qubit proposals to be brought down in size.

\subsection{Prior work}

We believe that ours is the first proposed quantum computing architecture that requires only geometrically local quantum gates in a two-dimensional layout, achieves universal fault-tolerant quantum computing with a proven accuracy threshold and polylogarithmic space overhead, and is programmable by instructions streamed through a constant number of external wires. But our work has important antecedents. 

Bacon, Flammia, and Crosswhite~\cite{bacon2013adiabatic} proposed a scheme in which a quantum computation is performed on data flowing through a ``material'' governed by a fixed many-body local Hamiltonian, driven by an applied time-varying control field. Their device is analogous to a classical ASIC, in which the program to be executed is set at the time the chip is manufactured. Alternatively, one could envision tuning the Hamiltonian to adjust the program, making the quantum chip reconfigurable like an FPGA, but then local control could be needed to change its functionality. Our architecture instead resembles a CPU, in which the chip's configuration is fixed, and the program is determined by streamed instructions. Furthermore, programs run by the device in \cite{bacon2013adiabatic}, encoded in the chip's Hamiltonian, lack the robustness achieved by our program code. 

Patom{\"a}ki \textit{et al.}~\cite{patomaki2024pipeline} propose a gate-based scheme resembling ours, in which semiconductor spin qubits are shuttled through a hard-coded circuit, and they discuss its realization in a silicon device. Their analysis, however, does not include error correction. Our work shows that their model is nearly compatible with fault-tolerance; what needs to be added are resets within the array, and interleaved vertical and horizontal propagation of electrons. With streamed instructions, the scheme becomes programmable as well.

Balasubramanian, Davydova, and Lake~\cite{balasubramanian2024local}
constructed a classical cellular automaton (CA) decoder for the 2d toric code, and a corresponding 3d CA that is time-translation invariant. Their system is a robust quantum memory on a $L \times L \times L $ grid of qubits encoding $O(L)$ logical qubits with a memory time exponential in $L$.  In comparison, our fault-tolerant quantum memory, in which a depth-3 circuit is repeatedly applied, lives on an $L \times L$ 2d grid, and encodes $L^2 / w$ logical qubits with memory time exponential in~$w$, where $w$ is a tunable parameter. In addition, our memory can be upgraded to a universal quantum processor robustly programmed by a stream of encoded classical data.

D\"unnweber, Styliaris, and Trivedi~\cite{DST2026quantummemoryautonomouscomputation} constructed a quantum cellular automaton (QCA) with a fixed local translation-invariant update rule which functions as an autonomous fault-tolerant 2d quantum memory, and can also be operated as fault-tolerant quantum computer. (Earlier constructions of universal translation-invariant QCA rules \cite{raussendorf2005quantum,shepherd2006universally} were not fault tolerant.) In this computational scheme, the entire program to be executed is stored in the device when it is initialized. In contrast, in our scheme the classical instructions are streamed as the computation proceeds.

In related independent work, Selub, Bhardwaj, and Lake~\cite{selub2026local} provide another route to fault-tolerant quantum computation with strictly local operations in two dimensions. They couple a topological code to a time-translation-invariant classical cellular-automaton decoder, using only a constant density of quantum and classical resources. They assume that the classical automaton is noiseless, and as a result find a high accuracy threshold for the noisy quantum components in their scheme. In contrast, in our model every element is noisy, including classical processing.

\subsection{Reader's guide}
The remainder of the paper is structured in two parts.
\zcref{sec:construction} gives a high-level view of the construction without regard to analysis of fault-tolerance or disambiguation.
\zcref{sec:fault-tolerance} to \zcref{sec:FastProcessorWithClassicalWiresFurtherDetail} give further details and proofs.
These sections stand alone from \zcref{sec:construction} and may be read independently.
However, regardless of the reading order, readers may find the remarks in \zcref{sec:discussion} useful along with \zcref{rem:one_qubit} (one controlled qubit), \zcref{rmk:fabrication-defects} (fabrication defects), \zcref{rem:const-rate} (constant rate memory), and \zcref{rem:more-control} (faster computation).
Readers may skip \zcref{sec:SkeletonOfArgument} on a first reading and return to it when reading the proof of \zcref{thm:memory}.
The proofs of \zcref{thm:memory}, \zcref{thm:all-quantum-processor}, \zcref{thm:fast-quantum-processor} are intended to be read in order as each proof builds on the previous one.

\subsection{Acknowledgments}
The authors would like to thank Nathaniel Bishop, Chi-Fang (Anthony) Chen, James Clarke, Andrew Doherty, Louis Golowich, Jim Harrington, Anne Matsuura, Quynh Nguyen, David I. Schuster, and Umesh Vazirani for useful discussions over the long writing process.

CAP is currently a Simons-CIQC postdoctoral fellow at the Simons Institute for the Theory of Computing, supported by NSF Grant CCF-231173, NSF QLCI Grant 2016245, and DOE grant DE-SC0024124.
CAP acknowledges funding from the U.S. Department of Energy Office of Science, DE-SC0020290.
This work was done in part while JH was visiting the Simons Institute 
and participating in the Challenge Institute for Quantum Computation at UC Berkeley. 
JH acknowledges funding from the U.S. National Science Foundation grant NSF 24-599.
JP acknowledges support provided by the Institute for Quantum Information and Matter, an NSF Physics
Frontiers Center (NSF Grant PHY-2317110), and the U.S. Department of Energy, Office of Science,
National Quantum Information Science Research Centers, Quantum Systems Accelerator.

No AI tools were used in the construction or proofs of the main results in Sections 3–6, or in writing those sections. AI models from OpenAI and Anthropic generated the source for Figures~\ref{fig:noiseless-routing}, \ref{fig:control-pipeline}, \ref{fig:layoutWithClassicalWire}, 
and assisted the editing of Sections 1 and 2.

Competing interest: JP is a shareholder and part-time employee of Oratomic,
Inc., which is developing fault-tolerant quantum computers.

\clearpage
\tableofcontents

\section{Construction}\label{sec:construction}

\begin{table}[h!]
    \centering
    \begin{tabular}{r | p{0.7\textwidth}}
        \hline\hline
        Term & Meaning
        \\ \hline\hline
        
        physical qubit & 
        The lowest level of a physical degree of freedom whose Hilbert space is~$\mathbb C^2$.
        \\ \hline
        
        code block &  
        A set of physical qubits that constitute a quantum code.
        \\ \hline
        
        computation code & 
        A quantum error-correcting code that is implemented in a code block.\\ \hline
        
        program code & 
        A classical code to encode a program (a sequence of logical gates).
        To run a program on a quantum processor, we send a stream of bits,
        a chunk of which is a codeword of a program code. \\ \hline
        
        readout code & 
        A classical code to encode an output. 
        Any classical code with sufficient error-correcting capability may be used. \\ \hline
        
        tile & A region of the quantum processor that implements a map from an input to an output. 
        This is analogous to ``pure functions'' in a functional programming language; 
        a tile may be composed of several subtiles. \\ \hline
        
        primitive tile & A tile constructed directly out of a single fault-tolerant gadget. \\ \hline
        
        tape & A narrow and long rectangular array of qubits. Most tapes have periodic boundary conditions along the long direction.\\ \hline
                
        slot & A propagating block of the computation code. At a given time, a slot occupies $w \times 1$ array of qubits on a tape. \\ \hline

        lane & If a tape is a $w \times L$ array of qubits, where \(w\) is the width of a slot, a lane is a $1 \times L$ array of qubits.
        \\ \hline
        
        tick & A point in time after the application of the clocked circuit. \\ \hline
        
        cycle & A time duration between consecutive ticks. \\ \hline
        
        station & Spatial locations of a slot where gates of the originating 1D circuit are applied. Each slot moves on to the next station in a cycle.\\ \hline
        
        delay & The number of ticks required for a slot to exit a tile after entering it. Equivalently, the number of slices in a tile. \\ \hline
    \end{tabular}
    \caption{Terminology used in this paper.}
    \label{tab:terminologies}
\end{table}

We give the construction in this section,
deferring the analysis of fault tolerance to later sections.
Our design starts with the idea that 
quantum information 
should pass through an array of physical qubits
while experiencing certain transformations,
just like how classical information processors work.
We begin explaining our design with the simplest setting,
gradually adding components towards quantum computational universality 
and relaxing assumptions.
Our fundamental picture is that 
every logical qubit moves along a tape, which is a cyclically ordered collection of stations,
with one operation performed per station.

\subsection{With noise-free operations}
\label{subsec:noise-free}

For now, every operation is assumed to be noise-free,
and hence there is no quantum error correction 
and every qubit is a legitimate logical qubit.
This will set the stage for understanding the fault-tolerant construction described in Sec.~\ref{subsec:noisy-ops} and Sec.~\ref{subsec:robust-control}.

\subsubsection{Applying $U(2)$ on logical qubits}

As a warm-up, we construct a gadget that applies a unitary circuit 
consisting of $H$ and $T$ gates to $W$ logical qubits.

Consider a ring of $2W$ qubits, which we call a {\bf tape}. Let $W$ be sufficiently large, say $100$.
Initialize the qubits that are indexed $1,3,5,\ldots$ in some fixed state;
we will not be interested in this state. 
Suppose that the remaining even-indexed qubits are in a desired state.
Apply a depth-2 unitary circuit
where the first layer is to swap qubit~$2i$ with qubit~$2i+1$
and the second layer is to swap qubit~$2i+1$ with qubit~$2i+2$ for all~$i$.
Since the qubits are arranged on the ring,
the qubit indices are read modulo~$2W$.
The effect of this depth-2 swap circuit is 
to cyclically permute all even-indexed qubits along the ring in one direction.
Simultaneously, all odd-indexed qubits are moved in the opposite direction.
The odd-indexed qubits are there just to allow for this circular transport to occur.
This simplistic setting mimics a classical processor
in which logical bits propagate through the chip under the action of a periodic drive.
In our case, the propagation speed is set by the frequency of the depth-2 unitary circuit.
We will call each qubit's location a {\bf station} on the tape.

Now modify the unitary circuit to have a third layer,
where we have a $T$ gate at station~$a$ and a Hadamard unitary~$H$ at station~$b$,
where $a$ and $b$ are located at, say, the antipodal points of the ring.
The first two layers implement the cyclic permutation as before.
Upon the application of this depth-3 circuit~$U$,
a logical qubit on station~$a-2$ will be transported to station~$a$ and then acted on by~$T$.
A logical qubit on station~$b-2$ will be transported to station~$b$ and then acted on by~$H$.
After $m W$ applications of~$U$, every logical qubit residing on an even-indexed station 
will be acted on by unitary~$(HT)^m$ where $m \ge 1$ is an integer.

Next, we upgrade the one-tape ring to a two-tape ring:
there are now two parallel identical tapes, 
and each station consists of two qubits.
We will call one of the tapes a {\bf memory} tape
and the other an {\bf auxiliary} tape.
The tape continues to have the periodic boundary condition.
For the auxiliary tape we keep the depth-3 circuit~$U$,
and for the memory tape we have a circular shift circuit~$I$ 
consisting of two layers of swap gates.
The circuit~$I$ has the same first two layers of swap gates as~$U$
but the third layer is empty.
So, $U \otimes I$ implements the circular shift followed by $H$ or $T$ on the auxiliary tape depending on the station.

\begin{figure}[t]
    \centering
    \includegraphics[width=\linewidth]{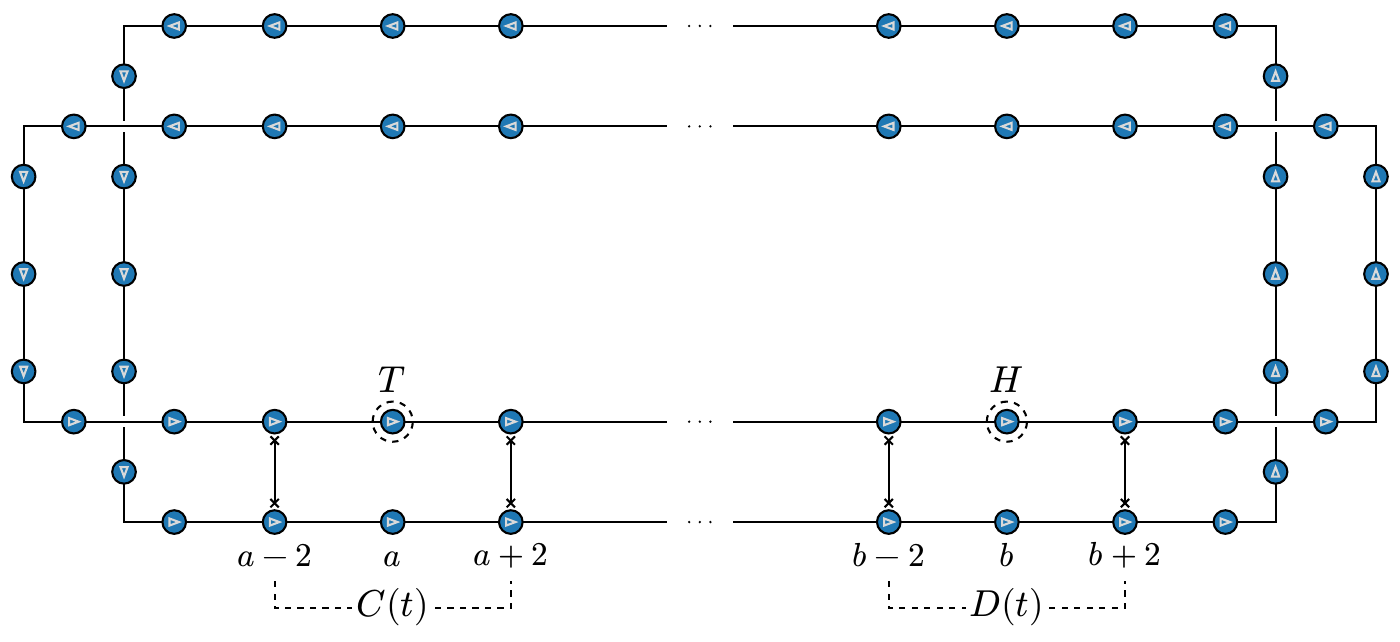}
    \caption{Controlled gates by routing on two tapes.
    Fixed $T$ and $H$ gates act only on the auxiliary tape.
    The bit $C(t)$ controls both swaps at $a-2$ and $a+2$, while $D(t)$ controls both swaps at $b-2$ and $b+2$.
    Only even stations are drawn (blue circles).
    Qubits at these stations propagate counterclockwise as indicated.
    Here, we illustrate our method of turning multiple tapes while maintaining a constant propagation speed.
    }
    \label{fig:noiseless-routing}
\end{figure}

Figure~\ref{fig:noiseless-routing} illustrates the transport and the controlled routing introduced next.

We introduce four swap gates, one for each pair of qubits across the two tapes 
in stations~$a-2$, $a+2$, $b-2$, and $b+2$.
These locations are chosen to sandwich the stations that contain $T$ and $H$ on the auxiliary tape.
These extra swap gates are controlled by two external control bits:
the swaps at the station~$a-2$ and~$a+2$ are applied if the first external control bit~$C$ is~$1$,
and nothing is applied if $C=0$;
the swaps at the stations~$b-2$ and~$b+2$ are applied if the second external control bit~$D$ is~$1$,
and nothing is applied if $D = 0$.
Given a control bit stream $C(1), D(1), C(2), D(2), \ldots$,
the qubits on the two tapes are acted on by
\begin{equation}
    \begin{split}
    &\cdots (\gateSWAP_{a-2} \gateSWAP_{a+2})^{C(t)} (\gateSWAP_{b-2} \gateSWAP_{b+2})^{D(t)}(U \otimes I)\\
    &\quad \cdots \\
    &\qquad (\gateSWAP_{a-2} \gateSWAP_{a+2})^{C(1)} (\gateSWAP_{b-2} \gateSWAP_{b+2})^{D(1)}(U \otimes I)
    \end{split}
\end{equation}
where $\gateSWAP_j$ denotes the swap of the two qubits on station~$j$.

One round of $(\gateSWAP_{a-2} \gateSWAP_{a+2})^{C(t)} (\gateSWAP_{b-2} \gateSWAP_{b+2})^{D(t)}(U \otimes I)$
defines the unit of time.
If the external control bits~$C(t)$ and~$D(t)$ are zero,
then the logical qubits on the memory are just moving along the tape and experiencing nothing.
If $C$ is set to~$1$ when a logical qubit on the memory tape arrives at the station~$a-2$,
the logical qubit will be pushed to the auxiliary tape and will experience $T$ in the next time step.
If $C$ continues to be~$1$ by the time the logical qubit arrives at the station~$a+2$,
then the logical qubit will be pushed back to the memory tape and resume the usual circular travel.
When the logical qubit approaches station~$b$, 
the control bit~$D$ will set the course for the logical qubit,
deciding whether $H$ acts on it.

In summary, every $W$ time steps, a logical qubit passes station~$a$ and~$b$,
and depending on the external control bits,
the logical qubit experiences a sequence of $H$ and $T$ gates.
Since $H$ and $T$ generate a dense subgroup of~$U(2)$ (up to global phases),
we have constructed a gadget that can accurately approximate any single-qubit unitary applied
to any logical qubit on the memory tape by choosing a suitable stream of control bits.
Microscopically,
there is a quantum circuit of depth~$3$ that is applied periodically,
interspersed by externally controlled unitaries at exactly 4 stations.
Similar to what happens to classical bits in a modern CPU,
there is a designated region in the tape, called a {\bf tile}, devoted to a specific unitary.
The external control bit streams do not directly control which unitary gate to apply,
but manipulate the \emph{path} that a logical qubit must follow.

Since every logical qubit travels along the tape,
it will be convenient to label each traveling block of logical qubits,
called a {\bf slot}.
The spatial location of a slot varies in time,
but a slot carries a specific logical state
if all the control bits are turned off.
Below we will make a slot carry a quantum error-correcting code
together with some ancilla qubits.
At present, a slot contains just two qubits.

\subsubsection{Applying $U(2^W)$}

The construction above is sufficient 
to implement an arbitrary element of $U(2)^{\times W}$ to $W$ logical qubits 
up to small error in operator norm.
The time complexity is $\mathcal O(W \log \frac 1 \epsilon)$
where $\log \frac 1 \epsilon$ is the single-qubit unitary compilation complexity
for error $\epsilon$ in operator norm in each single-qubit unitary~\cite{ross2016optimal}
and $W$ is the time for a slot to travel one full circle along the tape.

There are many ways to achieve a universal gate set for $W$ logical qubits.
One way is to implement an entangling two-qubit unitary gate, 
say~$CZ = \diag(1,1,1,-1)$,
between two consecutive slots.
That is, in addition to the $H$ and $T$ tiles near stations~$a$ and~$b$ as explained above,
we put a tile over stations~$c,c+2$ 
that applies the unitary on two stations~$c,c+2$ on the auxiliary tape,
where $c$ is sufficiently far from~$a$ and~$b$.
A pair of logical qubits will or will not experience $CZ$,
depending on another external control that shepherds the logical qubits.
This is computationally universal but comes with additional complication in compilation
(the task of decomposing a quantum algorithm down to elementary gates).
Every two-qubit unitary on an $W$-qubit logical circuit
must be re-expressed in terms of \emph{nearest neighbor} $CZ$ gates.

To facilitate the compilation,
we can introduce an additional logical swap between two consecutive slots.
Then, any permutation of logical qubits can be implemented
by a brickwork sorting network that only uses geometrically local swap gates
and has depth linear in the number of logical qubits.
Note that the swap tile is in fact redundant up to a constant overhead.
Any depth-1 circuit of geometrically local swap gates
can be written as depth-3 circuit of geometrically local CNOT gates,
and a CNOT gate is a product of two Hadamard gates and a $CZ$ gate.

In summary, any circuit of depth~$D$ on~$W$ logical qubits
consisting of one- and two-qubit logical gates (without geometric locality)
can be implemented on our two-tape ring of qubits with three tiles 
($H$, $T$, $CZ$)
in time $\mathcal O(W^2 D\log \frac 1 \epsilon )$, where $\epsilon$ denotes the error per gate, and 
 $\log \frac 1 \epsilon$ is the complexity of approximating 
arbitrary element of~$U(4)$ or~$U(2)$ using $H$, $T$, $CZ$~\cite{ross2016optimal}.
A factor of~$W$ in this upper bound is for worst cases;
if the circuit is already geometrically local with respect to a ring geometry,
the time complexity on our quantum processor is $\mathcal O(WD \log \frac 1 \epsilon)$.

The gate set $\{ H, T, CZ\}$ is simple, but there is a smaller universal gate set:
$H$ and $CS=\diag(1,1,1,\sqrt{-1})$ which also generates a dense subgroup of (projective)~$U(2^W)$.%
\footnote{See~\cite{Beverland_2020,kissinger2024catalysingcompletenessuniversality} for catalytic conversion techniques.}
From now on, we take $\{H, CS\}$ as our universal unitary gate set
and we will have just two logical unitary tiles.

\subsubsection{Program}

A program is a sequence of logical instructions that any quantum information processor can execute.
In a typical quantum architecture, 
the logical instruction is a sequence of tuples of the form
$(H,j),(CS,j,j')$,
where $j,j'$ are logical qubit indices.
In our case, the logical qubit index must be replaced by a time index
since the spatial positions (stations) of the tiles that apply the unitaries are fixed.
The external control keeps track of the logical qubit index~$j$ at each station
as a function of time,
and sends appropriate signals to each tile
when a target logical qubit (a slot) arrives at the tile. This control sequence is chosen by the external agent who operates the computer.

\subsubsection{Readout}

To retrieve the result of the quantum computation,
we introduce a readout tile located at station~$m$ on the auxiliary tape.
(Recall that $H$ and $CS$ tiles are located on the auxiliary tape as well.)
The readout tile is connected to two external control lines.
One determines whether the readout tile is activated
in the same manner as the $H$ and $CS$ tiles,
and the other is used to send the measurement outcome to the computer's user.
The readout tile always projectively measures the logical qubit on the slot in the $Z$-basis.

\subsubsection{Initialization}

It is convenient to introduce a reset tile on the auxiliary tape 
that resets a logical qubit in a slot to the logical state~$\ket 1$.
An external control line is attached to this reset tile.
There is a gadget inside this reset tile which always prepares~$\ket 1$,
but whether a slot is reset is determined by the external control
that dictates if the swap between the auxiliary and memory tape is activated.

When our quantum processor first starts,
this reset tile is activated to initialize all $W$ logical qubits.
The external control attached to the reset tile streams the ``on'' signal repeatedly
while all other external control signals are ``off,'' until all qubits have been initialized in the state $\ket 1$.

\subsubsection{Control tape and control tile}

In the construction described so far, each tile has its own external control line(s). There are five such lines, one each to activate the $H$, $CS$, reset, and readout tiles, and one to transmit the measurement outcome to the user. We will now modify the construction, not because this modification is necessary in the noiseless case, but rather to ease the way toward describing the fault-tolerant construction in the following sections. 

In this modified construction, five additional tapes are introduced, 
so that each slot now contains seven logical qubits that propagate along the tape, 
five of which carry only classical information. 
The swap gates straddling the memory and auxiliary tapes,
which were triggered by external control signals in the original scheme,
now become controlled swap gates acting across three tapes.
If the three tapes are not next to each other, we use swap gates straddling neighboring tapes
to bring three qubits in the support of the controlled swap gate to neighboring positions,
apply the controlled swap,
and then swap them back to the original positions.
This effectively increases the spatial and temporal extent of the controlled swap gate.
The external control lines may now be attached to the quantum processor
at a single \textbf{control tile} at one spatial location,
where control bits are copied from four incoming control lines to the corresponding control tapes,
and measurement results are copied from a tape to produce outgoing classical signals.

\subsubsection{Summary}

There are seven tapes in the system, each containing $2W$ qubits.
Every other qubit along each tape is a garbage qubit,
which exists only to implement circular transportation of slots. 
There are tiles at designated spatial locations,
implementing unitary~$H$, unitary~$CS$, preparation of~$\ket 1$, and
logical measurement in the~$Z$ basis.
Every tile implements the given operation repeatedly on the auxiliary tape
via a depth-3 unitary circuit,
interspersed by swap gates between the auxiliary and memory tapes near the tiles;
these tape swap gates are activated by the bits on the control tapes.
Hence, there are only four external control lines and one readout line,
all attached to the control tile.
Except at the control tile, where incoming bits are written onto the control tapes and outcomes read off, all operations are performed by a periodic depth-3 circuit
which consists of geometrically local gates in a 1d ring geometry.
Below, we will refer to a periodic circuit 
interspersed by externally controlled unitaries
as a {\bf controllable circuit}.

\subsection{With noisy operations}
\label{subsec:noisy-ops}

\begin{table}[h!]
\centering
\begin{tabular}{p{0.43\linewidth} | p{0.45\linewidth}}
\hline\hline
Noise source & Fault-tolerance mechanism \\
\hline\hline
Inaccurate physical gates
    & Computation code: a fault-tolerant circuit
      transcribed onto the tape
      (Secs.~\ref{sec:ab97}, \ref{sec:memory-2d}) \\[2pt] \hline 
Unreliable measurement and reset
    & Measurements are coherently controlled and corrected; entropy is removed by fresh ancillas
      (Sec.~\ref{sec:ab97}) \\[2pt] \hline
Noisy instruction stream and interface
    & Program code, decoded on-chip
      by logical gates of the computation code
      (Secs.~\ref{sec:interface-tile}--\ref{sec:control-tile}) \\[2pt] \hline
Unknown program codeword boundaries %
    & Metronome sequence and metronome tile
      (Sec.~\ref{sec:metronome-tile}) \\[2pt] \hline
Noisy output stream
    & Readout code, encoded by logical gates of the computation code before ejection, decoded externally by the user (Sec.~\ref{sec:readout-tile-intro}) \\[2pt] \hline
Fabrication defects
    & On-chip qubit defects are corrected by the computation code. The constant number of qubits attached to external control must be fabricated correctly\\
\hline
\end{tabular}
\caption{Noise sources on the quantum chip and mechanisms that neutralize them.}
\label{tab:noise-vs-mechanism}
\end{table}

We now allow noise in every constituent of our quantum chip.
We must modify the design above so that it is fault tolerant against 
all of the following noise sources.
\begin{itemize}
    \item Physical unitary gates (swap, $H$, $CS$) are not accurate.
    \item Physical measurement results may be wrong, 
    and hence the reset and readout are not reliable.
    \item The instruction bit stream and output bit stream cannot be trusted.
    \item Since the instruction bit stream is noisy, 
    it is not obvious how the quantum chip infers the beginning moment of computation.
    \item Some components in our quantum chip may be permanently damaged
    (fabrication defects). %
\end{itemize} 
There is one element that we assume is perfect:
\begin{itemize}
    \item Physical level gates are synchronized throughout our quantum processor.
\end{itemize}
This mimics modern classical information processors
that have a global ``clock'' which ticks a few billion times a second.

\begin{figure*}
    \centering
    \includegraphics[width=\textwidth]{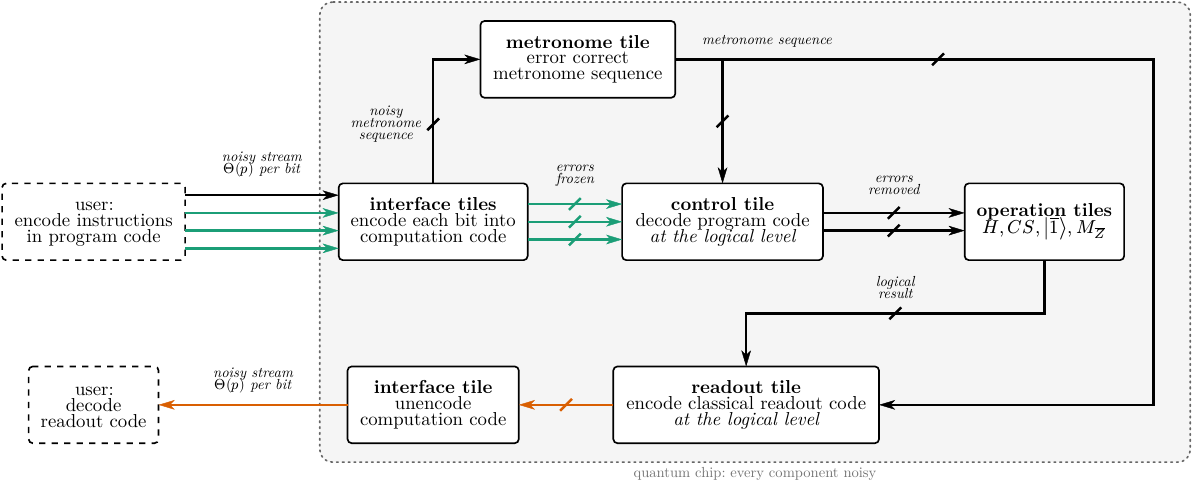}
    \caption{
        Pipeline for classical bits entering and leaving the quantum chip.
        Data encoded in the computational code is indicated as a slashed arrow.
        Data encoded in the program code (readout code) is indicated in green (orange).
        Top row, left to right (inbound):
        the user encodes the instruction stream in the program code;
        the noisy stream reaches the interface tile
        through a constant number of external wires,
        acquiring an error of strength~$\Theta(p)$ per bit;
        the input interface tile encodes each bit into the computation code,
        which \emph{freezes} the accumulated errors;
        the control tile decodes the program code
        at the logical level
        (i.e., entirely by logical gates of the computation code,
        which fail only at the suppressed rate),
        \emph{removing} the frozen errors
        and writing a reliable instruction stream onto the control tapes
        that route logical qubits through the operation tiles.
        In parallel, the metronome tile recovers the codeword framing (metronome sequence) by majority voting from a dedicated periodic signal which also enters from the interface tiles.
        Bottom row, right to left (outbound, the mirror image):
        a logical measurement result is first redundantly encoded
        by a classical readout code at the logical level;
        each bit is then unencoded down to a single physical qubit
        and measured by an output interface tile, incurring $\Theta(p)$ error per bit;
        the user decodes the readout code.
        Classical information is thus encoded before crossing the noisy interface in either direction, and decoded only after reaching safety: on-chip for inbound bits, and off-chip for outbound bits.}
    \label{fig:control-pipeline}
\end{figure*}

To achieve robustness against noise, three problems must be solved. First, the logical qubits traveling along the tape must be protected. Hence, each slot will now carry a block of a quantum error-correcting code, which we call the {\bf computation code}, and a fault-tolerant circuit that uses no measurement or feedback will be transcribed onto our periodic circuit, so that error correction occurs during transport.
(Sec.~\ref{sec:ab97}--\ref{sec:gate-tiles}). Second, the classical instructions entering the chip must be protected.
Only a constant number of external wires touch a constant number of physical qubits,
so every incoming bit unavoidably suffers an error of strength~$\Omega(p)$ at the interface,
where $p$ is the physical noise strength, no matter which quantum code we use on-chip. We will therefore protect the instructions by concatenating two codes: the computation code below, which \emph{freezes} the interface errors, and a classical {\bf program code} on top,
decoded on-chip at the logical level, which then \emph{removes} them
(Section~\ref{subsec:robust-control}).
Third, the readout must survive the same interface bottleneck in reverse: before a logical measurement result is ejected from the protection
of the computation code, it is redundantly encoded by a classical code at the logical level, and the redundancy is decoded outside the chip
(Section~\ref{sec:readout-tile-intro}). Table~\ref{tab:noise-vs-mechanism} matches each noise source with the mechanism that neutralizes it, and Figure~\ref{fig:control-pipeline} shows the resulting pipeline for classical bits into and out of the chip.

\subsubsection{Feedback-free fault tolerance in 1+1d}
\label{sec:ab97}

Our fundamental idea is that a fault-tolerant version of a logical circuit
is itself a circuit of physical gates,
all of which can be implemented in our scheme of traveling qubits and static tiles.
An important ingredient for this is the fault-tolerant scheme 
by Aharonov and \mbox{Ben-Or}~\cite{aharonov1997fault},
which is a little different from typical modern designs.
A typical fault-tolerant architecture
uses measurements (syndrome extraction) and feedback (Pauli correction), 
where the feedback is determined by a classical decoding algorithm.
Often this decoding algorithm uses all available syndrome information
in a centralized manner,
and hence it is difficult to implement with strictly local operations.%
\footnote{This is why the result of~\cite{balasubramanian2024local}, which constructs a local cellular automaton decoder for the toric code, is interesting.}
However, the Aharonov--Ben-Or scheme~\cite{aharonov1997fault} uses
operations that are geometrically local in a 1d array of qubits.
Crucially, this does not use any measurements but only unitary gates
and fresh ancillas, which our controllable circuit can provide.

The high level mechanism of the Aharonov--Ben-Or scheme is simple.
Imagine that we implement a Pauli stabilizer code with parameters, say, $[[n,k,d]]=[[17,1,5]]$.
The syndrome measurement and error correction can be done by some circuit~$C$
consisting of standard Clifford operations and Toffoli gates.
If a conventional error correction circuit uses measurements and feedback,
we implement it using isometries and controlled unitaries
where the decoding or error correction logic is implemented by Toffoli gates, with entropy removed continually by introducing fresh ancillas.
This might seem complicated, as the error correction circuit will contain many non-Clifford operations,
but $C$ is anyway a constant-sized circuit
because it is a code on $17$ qubits.
To have a quantum memory, we apply~$C$ repeatedly,
protecting one logical qubit for a time that is
longer than the coherence time of any physical qubit
(provided, of course, the physical error is sufficiently weak).
Then, we recursively simulate this constant-sized circuit
by concatenating quantum error-correcting codes.
At the $\ell$-th level of concatenation,
we need $\exp(\ell)$ qubits for a circuit~$C_\ell$, and
the error rate of the top level logical qubit is suppressed
doubly exponentially in~$\ell$.
The constraint of 1d geometric locality
entails overhead that is polynomial in the size of~$C_\ell$, 
which is still $\exp(\ell)$,
and increases the logical error rate compared to a circuit with nonlocal gates.
However, such damage and overhead is benign in view of the doubly exponentially decreasing 
logical failure rate.
The overall space and time overhead is $\polylog(1/\epsilon)$
for an idling logical qubit with logical failure rate~$\epsilon$
since we take $\ell \sim \log \log (1/\epsilon)$.

While conceptually simple,
this scheme may not have a competitive accuracy threshold for quantum error correction.
It would be possible to obtain higher thresholds
by local measurement and processing in classical degrees of freedom,
which are much more reliable than their quantum counterpart.
However, we do not consider this optimization here as the asymptotics will not change.
We comment on such optimizations in~\zcref{sec:discussion}.

\subsubsection{Quantum memory: 1+1d $\to$ 2+0d}
\label{sec:memory-2d}

We implement the 1+1d circuit~$C_\ell$
of unitary gates and fresh qubit preparation gates on $n_\ell$ qubits
which keeps one logical qubit for a long time
using the ever-moving slots.
Let $d_\ell$ be the depth of the circuit~$C_\ell$;
by design, $d_\ell = \exp(\ell) = \polylog(1/\epsilon)$.
We protect the logical qubit by repeating~$C_\ell$
with failure probability~$\epsilon$ every $d_\ell$ time steps. 

We choose the level~$\ell$ of the concatenation%
\footnote{Though we use a concatenated scheme here, 
any 1+1d fault-tolerant circuit for quantum memory can be used.}
when we fabricate our chip,
and put a level-$\ell$ concatenated code, which is our {\bf computation code},
on each slot of the tape.
Each slot contains any necessary ancilla qubits for error correction.
In the noise-free construction of quantum memory above,
a slot contains just one qubit,
but in this fault-tolerant version we need $n_\ell = \exp(\ell)$ physical qubits on each slot.
Note that $n_\ell$ includes all ancillas needed to implement the concatenated code.
Our tape thus has width~$n_\ell = \polylog(1/\epsilon)$.
For the moment, we assume the tape consists of $d_\ell$ stations
so the tape consists of exactly $2 n_\ell d_\ell$ physical qubits in total.

We construct a depth-3 (non-unitary) circuit~$U$ to implement $C_\ell$ along our tape.
The first two layers of~$U$ consist of unitary swap gates across neighboring stations,
propagating the slots forward;
they have nothing to do with the computation code.
If $G_t$ is the set of all gates of~$C_\ell$ at time~$t$,
which must be all geometrically local in 1d and pairwise nonoverlapping,
then the third layer of~$U$ consists of~$G_t$ placed at station~$t$.
Since our slots move along the tape,
the station index~$t$ \emph{is equal to} the time index~$t$ of~$C_\ell$
and every slot is acted on by~$C_\ell$ in every round.

For the proof that our scheme yields a fault-tolerant quantum memory,
detailed in Sec.~\ref{subsec-1DFT},
one must verify that faults in the 2d layout map to correctable errors in the 1+1d scheme of~\cite{aharonov1997fault}.
For this purpose, it is helpful to note that
each swap gate in the first two layers of~$U$ acts not only within a single tape,
but also within a single \emph{lane} in the tape.
Hence faults in these swap gates produce only single-qubit errors in the corresponding 1+1d circuit,
errors that are handled deftly by the error correction circuit~$C_\ell$.
Faulty swaps may generate correlated errors across different slots,
but these are corrected independently in each slot.

\begin{figure}
	\centering
    \begin{subfigure}{0.45\textwidth}
        \centering
        \includegraphics[width=\linewidth]{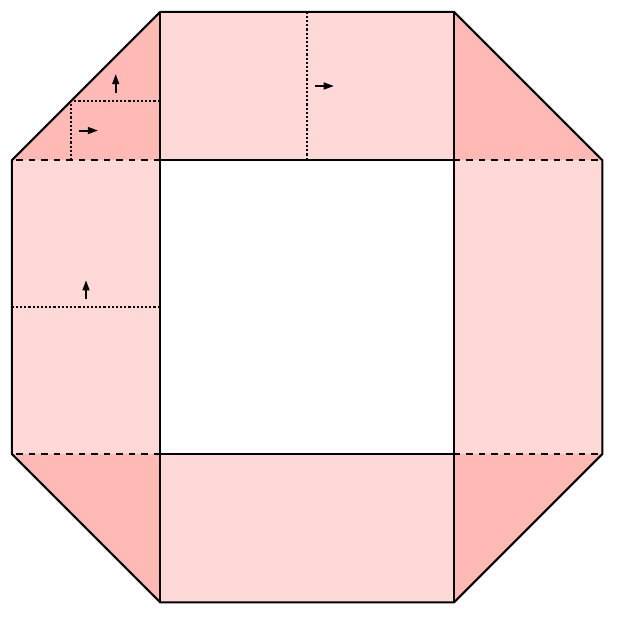}
	    \caption{Slots (small dashed lines) propagate through a tape (red shaded region) in a circle.
		The tape is folded over at the corners to implement the turn.
        }
        \label{fig:ribbon-loop}
    \end{subfigure}
    \hfill
    \begin{subfigure}{0.45\textwidth}
        \includegraphics[width=\linewidth]{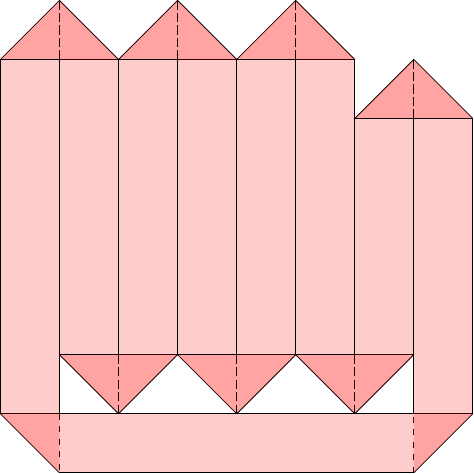}
        \caption{Layout of memory tape arranged into a nearly square configuration.
        }
        \label{fig:mem-layout}
    \end{subfigure}
    \caption{Immersion of a wide tape with a periodic boundary condition in 2d.}
\end{figure}

We have chosen the length~$W$ of the tape to be $d_\ell$ for clarity,
but we can take $W$ to be any integer multiple of~$d_\ell$.
The number of logical qubits is exactly~$W$ since each slot encodes one,
so we retain the encoding rate of the 1+1d fault-tolerant quantum memory.
The overall dynamics of the tape under~$U$ is time periodic with period~$3$,
independent of the tape length~$W$ or the choice of the computation code.
To make the dynamics geometrically local in a plane
we can ``fold'' some part of the tape as in 
Figures~\ref{fig:ribbon-loop} and~\ref{fig:mem-layout}.
Thus we arrive at a time-periodic circuit 
where one period consists of 3 layers of gates with 2d geometric locality,
implementing a quantum memory with memory time~$\tau$
and encoding rate $1/\polylog(\tau)$.

\subsubsection{Logical gate tiles}
\label{sec:gate-tiles}

The Aharonov--Ben-Or scheme is not just for quantum memory,
but also for logical operations.
In fact, for a concatenated scheme to function as a quantum memory
\emph{without feedback},
classical logic that decodes error syndromes needs to be simulated within the scheme.
This is most easily achieved 
by designing computationally universal logical gates
at every level of concatenation.

Hence, from the quantum computational universality of the Aharonov--Ben-Or scheme,
we can certainly implement the following operations:
\begin{itemize}
    \item[(i)] logical unitary $H$ on a slot on a tape,
    \item[(ii)]
    logical unitary $CS$ on a pair of neighboring slots on a tape,%
    \footnote{%
        The Aharonov--Ben-Or scheme explicitly constructs gadgets for \(\gateTOFF\), \(\gateS\), and preparation of logical \(\ket{0}\) which suffice to implement \(\gateCS\) acting on a pair of slots
        side-by-side on two different tapes.
        This can be easily adapted for our version in which
        we implement $CS$ on a pair of neighboring slots on a tape:
        Swap out two slots from the memory tape to (short) auxiliary tapes,
        let one of the tapes go along a longer route so that the two slots meet side-by-side,
        apply $CS$ across the two tapes,
        and finally return the slots to their original positions on the memory tape.
        We elaborate on this in~\zcref{sec:operations}.
    }
    \item[(iii)] preparation of a logical state~$\ket 1$ on a slot on a single tape,
    \item[(iv)]
    logical $Z$-measurement of a slot on a tape.
\end{itemize}
In addition, if we have multiple identical tapes, we can implement
\begin{itemize}
    \item[(v)] a logical Toffoli gate on 3 slots, each on a different tape.
\end{itemize}
Each operation is a 1+1d circuit with physical unitaries and measurements,
transcribed onto our 2d tapes using a periodic depth-3 nonunitary circuit. 

As in the noise-free version described in \zcref{subsec:noise-free},
we introduce an auxiliary tape in addition to a memory tape,
and the operations~(i), (ii), (iii), (iv) are implemented on 
designated collections of stations on the auxiliary tape.
A logical qubit on the memory tape experiences a logical operation if and only if
it is swapped to the auxiliary tape.
We control this swap by a control bit propagating on a control tape
that is parallel to the memory and auxiliary tape.
The controlled swap gate is a composition of three Toffoli gates in~(v).
There are 6 tapes so far: the memory, the auxiliary, and the four control tapes.
There must be one more tape to carry the measurement outcome from~(iv).
This measurement outcome will require an additional detail described below,
which uses~(iii) and~(v).

\subsection{Robust control and readout}
\label{subsec:robust-control}
The core of our quantum processor is now constructed. We have a fault-tolerant memory tape and logical operation tiles that provide quantum computational universality, where a tile acts on a logical qubit if and only if a control bit on a control tape routes the logical qubit through the tile.
In the noisy setting, these control bits must themselves be logical bits encoded in the computation code; otherwise a single physical fault on a control tape could misroute a logical qubit, producing a logical error. The remaining tasks, therefore, are ($a$) converting the classical bits streamed down the external wires into logical bits protected by the computation code, and ($b$) extracting qubit measurement outcomes that can be reliably transmitted to the user. 

This conversion poses a challenge for our architecture. We insist on a \emph{constant} number of external wires, each attached to a single physical qubit. A block of the computation code, however, is asymptotically large, occupying $n_\ell = \polylog(1/\epsilon)$ physical qubits.
Each external bit enters the chip as the state of one bare physical qubit. This vulnerable single qubit can be encoded into a robust qubit protected by the computation code, but the encoder is exposed to noise and is therefore imperfect. Consequently, every bit entering the quantum chip suffers an error of strength~$\Omega(p)$, where $p$ is the physical noise rate.

Our solution is to concatenate a classical code
on top of the computation code.
The {\bf user} ---
the external party that supplies the program and receives the results,
assisted by an ordinary classical computer ---
encodes the instruction sequence
in a classical {\bf program code} before streaming it.
Each arriving bit is immediately encoded
into the computation code by an input {\bf interface tile}.
This encoding is itself noisy
and cannot correct the $\Omega (p)$ error
that the bit has already acquired,
but it \emph{freezes} the error:
once a bit lives in the computation code,
it is no longer degraded as it travels along the tape.
A {\bf control tile} then decodes the program code,
operating entirely at the logical level of the computation code,
which removes the frozen errors
and writes a reliable instruction stream onto the control tapes.
In effect, the instructions are protected by two codes:
the computation code below handles all noise
\emph{after} the interface,
and the program code on top handles all noise
\emph{at and before} the interface.

One more ingredient is needed before the program code can be decoded.
The bits arrive in a stream with no delimiters,
so the decoder does not a priori know
where one codeword ends and the next begins;
nor does the chip know when the computation is supposed to begin.
Both problems are solved by a {\bf metronome tile},
which recovers a periodic framing signal
from a dedicated external wire by majority voting.
The beginning of the computation needs no separate signal:
until the program begins, the user streams encoded instructions directing the chip to idle.
The metronome tile should not be confused with
the global clock, which we assumed to be perfect:
the clock synchronizes physical gates at the sub-nanosecond scale,
whereas the metronome merely marks
where program codewords begin in the bit stream,
and is itself implemented by noisy components.

Readout is the reverse of the input path.
A logical measurement result must eventually leave
the protection of the computation code
and exit through the same narrow interface,
again suffering an $\mathcal O(p)$ error per bit.
So, before ejection, the result is redundantly encoded
by a classical {\bf readout code},
with all encoding operations performed at the logical level
where they are reliably protected by the computation code.
Only then is each bit unencoded down to a single physical qubit,
measured, and streamed out by an output {\bf interface tile};
the user decodes the readout code
to recover the result.

In short:
inbound bits are encoded at the physical level
and decoded at the logical level;
outbound bits are encoded at the logical level
and decoded outside the chip.
Errors are incurred only at the two physical-level ends,
and the classical codes are decoded only where it is safe to do so.
Figure~\ref{fig:control-pipeline} summarizes the pipeline.
The following subsections describe each tile in turn.

\subsubsection{Interface tile: encoding classical bits into the computation code}
\label{sec:interface-tile}

Every classical bit entering the chip ---
whether an instruction bit or a metronome bit ---
must be converted into a logical bit of the computation code
before it can participate in any logical operation.
The {\bf interface tile} performs this conversion.
Each incoming external line is attached to a single physical qubit
at a fixed {\bf entrance station};
we describe the tile serving one line,
and there is one such tile for each incoming line.

The conversion reuses our mapping of time to space.
Associated with the computation code there is a 1d local
{\bf encoding circuit},
of depth $\polylog(1/\epsilon)$ depending on the concatenation level~$\ell$,
that maps a single-qubit state
$\ket \psi = \alpha \ket 0 + \beta \ket 1$,
padded by fresh ancillas as
\begin{equation}
    \underbrace{\ket{\psi} \otimes \ket 0 \otimes \cdots \otimes \ket 0}_{n_\ell}\,,
\end{equation}
to the logical state $\alpha \ket{\bar 0} + \beta \ket{\bar 1}$.
Just as the memory circuit~$C_\ell$ was transcribed onto the tape,
the encoding circuit is transcribed onto the third layer
of the period-3 circuit over the tile:
the gates at time step~$t$ of the encoding circuit
are placed at the $t$-th station of the tile,
so a slot is acted on by the entire encoding circuit
as it travels through.
The tile thus spans a number of stations equal to
the depth of the encoding circuit, $\polylog(1/\epsilon)$.

The tile operates as a pipeline, one bit per tick.
When a slot arrives at the entrance station,
its first physical qubit is prepared in $\ket 0$ or $\ket 1$
according to the bit currently on the external line ---
this classically controlled preparation is the only point in the design
where an incoming external signal acts on a physical qubit ---
and the remaining $n_\ell - 1$ qubits of the slot are supplied
in the state~$\ket 0$ by reset gates at the entrance station.
The slot then propagates through the tile
and emerges carrying the logical state $\ket{\bar 0}$ or $\ket{\bar 1}$,
while the slot immediately behind it carries
the next bit of the stream.

Both the preparation and the encoding circuit
consist of noisy physical gates.
Nevertheless, as we state in \zcref{assume:concatenated-codes},
if the physical error rate is~$p$,
the emerging slot carries the intended logical state
flipped with probability at most~$c p$,
followed by residual local stochastic noise
of the kind the computation code corrects during transport;
crucially, the coefficient~$c$ is independent of~$\ell$.
Since the incoming bits are classical
($\ket\psi = \ket 0$ or $\ket 1$),
the guarantee takes a simple form:
each bit enters its code block correctly,
except for a bit flip with probability~$\Theta(p)$.

The interface tile thus \emph{freezes} the errors on the incoming bits
rather than eliminating them:
the flip a bit has suffered by the time it is encoded is locked in,
but no further error accrues at the physical rate,
because from this point on the bit is processed only
at the logical level of the computation code.
The subsequent metronome and control tiles
are therefore best described as logical circuits acting on encoded bits;
they contribute only the suppressed failure probability~$\epsilon$
per logical gate, which is part of the overall error budget.
Removing the frozen flips is the task of the program code,
decoded by the control tile as described below.

\subsubsection{Metronome tile: where the program codewords begin}
\label{sec:metronome-tile}

Recall from the overview above that the instruction stream
arrives encoded in the program code,
a classical code of length~$\Nprog = \polylog(1/\epsilon)$
fixed when the chip is fabricated,
and that no codeword can be decoded
until the chip knows where codewords begin.
The metronome tile supplies this framing information.
The convention is set on the user's side:
the user streams every program codeword so that
its first bit coincides with a pulse of the metronome sequence
described below.
It is this convention that later lets the control tile trust
the frame bit accompanying each window of instruction bits.

The mechanism is as follows.
\begin{enumerate}
    \item On a dedicated external line,
    the user constantly streams the {\bf metronome sequence},
    the $\Nprog$-periodic bit sequence with one period
    \begin{equation}
        1 \underbrace{00 \cdots 0}_{\Nprog - 1}\, .
    \end{equation}
    Like every incoming bit, each metronome bit is encoded
    into the computation code at the interface tile,
    freezing its error,
    before reaching the metronome tile.
    \item For every position~$s$ of the incoming stream,
    the metronome tile takes the majority vote of the $\Nprog$ bits
    at positions
    \begin{equation}
        s, \quad s+\Nprog, \quad \ldots, \quad s + (\Nprog-1)\Nprog,
        \label{eq:markerLooksAt}
    \end{equation}
    which are all equal in the absence of errors,
    and writes the result at position~$s$ of the metronome tape.
    The output is therefore the error-free $\Nprog$-periodic sequence,
    with all votes computed at the logical level of the computation code.
\end{enumerate}

\begin{figure}
    \centering
    \includegraphics[width=0.7\linewidth]{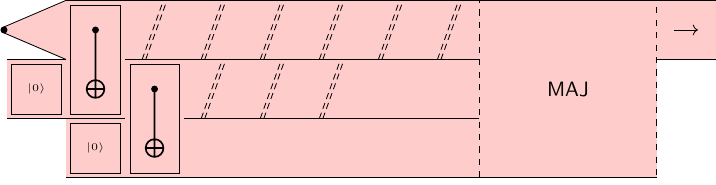}
    \caption{A stream of bits arrives at top left of the metronome tile,
    which computes the majority function on a subset of $\Nprog$ bits 
    specified in~\eqref{eq:markerLooksAt} out of $N_\text{prog}^2$ bits.
    In this case, $\Nprog=3$.
    The slanted dashed line in the figure denotes time delay/advance 
    whose role is to align time-separated slots 
    and make them arrive at the majority logic circuit
    at the same time.
    A similar alignment technique 
    is also used in the control tile
    that implements a decoder for the program code.
    }
    \label{fig:marker}
    \label{fig:metronomedecoder}
\end{figure}

\noindent
There are more qubit-efficient designs,
but for simplicity our metronome tile uses $N_\text{prog}^2= \polylog(1/\epsilon)$ internal tapes
whose delay lines bring the $\Nprog$ time-separated bits
of~\eqref{eq:markerLooksAt} to the majority circuit simultaneously;
see Figure~\ref{fig:marker}.

Each bit entering a vote carries an independent frozen error
of rate~$\mathcal O(p)$,
and no further error accrues at the logical level
(beyond the logical error budget~$\epsilon$),
so a vote outputs the wrong bit only if
at least $\Nprog/2$ of its $\Nprog$ input bits are corrupted,
which occurs with probability exponentially small in~$\Nprog$.
Taking the union bound over all votes during the computation,
the choice $\Nprog = \polylog(1/\epsilon)$ ensures that
every frame bit produced over the entire run is correct,
except with probability at most~$\epsilon$.

\subsubsection{Control tile: decoding the program code}
\label{sec:control-tile}

We use a program code of encoding rate~$1/4$.
The specific rate $1/4$ is not important,
but it is convenient to have an encoding rate that is the inverse of a fixed integer.
Many constructions are known with such a constant rate and linear minimum distance~\cite{SipserSpielman,zemor2002expander,GuruswamiIndyk}.
Then, by processing $4$ program code blocks in parallel 
(the inverse of the program code encoding rate)
we will be able to deliver one bit of logical instruction per tick.

The task of a control tile is to convert the encoded instruction stream
into reliable logical bits on the control tape;
these bits then route logical qubits through the logical operation tiles.
The control tile receives two inputs:
the instruction bit stream produced by the interface tile,
which the user has encoded in the program code,
and the framing bit stream produced by the metronome tile.

The main difficulty is alignment.
The action at each station is independent of upstream and downstream data;
it simply performs the same action over and over.
The instruction stream is just a sequence of bits with no delimiters,
so the control tile applies the decoding circuit
to \emph{every} window of $\Nprog$ consecutive bits.
Only one in every $\Nprog$ windows coincides with an intended codeword;
for the other windows, the decoder's output is meaningless.
The decoding circuit itself cannot distinguish the two cases ---
but the metronome bit can:
by construction, the metronome bit~$m$ accompanying a window
equals~$1$ exactly when the window is an intended codeword.

Concretely, as in \zcref{fig:marker},
the control tile uses $\Nprog+1$ internal tapes with advance tiles
so that the metronome bit~$m$ and the $\Nprog$ bits
$c_1, \ldots, c_\Nprog$ arrive at the decoding circuit simultaneously.
If $c = \mathsf{PROG}(d)$ is any codeword of the program code
corresponding to the message~$d$,
the decoding circuit implements the function
\begin{equation}
    (m, c_1, \ldots, c_\Nprog)
    \mapsto
    \begin{cases}
        \big(d_1, \ldots, d_{\Nprog/4}\big)&\text{if } m =1,\\
        \big(0, \ldots, 0\big) & \text{if } m=0.
    \end{cases}
\end{equation}
We now must turn the block of decoded bits exiting the decoder simultaneously
into a stream of consecutive bits.
To do this, we stagger the output with delay lines such that \(d_i\) receives a delay of \(i\).

Since we process 4 program code blocks in parallel,
there are 4 control tiles producing streams of decoded instructed bits
\begin{equation}
    \begin{matrix}
    (&d_1^{(1)}, \ldots, d_{\Nprog/4}^{(1)},& 0, \ldots, 0,& 0, \ldots,0,& 0 \ldots, 0&), \\
    (&0,\ldots, 0,& d_1^{(2)}, \ldots, d_{\Nprog/4}^{(2)},& 0, \ldots,0,& 0 \ldots, 0& ),\\
    (&0,\ldots, 0,& 0, \ldots, 0,& d_1^{(3)}, \ldots, d_{\Nprog/4}^{(3)},& 0 \ldots, 0& ),\\
    (&0,\ldots, 0,& 0, \ldots, 0,& 0, \ldots, 0,& d_1^{(4)}, \ldots, d_{\Nprog/4}^{(4)}& ).
    \end{matrix}
\end{equation}
(Note that each bit here is encoded in the computation code.)
These 4 streams can easily be combined into a single stream
by $\gateCNOT$ gates targeting the control tape.

The control tile spans $\mathcal O(\Nprog D d_\ell)$ stations,
where $D$ is the depth of the decoding circuit,
plus a few ancillary tapes internal to the tile.
To keep the tile small, we therefore require a program code
of constant rate and linear minimum distance
whose decoder corrects a constant fraction of errors
and is implemented by a circuit of size polynomial in~$\Nprog$.
The expander code~\cite{SipserSpielman} already fulfills the requirement.
Since each bit entering the decoder carries an independent error
of rate~$\mathcal O(p)$,
the probability that a codeword is decoded incorrectly
is exponentially small in~$\Nprog$,
and choosing $\Nprog = \polylog(1/\epsilon)$
makes the overall failure probability at most~$\epsilon$.

\subsubsection{Readout tile}
\label{sec:readout-tile-intro}

It remains to implement the logical $Z$-measurement~(iv)
and to deliver its outcome to the user.
Outcome bits exit the chip through the readout line
as bare physical bits, and,
just as at the input interface,
each exiting bit suffers an error of strength~$\Omega(p)$.
The remedy mirrors our solution for the input path:
the outcome is redundantly encoded by a classical code
\emph{before} it leaves the protection of the computation code,
and the redundancy is decoded by the user outside the chip.
For concretness we may take this {\bf readout code} to be the program code,
though nothing forces the two codes to coincide.
Our protocol never performs a logical measurement on-chip;
instead, the logical $Z$-measurement is realized as follows.
(Here, we also use the metronome stream to do the same process as the control tile in reverse;
otherwise, we will not be able to have delineated frames.)

\begin{enumerate}

    \item At the logical level of the computation code,
    apply the encoding circuit of the readout code
    to the state\footnote{%
        In \zcref{sec:quantum-only-processor},
        we will actually encode many logical qubits per block,
        but we describe here a simplified version.
    }
    $\ket{\bar \psi} \otimes \ket{\bar 0} \otimes \cdots \otimes \ket{\bar 0}$
    of $\Nprog/4$ logical qubits,
    where $\ket{\bar \psi}$ is the slot to be measured
    and the fresh $\ket{\bar 0}$ slots are supplied by
    the preparation operation~(iii).
    Here, only one of the $\Nprog/4$ information positions is used;
    the others are set to~$\bar 0$.
    Since the encoding circuit consists of logical CNOT gates,
    it fans out the $Z$-basis information of~$\ket{\bar\psi}$
    across $\Nprog$ slots without violating no-cloning:
    the resulting state is entangled,
    and it is only the eventual $Z$-measurement of every slot
    that renders the outcome classical.
    
    \item The $\Nprog$ slots carrying the encoded outcome
    travel in order along the memory tape
    toward the exit region of the chip.
    Much like how the decoder must use the metronome
    to delineate blocks and recombine the output into a single stream of data,
    we must do the same here.
    The output is \(0\) when the metronome is \(0\)
    and otherwise it is the encoded data.
    Different tapes receive different delays,
    and the results are combined together such that
    one bit of the encoded codeword
    is presented to the output at a time.
    
    \item Each slot in the combined train passes through the unencoding
    ({\bf eject}) circuit of the computation code,
    which acts at the physical level
    and leaves the logical state of the slot
    on the first physical qubit of the slot. 
    
    \item The qubit is then immediately measured in the $Z$-basis
    at the station attached to the external readout line,
    so one physical bit exits per slot,
    in a fixed order and at ticks known to the user.
\end{enumerate}

\noindent
Every operation in Steps~1 and~2 is a logical operation
of the computation code
and thus fails only with probability~$\epsilon$ per logical gate,
which is part of the overall error budget.
The eject circuit in Step~3 is noisy,
but the computation code guarantees that
the exposed qubit carries error at most~$c p$,
where $p$ is the physical noise strength
and the coefficient~$c$ is independent of the concatenation level~$\ell$;
since the exposed qubit travels only a constant number of stations
before being measured,
its total exposure adds only $\mathcal O(p)$ more.
The $\Nprog$ exiting bits therefore form a codeword of the readout code
in which each bit is flipped with probability~$\mathcal O(p)$,
so the user's decoding fails with probability
exponentially small in~$\Nprog$.
The user identifies which exiting bits belong to which measurement
by counting clock ticks relative to the metronome signal applied by the user,
since the schedule determines when each activated readout completes.

\subsubsection{Summary}

Classical bits flow through the chip as follows (Figure~\ref{fig:control-pipeline}).
Every external control line is attached at the interface tile,
where each incoming bit is encoded into the computation code.
The metronome tile, which is fully autonomous,
recovers from a dedicated periodic input stream
the framing bit that marks where program codewords begin.
The control tiles decode the program code at the computation code's logical level
and write the decoded instruction bits onto the control tapes.
There are four logical operation tiles
($H$, $CS$, $\ket{\bar 1}$, $M_{\bar Z}$);
the bit on the corresponding control tape determines whether
a logical qubit is routed through each of them.
The readout tile encodes each measurement outcome
in the readout code at the logical level,
ejects it to the physical level,
and sends it out on the readout line,
where it is decoded by the user.

The external lines are counted as follows.
A slot arrives at each operation tile once per tick,
so each of the four operation tiles consumes one routing bit per tick.
Since the program code has rate $1/4$,
each operation tile is served by four instruction lines
carrying program-encoded streams in parallel,
each decoded by its own control tile:
hence there are 16 instruction lines and 16 control tiles in total.
One further line streams the metronome sequence,
which is shared by all control tiles,
and one outgoing line carries the readout stream,
for a total of 18 external lines.
(In the later sections,
we have 4 lines for measurement, 1 metronome line,
and we encode the instruction in binary
so that there are only $3 \times 4$ lines used for control tiles.)

The overhead is as follows.
Every tile occupies $\polylog(1/\epsilon)$ stations,
where $\epsilon$ is the target error rate per logical gate,
and every tape has width $n_\ell = \polylog(1/\epsilon)$.
A slot on the memory tape carries one logical qubit,
and there are $\mathcal O(1)$ tapes in total ---
one memory, one auxiliary, four control tapes,
and one tape carrying measurement outcomes.
One slot of the memory tape carries one logical qubit accessible to the user. The chip therefore stores $W$ logical qubits in $\Theta(n_\ell W) = W \polylog(1/\epsilon)$ physical qubits, an encoding rate of $1/\polylog(1/\epsilon)$.

\subsection{Faster quantum processor with classical broadcast wires}

Since only a constant number of external lines control the processor, instructions arrive at a constant rate. Therefore, a circuit with $G$ logical gates must be executed serially, requiring depth $\Omega(G)$. Furthermore for each logical gate, naming which of $W$ stored qubits to act on requires $\Theta(\log W)$ bits delivered through the constant-rate lines.
Thus, the time overhead for each logical operation is \(\Omega(\log W)\) ticks.\footnote{Here, we ignore the \emph{latency} of \(\polylog(W/\epsilon)\) required to \emph{complete} each logical operation, as the number of gates per circuit layer is usually much more than this. In this case we can pipeline logical operations to hide the latency.}

In fact, though, the time overhead of the processor we have constructed so far is much worse than that, because all the logical qubits are stored in a single long one-dimensional memory tape. To perform a logical operation on a particular slot, one might need to wait time $\Theta(W)$ for that slot to propagate all the way around the tape and reach the operation tile.
Fortunately, the undesirable factor of $W$ can be substantially reduced if we make physically well motivated modifications to our hardware assumptions and alter the architecture accordingly, while maintaining a constant number of external control lines.

A more time-efficient quantum architecture follows the layout of a classical computer, in which a CPU interfaces with a random-access memory. The long memory tape is broken into a square array of smaller \textbf{memory cells}, each a fault-tolerant memory on a small circular tape holding~$\polylog(W/\epsilon)$ logical qubits. One distinguished cell in the corner of the array, the \textbf{primary cell}, contains the machinery discussed previously: the interface, metronome, control, and logical operation tiles. This primary cell is the only place where logical gates are applied and external lines are attached. 

To execute a logical gate, the target logical qubit(s)
are teleported from their memory cells to the primary cell,
acted on by operation tiles, and then teleported back.
To fuel the teleportation,
Bell pairs shared between neighboring memory cells are continually prepared,
and Bell measurements are regularly performed to generate long-range entanglement
via entanglement swapping and to complete the teleportation steps.
In contrast to our architecture with just one long memory tape,
the communication time between memory locations and operation tiles
depends only polylogarithmically on the total memory size~$W$
(and the target accuracy~$\epsilon$ per logical operation).

Preparing logical Bell pairs shared between neighboring cells,
and performing Bell measurement on neighboring cells,
can be achieved with the tools we have already developed.
The crucial new hardware requirement in this time-efficient architecture
is rapid long-distance classical communication between distantly separated cells,
as well as local classical processing, to execute entanglement swapping and teleportation.
To fulfill this requirement, we introduce \textbf{classical wires} in the plane of the chip
that broadcast classical bits.
A bit can be set at one point on the wire and then
becomes readable everywhere along the wire.
We assume:
\begin{itemize}
    \item Bits on classical wires are noiseless.
    \item Classical communication along wires is instantaneous.
    \item Classical circuits that read from classical wires and output on classical wires are perfect.
    \item When a wire is written to by multiple classical gates,
    the value of the wire is the \(\mathsf{OR}\).
\end{itemize}
Though surely idealized,
these assumptions are defensible in that
on-chip components of today's classical processors
are far more reliable than foreseeable quantum gates,
and signals can propagate across such chips
much faster than quantum-processor clock speeds.

In this time-efficient scheme,
geometric locality is violated only by the wires themselves, and mildly:
signals on a wire are available along its whole length,
but every wire occupies a fixed position,
no more than a constant number of wires pass through any unit square of the chip,
and all quantum operations remain strictly local.
The external interface is also unchanged:
we have the same constant number of external lines,
attached to the primary cell,
and the same concatenated protection of instructions and outcomes
as in the architecture we described previously.

\begin{figure}
    \centering
    \includegraphics[width=0.7\columnwidth]{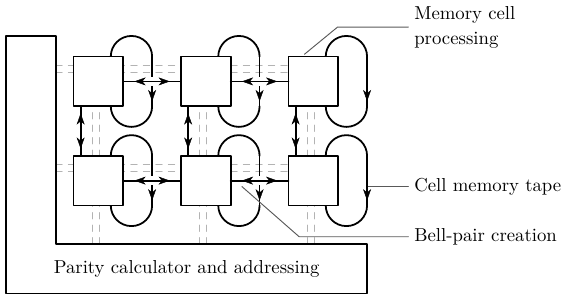}
    \caption{Sketched layout of the array of memory cells with addressing and parity calculations at the periphery. Classical wires are drawn as dashed lines. See Figure~\ref{fig:memory-cell} for more detail.
    }
    \label{fig:layoutWithClassicalWire}
\end{figure}

The remainder of this section describes
(i) the layout of the cell array and the classical wires (\zcref{fig:layoutWithClassicalWire}),
(ii) the creation of Bell pairs
between the primary cell and arbitrary memory cells,
and
(iii) the teleportation protocol.

\subsubsection{Grid of memory cells and logical operation tiles}
In the fault-tolerant construction in Sec.~\ref{subsec:noisy-ops} and~\ref{subsec:robust-control},
we used one long quantum memory tape and a few logical operation tiles.
For our time-efficient construction, we break up the long quantum memory tape
into $W' =W/\polylog(W/\epsilon)$ circular tapes as in Figure~\ref{fig:ribbon-loop},
each of which is called a {\bf cell} and has side length~$a = \polylog(W/\epsilon)$.
We put the cells in a square grid where the spacing between the cells is also~$a$.
So, the cells occupy $1/4$ of the total area of our quantum processor.
(The microscopic spacing between physical qubits is~$1 \ll a$.)
Each cell is a fault-tolerant quantum memory,
maintaining $\tilde {\mathcal O}(a^2) = \polylog(W/\epsilon)$ logical qubits independently.

We chose the spacing between cells to be the same as the cell size $a$ to be concrete, but the precise value of this spacing does not matter. It is important, though that $a$ be large enough for cells to function as a standalone fault-tolerant quantum memory and to house some simple classical logic, which requires $a \gtrsim \polylog(W/\epsilon)$. On the other hand, to minimize the time overhead we don't want $a$ to be any larger than necessary; a logical operation performed inside a cell requires enough time for logical qubits to circulate around the tape.

We interconnect the cells with two kinds of structure.
First, open-ended tapes of qubits, each of length~$\mathcal O(a)$,
run between nearest-neighbor cells, terminating at both ends with open boundary conditions.
These tapes supply logical Bell pairs
shared by neighboring cells (see Sec.~\ref{subsec:bell-creation}). 
Second, classical wires run along each row and each column, 
passing through the cells and extending across the full linear dimension of the grid. 
Each row (and column) carries a constant number of wires
to convey data needed for the entanglement swapping and teleportation protocols.
In addition, there will be $\polylog(W/\epsilon)$ wires along the boundary of the chip
to process the address of cells.
Hence the total number of classical wires is $\mathcal O(\sqrt {W'})$,
and the number of classical wires
passing through any unit square (of area~$1$ not~$a^2$) of our quantum processor
is less than a specified constant.

At the bottom and left boundaries of the grid,
we will have classical logic circuits
used for location indexing and parity computation (see Sec.~\ref{subsec:bell-creation}).
The number of these classical logical circuits is $\sqrt{W} \polylog(W/\epsilon)$; %
they occupy a small fraction of the space of our quantum processor.

We place a special cell at the bottom-left corner of the grid for logical operations.
This {\bf primary cell} contains 
logical Hadamard, $CS$, reset, and readout tiles,
as well as interface and control tiles.
This layout resembles a typical architecture of classical processors
where a CPU and memory are separated.
Each cell will also have its own small classical and quantum logical circuits,
as will be explained shortly.

Suppose that a row of quantum memory cells has
$L = \sqrt{W'}$ cells.
Let $(x,y)$ denote the coordinate of a cell.
By convention, $(0,0)$ is the primary cell at the bottom left,
and is bigger than any other cell.
However, the length of the memory tapes internal to the primary cell and other cells must be consistent, so that there is no complication in synchronization.
In each cell we have $\polylog(W/\epsilon)$ slots, circling along a memory tape.
So, a slot is indexed by a triple $(x,y,s)$ 
with the third component indexing a slot within the memory tape.

\subsubsection{Bell pair creation and cell addressing}
\label{subsec:bell-creation}

At the midpoint of each open-ended tape in between nearest-neighbor cells,
there is a tile that constantly produces Bell pairs across two slots,
which propagate in opposite directions toward the neighboring cells. 
Bell measurement at each cell can then generate Bell pairs shared between 
distantly separated cells via entanglement swapping.
If we have qubits $1,2,\ldots, 2n$, and Bell pairs shared between qubits $2j-1$ and $2j$ for~$j = 1,2,\ldots,n$, 
then we can make a Bell pair between qubit $1$ and $2n$ 
by Bell-measuring across qubits~$2j$ and $2j+1$ for~$j = 1,2,\ldots, n-1$,
and finally applying a Pauli correction at qubit~$1$.
The final Pauli correction depends on the parity of all Bell measurements in the middle of the chain,
which is an entirely classical computation.

As we will teleport logical qubits along the shortest Manhattan path
in the square grid of cells,
we explain how a logical Bell pair is created between cells at~$(x,y)$ and $(0,y)$.
Place an {\bf address router} (a classical logic circuit of size \(O(\log(L))\))
on the left of each cell on the left boundary of the grid
and also on the bottom of each cell on the bottom boundary of the grid,
and a {\bf parity calculator} at the bottom boundary of the grid.
An address router has coordinate $(-1,y)$ with $y \ge 0$ or $(x, -1)$ with $x \ge 0$.
Their role will be specified in the following protocol.
There are as many address routers as there are rows and columns of cells in the grid,
but there is just one parity calculator at the bottom of the grid.

\begin{enumerate}
    \item The primary cell chooses a cell at~$(x,y)$
    and broadcasts binary-encoded $x$ and~$y$ through classical wires
    to all address routers.

    \item The address router at~$(-1,y_a)$ sets a \textbf{row selection bit} to~$1$ 
    on a horizontal wire if and only if $y = y_a$;
    all other row selection bit is zero.

    \item The address router at~$(x_a,-1)$ sets two bits on two vertical wires,
    called \textbf{entanglement swapping participation bits},
    to~$01$ if $x_a = 0$
    to~$11$ if $0 < x_a < x$;
    and to~$10$ if $x_a = x$.
    These amount to ``left end'', ``middle,'' and ``right end.''\footnote{In Sec.\ref{sec:FastProcessorWithClassicalWiresFurtherDetail} we use a slightly different encooding that is a bit more efficient.}

    \item  Each cell, 
    if the row selection bit is~$1$ and if the entanglement swapping participation bits are~$11$,
    makes a Bell measurement across two halves of the adjacent Bell pairs
    on the open-ended tapes.
    This measurement is at the logical level, protected by the computation code.
    Each cell immediately broadcasts the two-bit Bell measurement outcome 
    through vertical classical wires.
    If a cell did not make a Bell measurement, then it broadcasts~$00$.
    
    Since the measurement is on the logical qubit,
    we make transversal destructive measurement, 
    from which the noisy outcomes
    are given to a classical decoding circuit of the computation code.
    This decoding circuit has size $\poly(n_\ell) = \polylog(W/\epsilon)$
    and needs to sit inside a cell.\footnote{We could perform the decoding outside the cell,
            but that would require more classical wires per cell.}
    This gives a requirement for the cell size: $a \gtrsim \polylog(W/\epsilon)$. 

    \item The parity calculator at the bottom end of the grid 
    reads bits from all vertical classical wires,
    calculates the parity using a tree of size \(L\log_2 L\), 
    and broadcasts the result through
    another set of vertical wires that can be read from all cells.
    The parity calculator must have classical wires that are horizontally extended 
    just to read intermediate parities in the sum and announce the global parity result.\footnote{In particular, the tree is very wide but very short with reduction operations occurring as the signals are brought towards the center of the tree. This avoids having too many horizontal wires.}
    
    There are $\mathcal O(L)$ bits as input,
    so this calculation circuit occupies much more space than a cell.
    However, this space cost is negligible compared to the space occupied by the quantum memory cells,
    and does not compromise the overall size of our quantum processor or its wire density.
    The parity calculation takes an ($L$-dependent) fixed time,
    so the cells know when to read the parity result.
    Specifically, using a classical circuit with the topology of a tree,
    the parity computation has depth~$\polylog(L)$,
    so the time that the cells wait for the parity result is
    smaller than or comparable to the time for a slot to travel around the memory tape in a cell.\footnote{In fact, we buffer the output of the parity calculator such that the parity appears at the correct time.}

    \item According to the announced parity result,
    a Pauli correction is applied at the leftmost cell.
    This completes the protocol to generate a Bell pair between the leftmost cell and a given cell 
    on the same row.
\end{enumerate}

The open-ended tapes linking neighboring cells are like other tapes in our architecture;
each slot carries a block of the computation code,
which is continually in motion.
Logical qubits, created in entangled pairs at the tape's midpoint,
travel to the cells at the endpoints,
where they either participate in Bell measurement or are discarded,
depending on the instructions relayed by the primary cell.
We require outcomes of logical Bell measurements to be decoded in less time than is needed for slots to circulate around a cell. In addition, the parity computation of depth $\polylog(L)$, which determines the Pauli frame of the long-distance Bell pairs prepared by the protocol, must be executed faster than this circulation time. Both requirements are fulfilled by setting $a = \polylog(W/\epsilon)$.

We have now described how to prepare a Bell pair shared 
between the leftmost cell and an arbitrary cell in a given row.
Through each memory cell we have a constant number of wires:
one row selection bit, 
two entanglement swapping pariticipation bits,
two Bell measurement outcome bits,
and two parity result bits.
An analogous protocol prepares a Bell pair shared
between the bottom-most cell and an arbitrary cell in the same column.
In the latter case, a parity calculator resides on the left boundary.
Hence, the total number of classical wires crossing a given memory cell is a constant.
The wires are not all simultaneously needed,
so one can reduce the count further.

\subsubsection{Logical teleportation and overall design}

The Bell pair generation between cells in a row or a column described above
is a subroutine for logical teleportation.
We aim to transport a logical qubit on a slot that is circling eternally
on the quantum memory tape on one of the memory cells
to the primary cell at the bottom left.
We will describe a modified design in \zcref{sec:FastProcessorWithClassicalWiresFurtherDetail} that reduces the complexity at the cost of reduced conceptual simplicity.

Below is the protocol to teleport a logical qubit from $(x,y,s)$ to
the primary cell.
There is a special classical wire that determines
whether horizontal or vertical teleportation is going to be performed.
We always implement horizontal teleportation first.

\begin{enumerate}
    \item While no teleportation is occurring, all the classical wires are set to~$0$.
    
    \item At a specific time determined by~$s$,
        the main logical operation tile announces 
        the binary-encoded cell index tuple $(x,y)$
        through the classical wires passing the cell.
        
    \item By interaction with the address router and parity calculator as above,
    a Bell pair is created between the cells at $(0,y)$ and $(x,y)$.

    \item The cell at $(x,y)$ makes a Bell measurement between a half of the shared Bell pair
    and the logical qubit at $(x,y,s)$.
    The cell at $(x,y)$ does not know $s$, but
    the position of the specific slot is determined by the duration of the protocol so far,
    which is pre-computed by the user.
    The Bell measurement outcome is broadcast 
    through designated horizontal classical wires
    to the $(0,y)$-cell.

    \item The $(0,y)$-cell looks at the Bell measurement outcome
    from the $(x,y)$-cell, and applies a Pauli correction to complete the quantum teleportation.
    The logical qubit is now in $(0,y)$.

    \item The primary cell announces via a classical wire
    that vertical teleportation is going to be implemented.

    \item The Bell pair creation protocol and cell-local operations 
    complete the quantum teleportation from $(x,y)$ to $(0,0)$.
\end{enumerate}

The primary cell can thus summon one or two logical qubits from arbitrary cells in the quantum memory. 
These qubits are then routed through an operation tile to perform a logical operation. After the logical operation, the logical qubits are teleported back to their original memory cells.
Since the memory tape internal to a cell is short (of length~$\polylog(W/\epsilon)$)),
the time overhead incurred by the length of the internal memory tape is acceptable.
The time needed for the teleportation protocol is the sum of
the time to execute the parity computation and to apply Pauli corrections,
which is $\polylog(L)$.

In Section~\ref{sec:FastProcessorWithClassicalWiresFurtherDetail},
we modify the design so that there is no need to sequentially implement
the horizontal teleportation and vertical teleportation. We also combine entanglement swapping and teleportation into a single unified protocol.
These modifications do not alter any asymptotic space and time cost estimates described in this section.

\subsubsection{Summary}

We have introduced classical wires that can broadcast classical bits instantaneously and noiselessly.
We divided one long quantum memory tape into an array of small quantum memory cells,
each long enough to implement the computation code and so to maintain logical qubits fault tolerantly.
Routers and parity circuits at the boundaries of the cell array coordinate entanglement swapping, 
implementing fast teleportation of logical qubits between any memory cell and the primary cell, 
where all logical operations are performed. 
This architecture enabled by classical wires
has a time overhead improved by an $\tilde{O}(W)$ factor
compared to our architecture with a single long memory tape,
where $W$ is the number of logical qubits stored on the chip.

Crucially, this speedup does not compromise the minimal external control
that is the point of our processor design.
The classical wires and circuits require no instructions from outside the chip,
other than those relayed by the primary cell.
The external interface is the same as in the single-tape architecture,
with the same constant number of lines now attached to the primary cell,
and with incoming instructions and outgoing measurement outcomes
protected by the same program and readout codes.
What changes, and facilitates the speedup,
is that the classical wires allow classical bits
to travel at high speed across the chip.

\subsection{Optimization}\label{sec:discussion}

Our construction has been optimized for conceptual simplicity rather than practicality.
Here we will highlight a few important optimizations that are likely to greatly improve the practical relevance of the construction.

\paragraph{Threshold}

We have used a scheme that coherently performs decoding and correction in the noisy qubits.
If there is a compatible classical information processing technology (e.g. single-flux-quantum logic, cryo-CMOS), the decoding can be performed locally in classical degrees of freedom.
Then the processing remains unit-density, but because classical processing is generally substantially less noisy, fewer parts of the circuit are subject to noise. Furthermore, recall that two of the three layers in our periodic circuit consist only of swap gates that propagate qubits along the tapes; if transport or swapping of qubits has substantially higher fidelity than entangling gates, as is the case in semiconductor spin qubits~\cite{yoneda2021coherent,seidler2022conveyor,de2025high}, then the accuracy threshold of our scheme may approach that of more standard concatenated coding schemes~\cite{knill2005quantum,goto2024many,yoshida2025concatenate}.
Local processing is also still compatible with soft-decision decoding~\cite{poulin2006optimal,goto2013fault}, which may further improve the threshold.
The scheme is also naturally resilient to fabrication defects (\zcref{rmk:fabrication-defects}).

\paragraph{Space}
In our base construction without classical broadcast wires, classical logic used in error correction (syndrome processing, majority voting, conditional corrections) is executed coherently, compiled into non-Clifford quantum gate gadgets. Introducing classical wires allows many of these expensive non-Clifford quantum gates to be replaced by simple classical gates.
For example, the controlled-\(\gateSWAP\) operations used to control whether operations are applied to the tape can be replaced by classically-controlled transversal swap operations.\footnote{The operation is not literally transversal, because of routing to satisfy the geometric locality constraints.}
In fact, the number of locations where non-Clifford gates are applied can be reduced to \(1\) -- all other tiles perform Clifford operations.
Our construction is also compatible with high-rate concatenated code families (\zcref{rem:const-rate}), which substantially reduce the space overhead. For these codes, which house many logical qubits inside each code block, addressing particular qubits within the block simply becomes another instruction added to the gate set.

One can also reduce the minimum size of the construction.
For example, suppose that the concatenated code scheme used is a three-level scheme.
Tiles can be constructed out of level-2 gadgets such that a level-3 code block is formed from multiple consecutive slots on the tape.
Level-3 error correction would be performed by routing slots through level-2 gadgets in such a way as to reuse a level-2 tile for multiple operations in the level-3 gadget. Reusing tiles reduces the number of tiles that must be printed on the chip, saving space.

\paragraph{Time}

For the fast-processing variant of our scheme (\zcref{thm:fast-quantum-processor}), factors that limit speed are similar to those faced by classical processors, which also stream data through a fixed chip. Classical machines can carry heavy workloads at acceptable speed, thanks in part to many constant-factor optimizations developed over decades. Various ideas from classical computer architecture can be invoked to improve the runtime of quantum processors as well. 

For example, instruction sets in classical computers are tailored for rapid execution of frequently used operations~\cite{hennessy2025computer}; in the case of integer arithmetic, a single instruction triggers a sequence of thousands of gates~\cite{goto1997}. Similarly, one could identify small subcircuits that are repeatedly used by quantum circuits; these could be hard-coded on the chip and called by specialized instructions. 
This strategy simplifies control and improves speed at the acceptable cost of increasing the chip area. The runtime of a subcircuit with $c$ gates can improve by nearly a factor of $c$, where a natural value for $c$ might be 10-100.

Another technique that can be fruitfully imported from classical computer architecture is multiprocessing. 
A typical modern CPU runs tens to hundreds of processor cores in parallel with a shared memory. An analogous tradeoff between the complexity of control and the speed of execution applies to our quantum processor. 
If the program admits parallel execution, then increasing the number of external control streams by a factor $m$ can increase the computation speed by roughly a factor of $m$ (see \zcref{rem:more-control}).
In the other extreme, the number of controlled qubits can be brought down to just \(1\) (see \zcref{rem:one_qubit}).

\clearpage

\section{Fault tolerance}\label{sec:fault-tolerance}

We now state the result given 1D quantum fault-tolerance schemes~\zcref{assume:concatenated-codes} which can be instantiated by concatenated codes~\cite{aharonov2008fault,gottesman2000fault}.
\zcref{thm:all-quantum-processor} establishes a blueprint for a scalable, fault-tolerant, universal quantum processor with limited classical control and geometrically local quantum processing in a two-dimensional layout, the model from the previous section.
The processor can implement any quantum circuit from the Clifford+$\gateCS$ gate set with modest overhead in both space and time, yet only qubits in a set $S$ of constant size are directly controlled.
In this scheme, a fixed constant-depth circuit $C$ is applied repeatedly; in between repetitions of $C$, a control operation is applied only to $S$.
The program to be executed is encoded in a sequence of such control operations.
This construction tolerates both transient errors and permanent fabrication defects,\footnote{We will not explicitly prove tolerance to fabrication defects. See \zcref{rmk:fabrication-defects} for further discussion.} enabling robust computation when the rate of faults is below a constant threshold value.

In what follows, we will introduce the following parameter \(w_\ell = 2^{\Theta(\ell)}\) corresponding roughly to the number of qubits in a code block for the underlying quantum code used to ensure fault-tolerance.
The exact definition will be given in \zcref{assume:concatenated-codes}, but \(w_\ell\) can be thought of as logarithmic in the total number of qubits in the circuit to be simulated.
Here we reproduce the theorem statement verbatim at the cost of using some terms defined in later sections.
\getkeytheorem{quantumprocessor}

The starting point for this theorem will be 1D fault-tolerance schemes (with a threshold) based on concatenated codes that use no classical processing \cite{aharonov2008fault,gottesman2000fault}.
At the top level of recursive simulation, these fault-tolerance schemes construct 1D geometrically local gadgets for a universal set of operations.
These gadgets will have growing depth, so we will apply a transformation (\zcref{sec:11Dto20D}) to convert them into a primitive we call ``tiles.''
Tiles are period-3 2D geometrically local circuits (a controllable quantum circuit) that implement the original circuit by successively shifting the information through a second spatial dimension which encodes which gate layer of the original circuit to apply.
The data is simultaneously transported from an input spatial location to an output spatial location along the perimeter of the tile.
Faults in the execution of a tile correspond to faults in the original circuit, so fault-tolerance properties of the tile can be related to fault-tolerance of the original circuit.
The distinction is that the tile is ``hard wired'' and will only execute a fixed circuit.
They will be assembled into a universal circuit which leverages two ideas that tiles make possible: 1) Outputs of tiles can be connected to inputs of other tiles \emph{including itself.} 2) Operations on data can be controlled by controlling its path through different tiles.

The tile possesses \emph{slots} which should be thought of as pieces of the computation (registers of qubits) at various stages of processing.
The slots physically propagate through the tile as they are processed.
We refer to these spatial locations as \emph{stations}.
Data propagates from the input station to the output station at a rate of one station per tick (one application of the clocked circuit).
The time interval between a slot entering at the input station and appearing at the output station is the \emph{delay} of the tile.

Later, in \zcref{sec:FastProcessorWithClassicalWiresFurtherDetail}, we will extend our model of computational to allow for \emph{classical wires}.
Much like a physical wire in a digital processor, the classical wire will allow for fast propagation of classical information along its length.
In particular, we allow the information to be broadcast instantaneously to all positions along the wire.
We additionally introduce classical gates which are part of the clocked circuit.
After every application of the clocked circuit, the classical gate performs a basic operation on the input values (e.g. \(\mathsf{NAND}\)) and writes the output to the output wire.
Both the classical wires and classical gates must have unit density.

The runtime slowdown of the all-quantum processor (\zcref{thm:all-quantum-processor}) is due to the required time to allow quantum information to propagate.
In this extended model, quantum information can instead be teleported, allowing \emph{lower runtime}.
In particular, the time overhead is proportional to the number of gates in the input circuit to be simulated up to factors polylogarithmic in the size of the processor.
We expect the extended model to be practically more relevant since it achieves substantially better time costs and (likely) thresholds.
However, many of the fundamental concepts are present in the proof of \zcref{thm:all-quantum-processor} already.
Many optimizations in the extended model are ignored for conceptual clarity
We again reproduce the theorem statement verbatim at the cost of using some terms defined in later sections.
\getkeytheorem{fastprocessor}

\subsection{Controllable quantum circuit}

We now introduce a notion of a ``controllable'' quantum circuit which captures the degree to which a physical quantum computing device can be controlled.
A controllable quantum circuit can be thought of as a periodic quantum circuit where some \emph{fixed} subset of qubits can be operated on in an arbitrary way.
This subset should be thought of as very small (\(\le 30\)).
One can draw parallels with classical digital processors: Computation in the digital circuitry of classical processors is driven by a periodic \emph{clock signal}, and the device is programmed by applying a signal specifying a program to an external programming pin that is coupled to some internal part of the circuit.

\begin{definition}[Controllable quantum circuit]\label{def:controllable-circuit}
	Let \(C\) be a quantum circuit of depth \(t \in \mathbb{N}\) on \(n\) qubits.
	With respect to a subset of qubits \(S \subseteq [n]\), referred to as the \emph{controllable qubits}, the corresponding \emph{controllable quantum circuit} \(C_{\bullet}\) is the family of classical-quantum circuits parameterized by a length-\(\tau\) gate sequence \(\bar{M} \equiv M_0,\dots, M_{\tau-1}\), referred to as the \emph{control sequence}.
	The family element corresponding to a particular gate sequence \(\bar{M}\) is the classical quantum circuit given by the alternating application of \(C\) and the control sequence
	\begin{align}
		C_{\bar{M}} = M_{\tau-1} C \dots M_1 C M_0 C~.
	\end{align}
	In other words, except for the control sequence, the circuit is periodic with period \(t\).
	The control sequence is restricted to be supported on \(S\) and an auxiliary classical register referred to as the \emph{output register}.

	We say that \(C\) is the clocked circuit of \(C_{\bullet}\).
	The depth of the clocked circuit \(C\) is referred to as the \emph{period} of the controllable quantum circuit \(C_{\bullet}\).
	The layer of \(C_{\bullet}\) after an application of \(C\) is a \emph{tick}.
\end{definition}

A controllable quantum circuit \(C_{\bullet}\) is said to be \emph{geometrically local} if its clocked circuit is geometrically local.\footnote{Gates of the control sequence are required to be geometrically local when considering the controlled qubits, but no locality is required of the gate support on the output register.}
A controllable quantum circuit is said to be \emph{universal} for a set \(\mathcal{A}\) of quantum circuits (with classical input and output), if, for every quantum circuit \(A \in \mathcal{A}\), there exists a gate sequence \(\bar{M}(A)\) of length \(\tau(A)\), such that the output of \(C_{\bar{M}(A)}\) is set to a bitstring that is distributed identically to the output of \(A\).\footnote{Note the independence of the initial state.}
The \emph{runtime} of the simulation is defined to be \(\delta\cdot \tau(A)\) where \(\delta - 1\) is the depth of the clocked circuit.

There are two relevant extremes of the model in \zcref{def:controllable-circuit}.
The first extreme is that of a standard quantum computer.
\begin{example}[Standard quantum circuit model]\label{example:standard-model}
	A standard circuit model quantum computer is a controllable quantum circuit where every qubit is a controllable qubit and the clocked circuit is trivial.
\end{example}

The second extreme is a quantum processor whos operation is controlled by a small subset of qubits much like a classical processor.
The remainder of the paper will be devoted to realizing a fault-tolerant quantum processor along these lines.
\begin{example}[Autonomous quantum processor]\label{example:autonomuous-model}
	The constructions considered from here onwards will possess very few controllable qubits and the majority of the computation will be carried out by the clocked circuit.
	We refer to these as autonomous quantum processors (AQP).
	The computation starts with an arbitrary state of the quantum registers and the classical register initialized to a polynomially-sized bitstring of the input circuit to simulate.
	We will first efficiently and classically compile the input circuit to a time-sequence of instruction bits.
	During execution, at every tick, the (input) controlled qubits will be reset to \(\ket{0}\) or \(\ket{1}\) according to the time sequence of instruction bits.
	At every tick, the output controlled qubits will be measured in the computational basis and the result written to an output tape.
	The output tape will then be efficiently processed to compute the output bitstring.
\end{example}

\subsection{Conventions}

In this paper, to simplify some notation,
we will use a \emph{zero-based indexing} convention similar to the programming language Python.
We define \(\mathbb{N} \equiv \{0, 1, \dots\}\).
For two numbers, \(k, k' \in \mathbb{N}\) with \(k \le k'\), we define \([k,k'] \equiv \{k, \dots, k' - 1\}\)
so \([k,k]\) is the empty set.
When \(k = 0\), we use the notation \([k']\) as shorthand for the set \([0,k']\).
Hence, \([0]\) is the empty set.

In order to state the result, we will need to introduce some notation.
We will use the definition of a quantum circuit from \cite{he2025composable}.
We will reserve quantum circuit to refer to a circuit without any auxiliary classical storage or gates, and use classical-quantum circuit to refer to a circuit that includes classical gates and storage.
To reduce the time overhead of the scheme, in \zcref{sec:FastProcessorWithClassicalWiresFurtherDetail} we will introduce a classical (but geometrically-local) component to the 2D circuits.

Throughout the paper,
a {\bf fault} $\fault$ will mean a collection of concrete superoperators
inserted in an ideal circuit~$C$.
The circuit with $\fault$ inserted will be denoted by $C[\fault]$.
Our superoperators will be slightly unconventional in that
they are not required to be quantum channels
(completely positive trace preserving maps on density operators),
but are always of form $\rho \mapsto \alpha P \rho Q$ where $P$ and $Q$
are possibly different Pauli operators.%
\footnote{
For a quantum circuit \(C\) of depth \(D\), a \textbf{fault} of \(C\) is two length-\(D\) lists of superoperators \((\mathcal{E}_1^{\mathrm{pre}}, \dots, \mathcal{E}_D^{\mathrm{pre}})\), \((\mathcal{E}_1^{\mathrm{post}}, \dots, \mathcal{E}_D^{\mathrm{post}})\) where the circuit \(C = C_D \dots C_1\) subject to the fault \(\fault\) has each gate layer \(t\) replaced by \(\mathcal{E}_t^{\mathrm{post}} C_t \mathcal{E}_t^{\mathrm{pre}}\).
This is a specialization of~\cite[Definition~3.15]{he2025composable} to Pauli faults.
}

\subsection{Skeleton of fault tolerance argument}\label{sec:SkeletonOfArgument}

Given a construction for fault-tolerant gadgets (a fault-tolerant scheme),
one has to show that their arbitrary compositions
work as intended under some weak noise.
It is usually not too difficult to argue that a gadget is fault tolerant,
but it can be substantially more complicated to guarantee its performance
in the presence of other components because of interfaces between gadgets.
Further complication may arise when we concatenate fault-tolerant schemes.
The analysis for a fault-tolerant scheme does not always characterize
the effective noise channel at the logical level,
but the probability distribution of rare events at the logical level
will be needed to analyze the performance at higher concatenated levels.

In~\cite{nguyen2025quantum,he2025composable}
these complications are mitigated by
considering ``bad fault paths'' and associated ``weight enumerators.''
Rather than considering probabilities for some bad events at a gadget
and performing conditional probability estimates,
we consider combinatorial characterizations of bad events
and manipulate the combinatorial objects and weight enumerators.
In this subsection,
we recall relevant notions that will be used later in this paper.

\subsubsection{Fault paths, avoidance, and weight enumerators}
In the majority of the analysis, we will only consider the fault path instead of the fault itself.
Accordingly, we will introduce combinatorial notions.
Let $\Omega$ be a set of spacetime locations where an error may be introduced.%
\footnote{
	Though we declare that $\Omega$ is a set of spacetime locations where errors may occur,
	the definitions here are just operations on the set~$\Omega$.
}
Let $\BB = \{ 0, 1\}$.
A {\bf fault path}~$f$ on~$\Omega$
is a formal linear combination~$f = \sum_{x \in \Omega} f_x x \in \BB[\Omega]$ of spacetime locations
with binary coefficients~$f_x$.
We will not use any algebraic structure (group or field) of~$\BB$,
so it is perfectly fine to regard a fault path~$f$ as a subset of~$\Omega$,
consisting of all elements of~$\Omega$ with coefficient~$1$ in~$f$.
Put verbosely, the coefficient function~$x \mapsto f_x$ is the indicator function
of the subset~$\supp(f) \subseteq \Omega$.
When we explicitly regard a fault path as a set,
we will write $\supp(f)$.
For a set \(\calF\) of fault paths, we define $\supp(\calF) = \bigcup_{f \in \calF} \supp(f)$.
The fault path \(f\) associated with a fault \(\fault\) is
the spacetime support of~$\fault$;
the coefficient $f_x$ is $1$ on every spacetime location where \(\fault\) acts non-trivially,
and zero elsewhere.

We say a fault path~$x$ {\bf avoids} a set~$\calF$ of fault paths
if $\supp(x)$ does not contain $\supp(f)$ for any~$f \in \calF$.
In other words, for every ``bad'' fault path~$f \in \calF$,
the fault path~$x$ leaves out at least one error location of~$f$.
We will often say $x$ is {\bf $\calF$-avoiding}.

The {\bf weight} of a fault path~$f$ is defined to be the cardinality of~$\supp(f)$,
and is denoted by~$\abs f$.
This is consistent with a common notion ``weight of a Pauli error''
which means the number of nonidentity tensor factors of a Pauli operator.
A {\bf weight enumerator} of a set~$\calF$ of fault paths
is
\begin{equation}
	\weightenum{\calF}{y} = \sum_{f \in \calF} y^{\abs f} \, .
\end{equation}
This is a polynomial in a variable~$y$ where
the coefficient of~$y^w$ is the number of fault paths in~$\calF$ whose weight is~$w$.
We will substitute~$y$ with a positive real number or another expression
to bound error probabilities.
Here is a crucial example.
Let $x$ be a random fault path
where every point of~$\Omega$ is independently included in~$x$ with probability~$p$.
Then, for any set~$\calF$ of fault paths,
\begin{equation}
	\Pr[x \text{ is not $\calF$-avoiding}]
	\le
	\sum_{f \in \calF} \Pr[f \subseteq x]
	\le
	\sum_{f \in \calF} p^{\abs f}
	= \weightenum{\calF}{p} \, .
\end{equation}

\subsubsection{Sums and products}

Let $\calF_1$ and $\calF_2$ be some sets of fault paths.
They are not necessarily linear spaces of fault paths.
We define their {\bf sum} as
\begin{equation}
	\calF_1 \wtsum \calF_2 = \calF_1 \cup \calF_2 \, .
\end{equation}
It follows from definitions that for all~$y \ge 0$
\begin{equation}
	\weightenum{\calF_1 \boxplus \calF_2}{y} \le \weightenum{\calF_1}{y} + \weightenum{\calF_2}{y}
	\label{eq:wtsumInequ}
\end{equation}
where the equality holds if $\calF_1$ and $\calF_2$ are disjoint.
Note that $\calF_1$ may be disjoint from~$\calF_2$
even if $\supp(\calF_1)$ overlaps with~$\supp(\calF_2)$.

For fault paths $f_1 = \sum_{x \in \Omega} a_x x$ and
$f_2 = \sum_{x \in \Omega} b_x x$
where $a_x, b_x \in \BB$ for all $x\in \Omega$,
we define their {\bf product} as
\begin{equation}
	f_1 \wtprod f_2 = \sum_{x \in \Omega} [a_x \,\mathsf{OR}\, b_x] ~x
\end{equation}
where $\mathsf{OR}$ is the $\BB$-valued function that is zero
if and only if both arguments are zero.
This amounts to taking union of the paths~$f_1$ and~$f_2$,
but we use a different symbol to avoid confusion with the sum operation~$\wtsum$ between sets of paths.
We define the product of two sets of fault paths as
\begin{equation}
	\calF_1 \wtprod \calF_2 = \{ f \wtprod g ~|~ f \in \calF_1, g \in \calF_2 \}
\end{equation}
If $\supp(\calF_1)$ is \emph{disjoint} from~$\supp(\calF_2)$, then
\begin{equation}
	\weightenum{\calF_1 \wtprod \calF_2}{y}
	=
	\sum_{f \in \calF_1,~ g \in \calF_2} y^{\abs{f \wtprod g}}
	=
	\sum_{f \in \calF_2,~g \in \calF_2} y^{\abs f + \abs g}
	=
	\weightenum{\calF_1}{y} \cdot \weightenum{\calF_2}{y} \, . \label{eq:wtprodIneq}
\end{equation}

\subsubsection{Compositions}

In addition to the sum and product operations on fault path sets above,
we can also consider compositions.
Although composition is not used anywhere in our proofs below,
we discuss it here for completeness as it is useful for instantiating \zcref{assume:concatenated-codes}.

For each location $x \in \Omega$,
we consider a substructure at~$x$
and a set~$\omega_x$ of possible error locations in this substructure.
There is no required relation between~$\Omega$ and~$\omega_x$.
Define $\Omega^{(2)}$ as the \emph{disjoint union} of~$\omega_x$ over all~$x \in \Omega$.
This is an abstraction of concatenation,
and $\Omega^{(2)}$ amounts to the set of all possible error locations after the concatenation.
We remark on the disjoint union $\Omega^{(2)} = \bigsqcup_{x \in \Omega} \omega_x$:
if $\omega$ is a set of all possible error location in a gadget,
and if we replace every component of~$\Omega$ by $\omega$ as a result of concatenation,
then we would regard that $\omega_x = \omega$, but
$\Omega^{(2)}$ has $\abs \Omega \cdot \abs \omega$ locations in it;
elements from differently indexed sets in a disjoint union
are automatically treated distinct.

Let $\calF$ be some set of fault paths on~$\Omega$,
and, for each~$x \in \Omega$,
let $\calS_x$ be some set of fault paths on~$\omega_x$.
Naturally, every $\calS_x$ is a set of fault paths on~$\Omega^{(2)}$.
We define {\bf composition}
\begin{equation}
	\calF \bullet \{\calS_x ~|~ x \in \Omega\}
	=
	\bigwtsum_{f \in \calF} \bigwtprod_{x \in \supp(f)} \calS_x \, .
\end{equation}
An important case of concatenation is where $\calS_x$ is a copy of~$\calS$ for all~$x \in \Omega$;
that is, the inner gadgets are the same for all higher level locations~$x \in \Omega$.
In this case,
we simply write $\calF \bullet \calS$ instead of $\calF \bullet \{\calS_x ~|~ x \in \Omega\}$.
We can bound the weight enumerator for the composition:
for any~$y \ge 0$
\begin{align}
	\weightenum{\calF \bullet \{\calS_x\}_x}{y}
	                                                                  & =
	\weightenum{\bigwtsum_{f \in \calF} \bigwtprod_{x \in \supp(f)} \calS_x}{y} \nonumber                                                           \\
	                                                                  & \le
	\sum_{f \in \calF} \prod_{x \in \supp(f)} \weightenum{\calS_x}{y} & \text{by \eqref{eq:wtsumInequ} and \eqref{eq:wtprodIneq}}\nonumber          \\
	                                                                  & \le
	\sum_{f \in \calF} \prod_{x \in \supp(f)} p(y)
	                                                                  & \text{ where } p(y) = \sup_{x \in \Omega} \weightenum{\calS_x}{y} \nonumber \\
                                                                    & = \sum_{f \in \calF} p(y)^{\abs f} \nonumber \\
	                                                                  & = \weightenum{\calF}{p(y)} \, . \label{eq:WtCompositionIneq}
\end{align}
In typical analysis, $p(y)$ will be bounded by a polynomial in~$y$
for all $y \in [0,y_*] \subset \mathbb R$ with some $y_* > 0$.

\subsubsection{Fault-equivalent gadgets}
In many parts of this paper, we will construct new gadgets from old gadgets.
To aid the analysis, we will prove a lemma that relates sets of bad fault paths of gadgets given a relation between faults.
Given two circuits \(C\) and \(C'\) with spacetime locations \(\Omega\) and \(\Omega'\), respectively, we say that \(C'\) is \textbf{fault equivalent}\footnote{This notion is distinct from that of~\cite{rodatz2025fault}.} to \(C\) with respect to the bipartite \emph{attribution graph} \(G = (\Omega' \sqcup \Omega, E)\) if for all faults \(\fault'\) of \(C'\) there exists a fault \(\fault\) of \(C\) such that: i) the circuits have equivalent action as channels \(C[\fault] = C'[\fault']\). ii) The support of \(\fault\) is contained in the boundary of the support of \(\fault'\). I.e. \(\supp \fault \subseteq \Gamma_G (\supp \fault')\) where \(\Gamma_G(X)\) is the graph boundary of \(X \subseteq \Omega' \times \Omega\) in \(G\).

\begin{lemma}[Fault-equivalent gadgets]\label{lemma:fault-equiv}
	Let \(C'\) and \(C\) be quantum circuits with spacetime locations \(\Omega'\) and \(\Omega\) such that \(C'\) is fault equivalent to \(C\) with respect to an attribution graph \(G = (\Omega' \sqcup \Omega, E)\) with maximum left degree \(\delta_L\) and maximum right degree \(\delta_R\).
	For a set \(\badfaults\) of bad fault paths of \(C\), there exists a set of equivalent bad fault paths \(\badfaults'\) of \(C'\) such that
	\begin{itemize}
		\item If \(\fault'\) is a \(\badfaults'\)-avoiding fault of \(C'\), then there exists a \(\badfaults\)-avoiding fault \(f\) of \(C\) such that the two circuits are equivalent as superoperators \(\circuitMap{C}[\fault] = \circuitMap{C'}[\fault']\).
		\item For \(y \in [0,1]\), the weight enumerator of \(\badfaults'\) is at most
		      \begin{align}
			      \weightenum{\badfaults'}{y} \le \weightenum{\badfaults}{\delta_R~y^{\frac{1}{\delta_L}}}~.
		      \end{align}
	\end{itemize}
\end{lemma}
\begin{proof}
	For a fixed fault path \(x \in \BB[\Omega]\), we first construct a set of fault paths \(\badfaults'_x\) of \(\Omega'\) such that if a fault \(\fault'\) of \(C'\) is \(\badfaults'_x\)-avoiding, then there is an equivalent fault \(\fault\) of \(C\) that does not contain \(x\) in its support.
	That is, \(\circuitMap{C}[\fault] = \circuitMap{C'}[\fault']\) as superoperators and \(\supp(x) \not\subseteq \supp(\fault)\).

	A fault path \(w \in \BB[\Omega']\) is said to be \(x\)-minimal if \(x\) is contained in its attribution graph boundary \(x \subseteq \Gamma_G(w)\) and \(x\) is not contained in the boundary of any proper subset, \(z \subset w\), \(x \not\subseteq \Gamma_G(z)\).
	It follows that any superset of \(x\) must contain the boundary of an \(x\)-minimal set.
	Define \(\badfaults'_x \subseteq \BB[\Omega']\) to be the set of all \(x\)-minimal fault paths.
	We now bound the weight enumerator of \(\badfaults'_x\): For an \(x\)-minimal fault path \(w\), each vertex can be adjacent to at most \(\delta_L\) vertices in \(x\), so the size must satisfy \(|w|\delta_L \ge |x|\).
	The number of distinct \(x\)-minimal fault paths is at most \(\delta_R^{|x|}\),\footnote{This tight when every vertex has degree \(\delta_R\) and every neighbor of \(x\) is a unique neighbor.} so for \(y \in [0,1]\) the weight enumerator of \(\badfaults_x'\) is less than \(\weightenum{\badfaults_x'}{y} \le \delta_R^{|x|} y^{|x|/\delta_L}\).

	We will now show that the set of bad fault paths \(\badfaults'\) constructed as the union \(\cup_{x \in \badfaults} \badfaults_x'\) satisfies the theorem statement.
	First, let \(\fault'\) be a \(\badfaults'\)-avoiding fault of \(C'\) and (by fault equivalence) let \(\fault\) be the equivalent fault of \(C\) satisfying \(C[\fault] = C'[\fault']\) and \(\supp \fault \subseteq \Gamma_G (\supp \fault')\).
	Suppose, for a contradiction, that \(\supp \fault\) contains a bad fault path \(x \in \badfaults\).
	Then, \(x\) must be in the boundary of \(\fault'\), \(x \subseteq \Gamma_G (\supp \fault')\), so \(\supp \fault'\) contains an \(x\)-minimal fault path, contradicting the assumption that \(\fault'\) was \(\badfaults'\)-avoiding.

	The bound on the weight enumerator follows straightforwardly.
	\begin{align*}
		\weightenum{\badfaults'}{y} \le \sum_{x \in \badfaults} \weightenum{\badfaults'_x}{y} \le \sum_{x \in \badfaults}  \left(\delta_R~y^{\frac{1}{\delta_L}}\right)^{|x|} = \weightenum{\badfaults}{\delta_R~y^{\frac{1}{\delta_L}}}
	\end{align*}
\end{proof}

\section{Quantum memory}
\subsection{1D fault-tolerance}
\label{subsec-1DFT}
The construction of the main result will leverage results about 1D quantum fault-tolerance schemes~\cite{aharonov1997fault,gottesman2000fault}.
In this section, we will introduce some necessary pieces to prove the main result.

For convenience, let us introduce a special gate for bookkeeping that we label \(\xchg[i,j]\).
This gate will swap qubits \(i\) and \(j\), but qubits \(i\) and \(j\) are considered to be supported in different spacetime locations.\footnote{Within the framework of \cite{he2025composable}, we can consider this as a fake gate that simply relabels the qubits relative to a geometrically local embedding.}
This can also be thought of as swapping \(i\) and \(j\) without directly interacting them.
\(\xchg[i,j]\) can always be replaced by \(\swap[i,j]\), potentially at the cost of worsened fault-tolerance properties.\footnote{Further discussion of this point can be found in \cite{gottesman2000fault}.}
One can also think of \(\xchg\) as a physical \(\swap\) of the quantum degrees of freedom not achieved by quantum gates.

While intermediate 1D geometrically local quantum circuits may contain \(\xchg\) gates, the quantum circuits in the main results will not.

\subsection{Long-range gate simulation}
For later convenience, we will state the following proposition that allows us to simulate long-range gates using 1D geometrically local gates.
This proposition can be used in order to instantiate \zcref{assume:concatenated-codes}, although the primary usage will be to neglect embedding details when constructing new circuits in parts of the construction.
\begin{proposition}[1D circuit embedding]\label{prop:long-range-sim}
	Let \(C\) be a circuit using only \(k\)-local gates of depth \(D\) and width \(W\).
	Then, for any desired input and output ordering of qubits, there exists an equivalent circuit \(C'\) using \(\xchg\) that is geometrically local in 1D with width \(W\) and depth \((W-1)D\).
	\(C'\) is fault equivalent to \(C\) with respect to an attribution graph with maximum left degree \(k\cdot W\) and maximum right degree \(1\).
\end{proposition}
\begin{proof}
	Using an even-odd sorting network, any permutation of the line can be accomplished in \(W-1\) layers of swaps~\cite{knuth1997art}.
	For every gate layer \(M\) of \(C\), we route each qubit to an arbitrarily selected spatial location such that \(M\) can be implemented using geometrically local gates.
	\(C'\) can be constructed by alternating these permutations with a geometrically local implementation of each gate layer.
	This is done \(D\) times before each layer of gates.

	We now proceed to the fault analysis.
	Let \(\Omega'\) be the locations of \(C'\) and \(\Omega\) be the locations of \(C\).
	\(C'\) differs from \(C\) by the insertion of \(\xchg\) gates, so we may consider \(\Omega\) to be a subset of \(\Omega'\).
	For any fault \(\fault'\) of \(C'\), one can construct an equivalent fault \(\fault\) of \(C\) by first ``pushing'' the part of the fault supported on \(\Omega' \setminus \Omega\) to the next location (at a later circuit layer) in \(\Omega\) and then discarding locations in \(\Omega'\setminus\Omega\), rewiring gate inputs and outputs to match that of \(C\).
	The qubits \(\xchg\) acts on are considered to be in distinct locations,\footnote{Recall that this is the distinction between \(\swap\) and \(\xchg\).} so this procedure does not increase the fault weight.\footnote{In the case of \(\swap\) gates, the fault weight is at most doubled.}
	Additionally, each location in \(\Omega\) succeeds at most \((W-1)k\) \(\xchg\) gates.
	It follows that \(C'\) is fault equivalent to \(C\) with respect to the fault attribution graph \(G = (\Omega' \sqcup \Omega, E)\) where two locations \((\ell',\ell) \in \Omega' \times \Omega\) share an edge if a fault supported only on \(\ell'\) may be pushed to a fault with support containing \(\ell\).
	This graph has left degree at most \(kW\) and right degree at most \(1\) (2 if using \(\swap\) gates).
\end{proof}

\subsection{1D fault-tolerant schemes}
Here we introduce the required notion of 1D fault tolerance schemes which can be supplied using recursive simulation \cite{aharonov2008fault,gottesman2000fault}.
To avoid a large technical excursion, we will utilize the definitions from \cite[Sections 2-4]{he2025composable} summarized in \zcref{sec:SkeletonOfArgument}.
In particular, we will utilize notations and definitions required to state \cite[Definition 4.8]{he2025composable} of a fault-tolerant gadget.
Let \(\gateInject\) be an operation that encode a trivial qubit into a code block and \(\gateEject\) be its inverse.
\(\gatePREP_{\ket{0}}\) prepares a logical \(0\) state.

\cite{aharonov2008fault} provides gadgets that essentially instantiate \zcref{assume:concatenated-codes}.\footnote{We note that while \(\gateCS\) is not in the original gateset, it can be implemented using \(\gateTOFF\), \(\gateS\), a \(\ket{0}\) qubit by writing the product to the ancilla and applying \(\gateS\) to it: \(\ket{x}\ket{y}\ket{0} \mapsto \ket{x}\ket{y}\ket{xy} \mapsto i^{xy}\ket{x}\ket{y}\ket{xy} \mapsto i^{xy}\ket{x}\ket{y}\ket{0}\).}
To prove this fact within composable fault-tolerance, it is easiest to invoke the level reduction theorem for composable fault tolerance~\cite{he2025composable}, a generalization of \cite{aliferis2007level} along with \zcref{prop:long-range-sim}.
Later, it will be convenient to have a brickwork connectivity.
This can be enforced by first replacing physical-level Toffoli gates with one and two-qubit gates~\cite{barenco1995elementary} and then splitting each circuit layer into two circuit layers with each gate executed in one of the two layers.
By an identical argument to the long-range gate simulation \zcref{prop:long-range-sim}, this new circuit is a fault-tolerant gadget if the original one is at the cost of a constant prefactor in the weight enumerator of the family of bad fault paths.

\begin{figure}
	\centering
	\includegraphics[width=0.5\columnwidth]{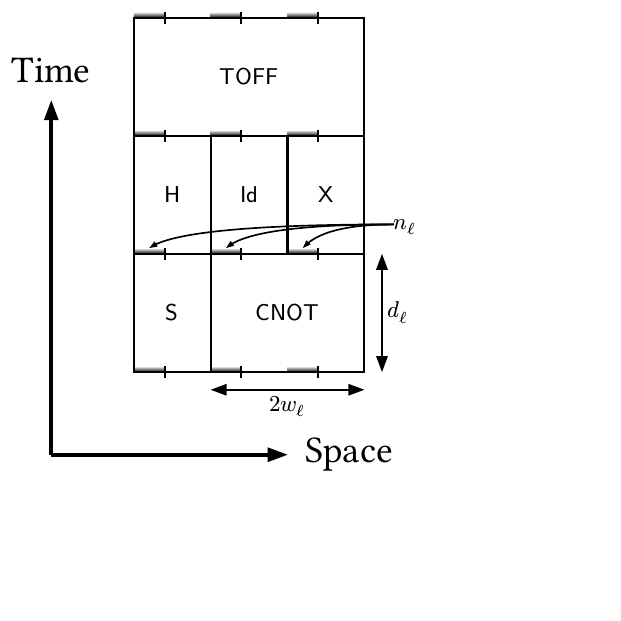}
	\caption{\label{fig:1D-gadget-embedding}1D gadgets and the compatibility constraints from \zcref{assume:concatenated-codes}.}
\end{figure}

Informally, \zcref{assume:concatenated-codes} states there exists a family of concatenated quantum error-correcting codes such that a universal set of fault-tolerant gadgets (in the sense of \cite{he2025composable}) can be constructed using only geometrically local quantum processing in one dimension.
This family admits a constant accuracy threshold; when the physical error rate is below threshold, the probability that a gadget fails falls doubly exponentially with the number of levels of concatenation, while the size and depth of the gadgets increases only exponentially.
Well-behaved gadgets for encoding and decoding can also be constructed.
The gadgets are also compatible in the sense that they can be composed to build larger circuits as shown in \zcref{fig:1D-gadget-embedding}.

\begin{ingredient}[Concatenated code fault-tolerance]\label{assume:concatenated-codes}
	There exists a constant \(p_* \in (0,1]\), a family of quantum codes \(\{\mathcal{C}_\ell\}_{\ell \in \mathbb{N}}\) on \(n_\ell\) qubits, corresponding code types (\cite[Definition 4.2]{he2025composable}) \(\{\compCodeType[\ell]\}_\ell\), a function \(\epsilon(\ell) = O(e^{-2^\ell})\), and for each \(\ell\), we have that:
	\begin{itemize}
		\item For each quantum operation \(\qGate \in \qOp = \{\gateI,\gateX,\gateH,\gateTOFF,\gateCS,\gatePREP_{\ket{x}}, \swap\}\), there exists a corresponding quantum circuit (quantum input and output) \(\gadget_{\qGate,\ell}\) that is a friendly fault-tolerant gadget with input and output code types \(\compCodeType[\ell]\) and family of bad fault paths \(\badfaults_{\qGate,\ell}\) such that for \(x \in [0,p_*]\), the weight enumerator is bounded as
		      \begin{align}\label{eq:weight-enum-1dgadgets}
			      \weightenum{\badfaults_{\qGate,\ell}}{x} \le \epsilon(\ell)~.
		      \end{align}
		\item For \(\qGate = \gateInject\) (\(\qGate = \gateEject\)), there exists a corresponding quantum circuit \(\gadget_{\qGate,\ell}\) that has trivial input (output) code type, non-trivial output (input) code type \(\compCodeType[\ell]\), and a set of bad fault paths \(\badfaults_{\qGate,\ell}\) such that for \(x \in [0,p_*]\) and some absolute constant \(c_{\mathrm{inject}} > 0\), the weight enumerator is bounded as
		      \begin{align}
			      \weightenum{\badfaults_{\qGate,\ell}}{x} \le c_{\mathrm{inject}} x~.
		      \end{align}
		\item Each \(\gadget_{\qGate,\ell}\) uses 1 and 2-qubit gates and is 1D geometrically local with an embedding into \(w_\ell \cdot |\supp \qOp|\) qubits where \(w_\ell = 2^{\Theta(\ell)}\).
		\item The embeddings are compatible in the sense that the input and output code blocks are ordered sequentially. I.e. for an operation \(\qOp\), \(k \in \{1, \dots, |\supp \qOp|\}\), the qubits \(\{(k-1) w_\ell + 1, \dots, (k-1) w_\ell + n_\ell\} \subseteq \left[w_\ell \cdot |\supp \qOp|\right]\) are the qubits of the \(k\)-th input (output) code block arranged in a fixed ordering. (See \zcref{fig:1D-gadget-embedding})
		\item The gadgets have uniform depth \(d_\ell = 2^{\Theta(\ell)} \ge w_\ell\).
		      We additionally require that the last layer of gates in the gadget for \(\gateI\) is identity.\footnote{This can be accomplished for example by padding all gadgets with a layer of identities at the end.}
	\end{itemize}
\end{ingredient}

\subsection{Tiles}\label{sec:11Dto20D}
We now describe a method of ``unrolling'' a 1D geometrically local circuit such that it can be implemented by a depth-3 clocked circuit in two spatial dimensions.
The clocked circuit encodes the layers of the circuit in the spatial position along a second axis.
We refer to this procedure as the \emph{tile transformation}.
A similar technique was used in a recent local decoder for the two-dimensional toric code~\cite{balasubramanian2024local} to obtain time-translation invariance with three spatial dimensions.

\begin{figure}
	\centering
	\includegraphics[width=\columnwidth]{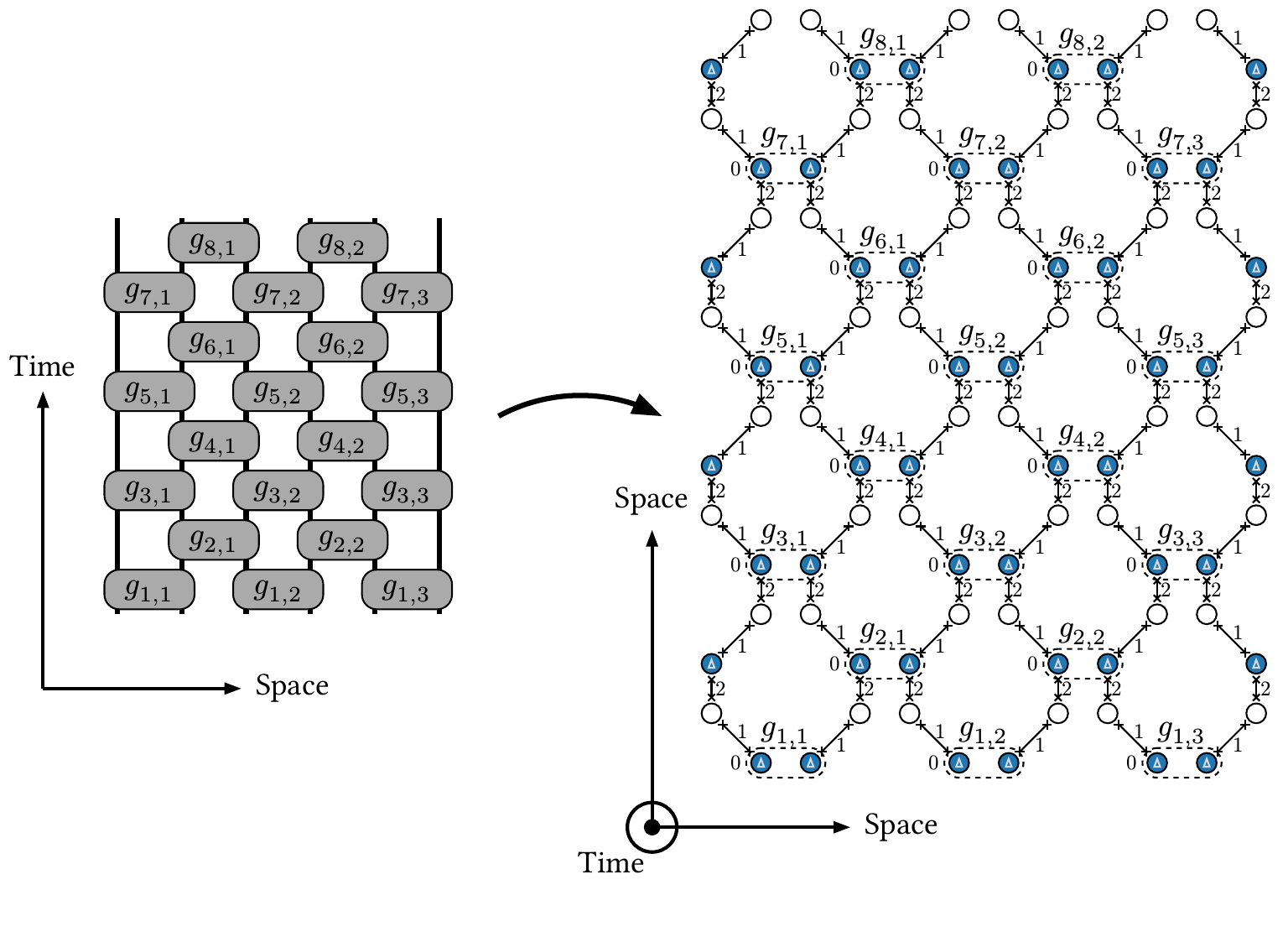}
	\caption{\label{fig:tile-transformation}
		Transformation of a brickwork-connectivity circuit into an upward propagating tile with inputs entering at the bottom and leaving at the top.
		Circles indicate the locations of qubits physically in a two-dimensional plane.
		Arrows overlaying the blue circuits indicate the direction of qubit data propagation.
		Shaded circles are qubit locations where two-qubit gates are applied while unshaded circles contain ancilla qubits required to implement the qubit data propagation using \(\swap\).
		The circuit has period 3: Layers \(i \equiv 0 \textrm{ mod } 3\) apply the indicated gate (or single qubit operations) to each pair of qubits. Layers \(i \equiv 1 \textrm{ mod } 3\) swap shaded locations with their downstream neighbors while layers \(i \equiv 2 \textrm{ mod } 3\) swap shaded locations with their upstream neighbors.
		The number adjacent to each gate indicates the layer in which it is applied.
		Adjacent qubits in the 1D line may be swapped without directly interacting by applying the \(\swap\) in an `X' pattern in round two immediately before the gate location. Importantly, though, at each physical location in the tile where layer-2 gates are applied, whether this X $\swap$ occurs or not is fixed, so that the circuit has period 3.
		This avoids certain correlated weight-2 errors stemming from a \(\swap\) failure that would otherwise be present.
		See also \cite[Section 3]{gottesman2000fault}.}
\end{figure}

\begin{construction}[Tile transformation]\label{prop:time-to-space}
	For any 1D geometrically local gadget \(C\) (with \(\xchg\)) of width \(W\) and depth \(D\), we can construct a depth-3 clocked circuit \(C'\), referred to as a \textbf{tile}, on a \(2D \times W\) grid of qubits that implements the gadget in the sense that after each application of the clocked circuit (a \textbf{tick}), data is transported one position (a \textbf{station}) forward and has had a layer of \(C\) applied. (See \zcref{fig:tile-transformation})

	Concretely, define the \(i \in [D]\)-th station of \(C'\) to be the row of qubits \(\{2i\} \times [W]\).
	The three layers of \(C'\) perform the following depth-3 sequence of gates in parallel for each \(i \in [D]\):
	\begin{enumerate}
		\item \label{prop:time-to-space:construction:gates}Apply the gate layer \(\gatelayer[C](i)\) on row \(2i\). That is, a gate acting on qubit \(j\in [W]\) of \(C\) is applied to qubit \((2i,j)\) of \(C'\).
		\item \label{prop:time-to-space:construction:swap1}Swap row \(2i\) with row \(2i+1\) column-wise.
		\item \label{prop:time-to-space:construction:swap2}For each column \(j \in [W]\), if \(\gatelayer[C](i+1)\) contains the gate \(\xchg[j,j']\) with \(j' \in \{j \pm 1\}\), swap qubit \((2i+1, j)\) with \((2i+2, j')\). Otherwise, swap qubit \((2i+1, j)\) with \((2i+2, j)\).\footnote{This implements an `\textsf{X}' pattern of \(\swap\).}
	\end{enumerate}
\end{construction}

We refer to the propagating data located at a single station as a \textbf{slot}.
We refer to the first and last stations of a tile as the input and output stations, respectively.
If the output ``type'' of a gadget \(C_1\) matches the input type of a gadget \(C_2\), then they can be composed into a new gadget \(C_2 \circ C_1\).
Likewise, the tiles corresponding to \(C_1\) and \(C_2\) can be composed such that the input station of \(C_2\) follows the output station of \(C_1\) with a corresponding insertion of \(\swap\) gates.
This is identical to applying the tile transformation to the gadget \(C_2 \circ C_1\).
However, not all compositions of tiles correspond to the tile transformation applied to gadgets.
In particular, consider a ``tail-biting'' configuration on a closed ribbon where the input of a tile is connected to its own output i.e. a loop.
To distinguish tiles arising from composition from the simpler building blocks, tiles constructed from the concatenated code gadgets from \zcref{assume:concatenated-codes} will be referred to as \textbf{primitive tiles}.
If the input or output of the gadget is non-empty (i.e. the gadget does not begin or end by resetting qubits), we refer to the corresponding tile stations as the input or output, respectively.

When the input circuit satisfies a brickwork connectivity constraint, we can slightly spread out the qubits at the cost of increasing locality diameter to \(\sqrt{2}\).
This allows for a vertically propagating tile to be placed on top of a horizontally propagating tile (see \zcref{fig:tile-transformation}).
Placing tiles with perpendicular propagation directions on top of each other serves two purposes: The first is a technique to change the propagation direction, while the second purpose is to allow two tiles to cross over each other without modification.

To compose tiles in a loop while maintaining a planar geometry, we will need to be able to change the direction of propagation of tiles.
However, in order to maintain logical information in a quantum code block, we must still be capable of applying gates during the turning process.
A naive attempt to turn a slot in the plane (like a car) will require a superconstant change in the propagation speed in order to maintain the unit spacing.
Instead, one can ``fold'' the path over itself like a ribbon which still maintains the unit density.
The macroscopic picture is illustrated in \zcref{fig:ribbon-loop}.

\begin{figure}
	\centering
	\includegraphics[width=\columnwidth]{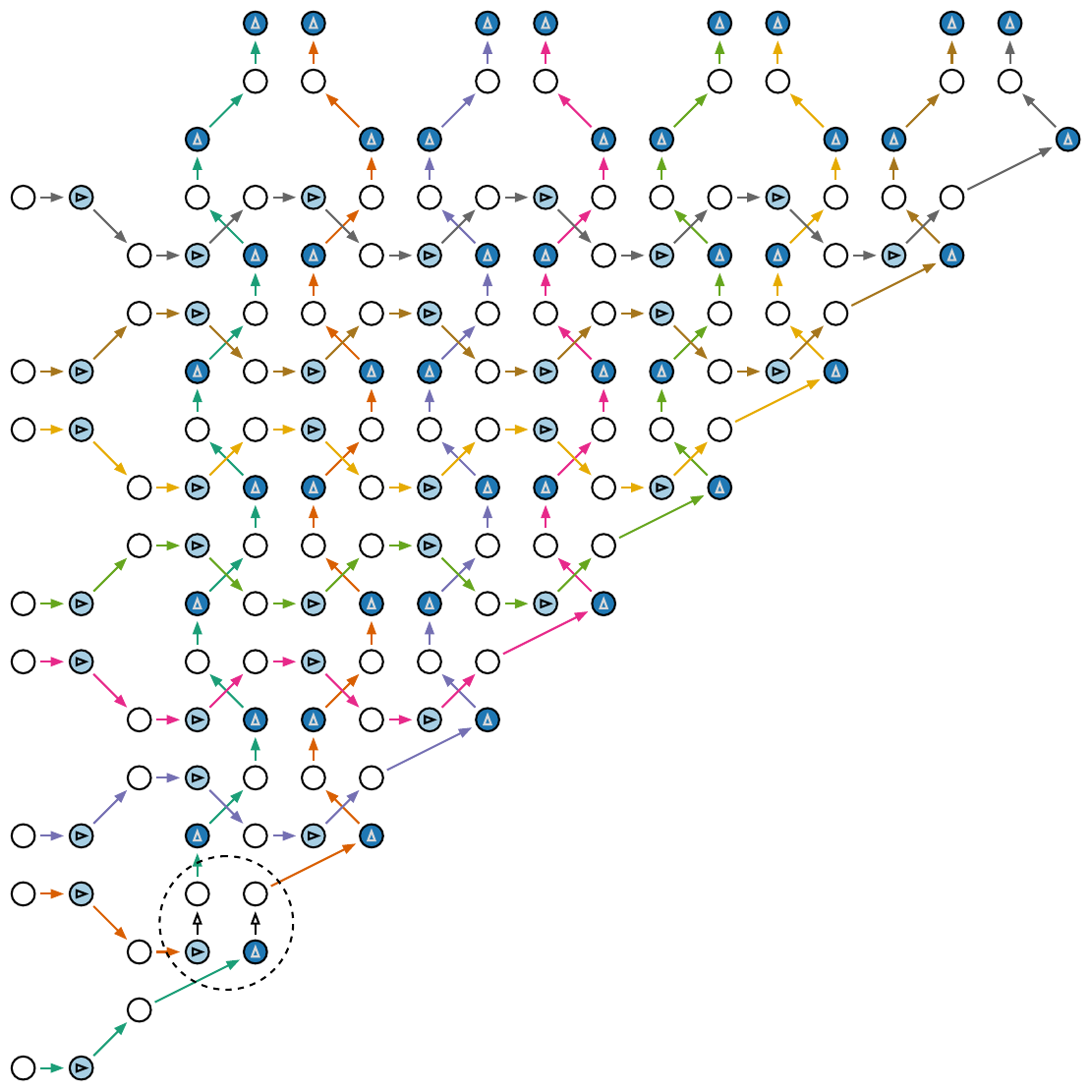}
	\caption{\label{fig:turning}Microscopic implementation of turning for a width-8 slot.
		Rightward propagating qubits are light blue while upward propagating qubits are dark blue.
		Ancilla qubits with no further purpose in the tile are white.
		Two-qubit gates acting on neighboring pairs of colored qubits are not explicitly indicated.
		Input qubits that are ordered from top to bottom are transformed to output qubits ordered from right to left.
		Gates and propagation are implemented by a period-3 circuit as in \zcref{fig:tile-transformation}. %
		Note that the length traveled from input to output is identical for all qubits, so that qubits entering the tile simultaneously also exit simultaneously.
		At the ``fold'' where data qubits transition from moving right to moving up, a two-qubit gate acts on one right-moving qubit and one up-moving qubit. A $\swap$ gate at the fold acts on one ancilla qubit and one data qubit that are separated by Euclidean distance $\sqrt{5}$.
		To perform gates on colored qubits crossing the ``fold'' \(\xchg\) is applied \emph{after} the gate instead of before while the gate itself is applied while one qubit is moving rightward and the other qubit is moving upward.
		This is illustrated within the dashed black circle in the bottom left.
	}
\end{figure}

A microscopic implementation with locality diameter \(\sqrt{5}\) is illustrated in \zcref{fig:turning}.
Qubits crossing the fold change the tile transformation of \zcref{fig:tile-transformation} in the following way: A gate may be applied between the rightward propagating qubit and upward propagating that is ``across the fold.''
Similarly, an \(\xchg\) gate acting on this same pair of qubits can be performed by modifying the crossing pattern of \(\swap\) to be non-crossing.
For a tile of width \(w\), the turn occupies a region of area \(w \times w\).
Likewise, to turn multiple \(m\) tiles propagating side-by-side, the turn will occupy a region \(mw \times mw\).

\begin{figure}
	\centering
	\includegraphics[width=\columnwidth]{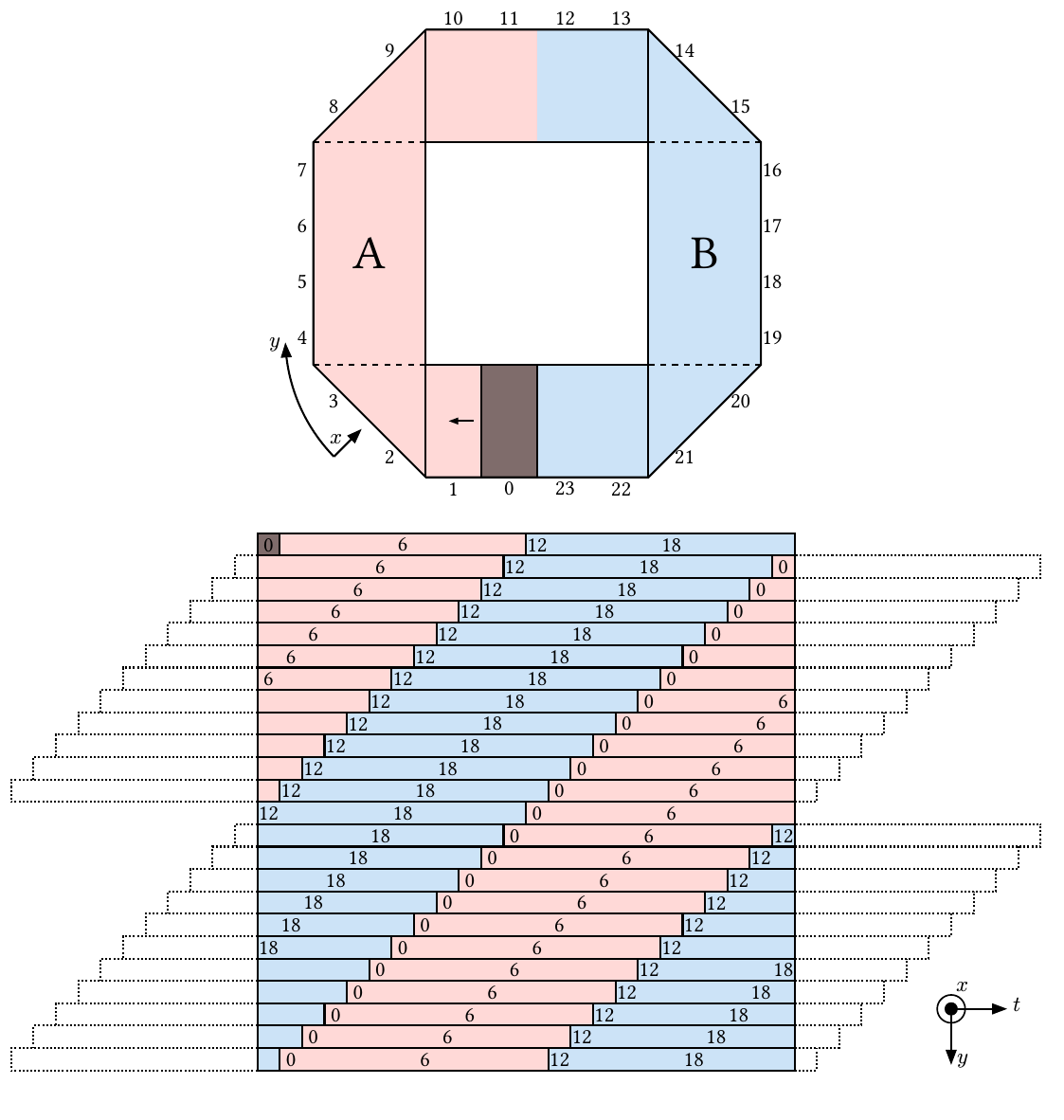}
	\caption{\label{fig:trace}
		A macroscopic view of:
		(a) A closed clocked circuit composed of two tiles (shaded red/blue) with \(12\) stations each (indicated \(0\) to \(23\)). Slots extend into the \(x\) direction and propagate in the \(y\) direction. The slot at station \(0\) corresponding to the first slot of the bottom diagram is shaded (black). (b) The execution trace of \(24\) applications (left to right) of the clocked circuit on 24 slots (top to bottom) with gadgets (shaded regions) indicated. This can be thought of as a ``comoving'' frame propagating through the clocked circuit on the left. The bottom diagram is \(3/5\) scale compared to the top diagram. Slots extend into the page. The additional gates of the extended execution trace are drawn as a dashed outline.
	}
\end{figure}

We now show that the tile transformation preserves the fault-tolerance of gadgets.
We refer to a clocked circuit composed of tiles as \textbf{closed} every tile input (output) is controlled qubits or connected to a tile output (input).
Consider \(D\) applications of a closed clocked circuit \(C^{(D)}\).
It is convenient to introduce a term to refer to be able to refer to the circuit that was executed:
The \textbf{execution trace}\footnote{This term is meant in analogy to the ``execution trace'' of a program.} \(C^{(D)}_{\mathsf{TR}}\) of \(C^{(D)}\) is a functionally identical quantum circuit where, at every layer of \(C^{(D)}\), qubits are relabeled such that slots are stationary instead of propagating.
This can be thought of somewhat as an inverse operation to the tile transformation \zcref{prop:time-to-space} or a sort of ``comoving'' frame.
Concretely, at tick 0, label the qubits of each slot by its station and position within the slot.
	After every tick, a qubit at station \(s\) takes the label of the corresponding qubit at the station preceding \(s\).
	For stations with no preceding stations (as would be for initialization gadgets), introduce a fresh label.
	The counter-propagating ancilla qubits from the tile transformation are labeled arbitrarily.
The \textbf{extended execution trace} of \(C^{(D)}\) is the execution trace of \(C^{(D)}\) with the minimal parts of additional clock circuit layers appended and prepended such that the first gate acting on each slot is that of the input (or output) station of a tile.
See \zcref{fig:trace}.

Let \(\Omega\) be the set of primitive tiles of the clocked circuit of \(C^{(D)}\).
Gates of the extended execution trace \(C^{(D)}_{\mathsf{TR}}\) can be labeled by the corresponding tile that applied them and the time at which the slot entered the tile, \(\Omega \times \mathbb{Z}\).
This induces a labeling of the gates of the execution trace \(C^{(D)}_{\mathsf{TR}}\).

Gates of the extended execution trace can be grouped together and implement gadgets.
\begin{proposition}[Gadgets of the execution trace]\label{prop:trace-gadgets}
	For each label \((\omega, t) \in \Omega \times \mathbb{Z}\), the corresponding subcircuit of the extended execution trace \(C^{(D)}_{\mathsf{TR}}\) is fault-equivalent to the gadget \(G_\omega\) from which the tile \(\omega\) was constructed.
	The fault-equivalence is with respect to an attribution graph with maximum left degree \(3\) and maximum right degree \(1\).
\end{proposition}
\begin{proof}
	Let \(G\) be the gadget and \(C\) be the subcircuit labeled by \((\omega, t)\) in the extended execution trace.
	By construction, after the relabeling of qubits in every layer, the implemented circuit \(C\) (see \zcref{fig:tile-transformation}) differs from the gadget \(G\) by the introduction of an ancilla qubit for each qubit from \(G\) and two layers of \(\swap\) gates supported between each pair of ancilla qubits and qubit from \(G\).
	These \(\swap\) gates cancel out due to the relabeling in the construction of the extended execution trace.
	Since the faults are Pauli faults, they can be commuted forward past the \(\swap\) gates to a location implementing a gate from \(G\).
	There are 2 such locations in addition to the gate from \(G\) for a total of 3 locations.
\end{proof}

\subsection{Geometrically local quantum memory}\label{sec:memory}
As an intermediate step, let us construct a memory using the tools introduced thus far.
Recall that \(w_\ell\) is the number of qubits in a slot encoding one or more logical qubits from \zcref{assume:concatenated-codes}.
\begin{theorem}[store={memory},note={Quantum memory}]\label{thm:memory}
	There exists an maximum permissible error rate \(p_* \in (0,1]\) and a family of geometrically-local depth-3 quantum circuits \(C_{(\ell,L)}\) parameterized by \(\ell,L \in \mathbb{N}\) such that for each \(L \ge 4 d_\ell\),
	\begin{itemize}
		\item \(C_{(\ell,L)}\) has \emph{locality diameter} \(\sqrt{5}\) and fits in a \emph{square} of side length\footnote{The constant is a small number \(\sim 4\).} \(O(w_\ell L)\).
		\item The total number of logical qubits is \(W \equiv d_\ell \lfloor\frac{4 w_\ell L^2}{d_\ell}\rfloor\) and the number of stored logical qubits per unit area is \(\Omega\left(\frac{1}{w_\ell}\right)\). That is, the rate of the memory is essentially the rate of the 1D scheme from \zcref{assume:concatenated-codes}.
		\item For any storage duration \(\tau \in \mathbb{N}\), there exists encoding and recovery channels \(\mathcal{E}\), \(\mathcal{D}_\tau\) such that for any fault \(\fault\) of \(C^{(\tau)} \equiv \left(C_{(\ell,L)}\right)^{\circ \tau}\) distributed according to a \(p\)-local stochastic noise model with \(p \in [0,p_*]\), the faulty circuit \(C^{(\tau)}[\fault]\) preserves \(W\) qubits of encoded data in the sense that encoded data is retrievable to high accuracy using the recovery channel
		      \begin{align}
			      || \mathcal{D}_\tau \circ \mathbb{E}_{\fault}\left(\circuitMap{C^{(\tau)}[\fault]}\right) \circ \mathcal{E} - \mathsf{Id}_{2^W} ||_{\diamond} \le O\left(\tau \cdot W e^{-2^\ell}\right)~.
		      \end{align}
	\end{itemize}
\end{theorem}
\begin{proof}
	Using the turning technique illustrated in \zcref{fig:turning}, one can lay out identity primitive tiles constructed from the identity gadget from \zcref{assume:concatenated-codes} to create a closed clocked circuit in the form of a space filling curve.

	Let \(w_\ell\) and \(d_\ell\) be the width and depth of the identity gadget of the \(\ell\)-th element of the family from \zcref{assume:concatenated-codes} with corresponding family of bad fault paths \(\badfaults_{\gateI}\).
	Consider the tile transformation (\zcref{prop:time-to-space}) applied to the identity gadget \(\overline{C}_{\gateI}\).
	The clocked circuit of the memory \(C_{(\ell,L)}\) will consist of the composition of \(\eta = \lfloor\frac{4 w_\ell L^2}{d_\ell} \rfloor\) identity tiles \(C_{\gateI,\mathsf{tile}}\) such that the output of tile \(i \in [\eta]\) is input of tile \(i + 1 \text{ mod } \eta\).
	Label the \(W = \eta d_\ell\) stations by coordinates \((i,j) \in [d_\ell] \times [\eta]\) referring to the \(i\)-th station of the \(j\)-th tile.

	Consider now \(\tau\) applications of the clocked circuit, \(C^{(\tau)}\).
	For convenience, we will work with its execution trace.
	The corresponding maps on the circuit can be found by appropriate qubit relabeling.
	Define \(\mathcal{E}_{\mathsf{block}}\) and \(\mathcal{D}_{\mathsf{block}}\) to be the encoder and decoder of the codes from \zcref{assume:concatenated-codes}.
	The encoding map \(\mathcal{E}\) first applies \(\mathcal{E}_{\mathsf{block}}^{\otimes W}\) to obtain \(W\) slots, with each slot forming a code block from \zcref{assume:concatenated-codes}.
	The second step of the encoding map ensures that each slot must be at the correct step of processing (corresponding to its initial station), so the encoding map applies the portion of the extended execution trace of \(C^{(\tau)}\) that precedes the execution trace.
	This map is independent of \(\tau\).
	The recovery map \(\mathcal{D}_\tau\) will apply the reverse operation: For each slot, apply the portion of the extended execution trace of \(C^{(\tau)}\) that follows the execution trace and then apply the decoding map of the code \(\mathcal{D}_{\mathsf{block}}\) to each slot.

	We now show the clocked circuit is a good memory.
	Consider the circuit given by encoding, storage, and recovery of the information \(\mathcal{D}_\tau \circ  C^{(\tau)} \circ \mathcal{E}\).
	Up to a relabeling of the qubits, this circuit is the extended execution trace of \(C^{(\tau)}\) with the encoding map of the quantum code before and after.
	Let \(\epsilon(\ell)\) be the weight enumerator bound of the bad fault paths of the gadgets on \([0,p_*']\) from \zcref{assume:concatenated-codes}.
	By \zcref{prop:trace-gadgets}, (and the composition lemmas from \cite{he2025composable}), the extended execution trace is a gadget implementing identity \(\gateI^{\otimes W}\) with a set of bad fault paths \(\badfaults_\tau\) where the bad faults are the sum \(\boxplus\) of bad faults from each individual gadget in the extended execution trace.
	There are at most \(\tau W\) gadgets in the extended execution trace.
	Using fault equivalence (\zcref{lemma:fault-equiv}) and the 1D fault-tolerance scheme weight enumerator bound from \zcref{assume:concatenated-codes}, on \(x \in [0,p_*'/3]\), the weight enumerator of \(\badfaults_\tau\) is bounded as
	\begin{align*}
		\weightenum{\badfaults_\tau}{x} \le \tau W \epsilon(\ell)~.
	\end{align*}

	Using this bound, we can show the recovery guarantee in diamond norm via a standard argument from \cite[Section 5]{he2025composable}.
	Consider a \(\badfaults_\tau\)-avoiding physical\footnote{A physical fault is one where fault superoperator is CPTP.} fault \(\fault\) of \(C^{(\tau)}\).
	Since \(\fault\) is also a fault of the extended execution trace \(C^{(\tau)}_{\mathsf{extTR}}\) of \(C^{(\tau)}\)
	\begin{align*}
		\mathcal{D}_{\tau} \circ C^{(\tau)}[\fault] \circ \mathcal{E} = \mathcal{D}_{\mathsf{block}}^{\otimes W} \circ C^{(\tau)}_{\mathsf{extTR}}[\fault] \circ \mathcal{E}_{\mathsf{block}}^{\otimes W}~.
	\end{align*}
	The fault \(\fault\) is \(\badfaults_\tau\)-avoiding with probability \(1-\tau W \epsilon(\ell)\), and in this case, we have that \(\mathcal{D}_{\mathsf{block}}^{\otimes W} \circ C^{(\tau)}_{\mathsf{extTR}}[\fault] \circ \mathcal{E}_{\mathsf{block}}^{\otimes W} = \gateI^{\otimes W}\) (applying the definition of a fault-tolerant gadget).
	Then, the result follows with \(p_* \equiv p_*'/3\) using subadditivity of the diamond norm
	\begin{align}
		|| \mathcal{D}_\tau \circ \mathbb{E}_{\fault}\left(\circuitMap{C^{(\tau)}[\fault]}\right) \circ \mathcal{E} - \mathsf{Id}_{2^W} ||_{\diamond} & \le \mathbb{E}_{\fault} \left(|| \mathcal{D}_\tau \circ \circuitMap{C^{(\tau)}[\fault]} \circ \mathcal{E} - \mathsf{Id}_{2^W} ||_{\diamond}\right) \\
		                                                                                                                                              & \le \mathbb{E}_{\fault} \left(\mathbb{I}\left[\fault\text{ is not }\badfaults_\tau\text{-avoiding}\right]\right)                                   \\
		                                                                                                                                              & \le \weightenum{\badfaults_{\tau}}{p} \le \tau W \epsilon(\ell)~.
	\end{align}

	We now turn our attention to the layout (\zcref{fig:mem-layout}).
	So far, the natural spatial layout of the clocked circuit \(C_{(\ell,L)}\) is that of a ``ribbon'' of width \( \Theta(w_\ell)\) and length \(\Theta(\eta d_\ell)\).
	We will use the turning method (\zcref{fig:tile-transformation}) to fold the ribbon into a square \(2D\) structure as in \zcref{fig:mem-layout} where \(L\) ``fingers'' extend from a ``backbone'' by at most \(O(w_\ell L)\).
	More precisely, each turn takes a number of stations \(w_\ell\) to complete and occupies an area with linear dimensions roughly \(w_\ell \times w_\ell\), so the \(i\)-th finger will have length \(2w_\ell + \delta_i\) where \(0 \le \delta_i \le \delta_{\mathrm{max}} \equiv 2 w_\ell L - 3w_\ell\).
	The total number of slots in a finger is \(4w_\ell + 2\delta_i\), so including the backbone \(2wL\) slots, the total number of slots required is \(6w_\ell L + \sum_{i \in [L]} 2\delta_i\).
	Thus, as long as the number of slots in the ribbon satisfies \(6w_\ell L \le W \le 6w_\ell L + 2L\delta_{\mathrm{max}} = 4 w_\ell L^2\), there is a choice of \((\delta_0, \dots \delta_{L-1})\) such that this is a valid layout i.e. arranging the tiles in this way leads to a closed loop of propagating slots on the \(2D\) plane.
	From the restrictions \(L \ge 4 d_\ell\) and \(d_\ell \ge w_\ell\), the number of slots \(W = \eta d_\ell\) satisfies \(W \le 4 w_\ell L^2\) and \(W \ge 4 w_\ell L^2 - d_\ell \ge 8 w_\ell d_\ell L \ge 6w_\ell L\), so there is a valid layout of the ribbon within a square of linear dimensions \(O(w_\ell L)\).\footnote{This implies the rate of our quantum memory is \(\frac{W}{\Delta^2} = \Omega(1/w)\) which asympotically matches the rate of the original 1D fault-tolerance scheme.}
\end{proof}

Having constructed a period-3 circuit that is a quantum memory, we remark on the possibility of obtaining constant coding rate.
\begin{remark}[Constant-rate 2D quantum memory]\label{rem:const-rate}
	One can construct a \emph{constant-rate} 1D quantum fault-tolerance scheme with local computation\footnote{It is worth noting here that \cite{gidney2025constant} does not suffice for our purposes, since the processing is not local.} yielding a constant-rate 2D geometrically-local period-3 circuit via the construction of \zcref{thm:memory}.

    \hspace{2ex} The general idea is to follow the concatenation ideas of \cite{yamasaki2024time}
    with sufficient amortization of resources (essentially parameter hacking);
    We have to reduce the space overhead coming from ancilla use.
    First note the following observation: Fix a sequence of quantum codes \(\mathcal{C}_i\) of length \(n_i\) labeled by \(i = 1, 2, \dots\) with rate \(\frac{k_i}{n_i} = 1-O(1/i^2)\).%
    \footnote{so that $\prod_{i}^{\infty} (1 - O(i^{-2})) > 0$.}
    Then, the family of concatenated codes \(\ell =  1, 2, \dots\), \(\mathcal{D}_\ell\) given by concatenating the codes \(1, \dots, \ell\) has constant rate.
    This constant rate can be straightforwardly preserved in the fault-tolerance setting.
    Suppose now that one knew a fault-tolerant\footnote{Here, we mean fault tolerant in the concatenated code sense.} error correction gadget for the code \(\mathcal{C}_i\) with \(m_i = \poly(n_i)\) ancilla qubits.
    Then, construct a code family \(\mathcal{C}_i'\) where the \(i\)-th member of the family is \(s_i = i^2 \frac{m_i}{n_i}\) blocks of \(\mathcal{C}_i\) and \(m_i\) qubits at the end of the block set to \(\ket{0}\).
    \(\mathcal{C}_i'\) also has rate \(\frac{s_i k_i}{s_i n_i + m_i} = 1-O(\frac{1}{i^2})\).
    The code family \(\mathcal{C}_i'\) has a fault-tolerant error correction gadget given by using the \(m_i\) qubits at the end of the block to execute one-at-a-time the error correction gadget on each of the \(s_i\) blocks of \(\mathcal{C}_i\).
    Furthermore, this error correction gadget is only polynomially larger than the one for the original code family and requires no extra qubits beyond those in the code block.

    \hspace{2ex} In fact, we may perform this amortization for \emph{any} logical operation gadget,\footnote{A simple method is to use a naive concatenated code scheme to perform a non-fault-tolerant logical gate.} and as observed by Gottesman~\cite{gottesman2000fault}, a polynomial blowup in the number of spacetime locations still allows for a constant threshold under concatenation.
    That is, the concatenated code scheme \(\mathcal{D}_\ell'\) given by concatenating the codes \(\mathcal{C}'_1, \dots, \mathcal{C}'_\ell\) has constant space overhead and constant threshold.
    Embedding the scheme into one-dimension as in \cite{gottesman2000fault} is just a further polynomial blowup in the number of spacetime locations.
    As for obtaining local noisy computation, it suffices to start with a sufficiently large distance code family\footnote{For example, concatenating a polynomially-growing sequence of quantum Hamming codes with themselves several times and then forget that this concatenation has been done.} and use the classical fault-tolerance techniques in \cite{cirel2006reliable,aharonov1997fault} for a further polynomial blowup.
\end{remark}
One might naturally wonder whether any difficulty with targeting gates in high rate codes arises, but this is only a problem when considering the practical or highly optimized spacetime overhead settings.
Already \cite{gottesman2013fault} shows that a non-constant space overhead scheme \emph{for computation} can be combined with a constant space overhead scheme \emph{for memory} to achieve a scheme for computation with overall constant space overhead.

\section{2D processor without classical processing}\label{sec:quantum-only-processor}

\begin{table}
    \centering
	\begin{tabular}{r | p{0.7\textwidth}}
		\hline\hline
		Parameter             & informal
		\\ \hline\hline
		\(\ell\)              & Number of levels of concatenation in \zcref{assume:concatenated-codes}.                                                                                                             \\
		\(\maxwidth\)         & Maximum number of qubits in any simulated circuit.                                                                                                                                  \\
		\(W\)                 & Number of stations in the memory tape                                                                                                                                               \\
		\(\Nprog\) & Block length of the program code.                                                                                                                                                   \\
		\(w_\ell\)            & Width of the gadgets from \zcref{assume:concatenated-codes}.                                                                                                                        \\
		\(d_\ell\)            & Depth of the gadgets from \zcref{assume:concatenated-codes}.\footnote{\(\poly(d_\ell)\) appears in many bounds and should be thought as logarithmic in the total computation size.} \\
		\(\epsilon\)          & Maximum permissible failure probability of simulated circuit.
        \\
        \hline\hline
	\end{tabular}
	\caption{\label{table:processor-params}Table of parameters used in \zcref{sec:quantum-only-processor}.}
\end{table}

\subsection{Fault-tolerant interfaces}\label{sec:ft-interface}
In order to provide a stream of noiseless instruction bits, we must construct a tile that receives a stream of bits forming noisy codewords of the \emph{program code} and outputs a noiseless stream of bits encoded in the computational code.
We begin with the following fact about the existence of efficiently decodable good binary classical codes.
\begin{ingredient}[Program code]\label{assume:program_code}
	There exists a family \(i \in {\mathbb{N}}\) of classical binary codes with block length \(\Nprogi = 2^{\Theta(i)}\) such that \(\log_2 \Nprogi \) is integer, rate \(\Kprogi/\Nprogi = 1/4\), polynomial (in \(\Nprogi\)) sized encoding circuit \(\mathcal{E}_{\mathsf{prog}} \colon \{0,1\}^{\Nprogi/4} \to \{0,1\}^{\Nprogi}\), and an polynomial sized decoding circuit \(\mathcal{D}_{\mathsf{prog}}\colon \{0,1\}^{\Nprogi} \to \{0,1\}^{\Nprogi/4}\) that can correct all errors of weight less than \(\alpha \Nprogi\) where \(\alpha \in (0,1/4]\) is some constant (e.g. \(\frac{1}{100}\)).
\end{ingredient}
This assumption can be instantiated using low-density parity check (LDPC) codes~\cite{SipserSpielman,zemor2002expander}.%
We fix a particular element of this sequence of length \(\Nprog\) which is the minimum element such that \(2^{-\alpha \Nprog} \le e^{-2^\ell}\) where \(\ell\) is the family index used for the computation code.
This implies that \(\Nprog = \Theta(\poly(d_\ell))\).
We define the \(K_{\mathsf{prog}}\) to be the number of bits encoded in this code.

\subsubsection{Advance tile}\label{sec:adv-tile}
\begin{figure}
	\centering
	\includegraphics[width=0.4\columnwidth]{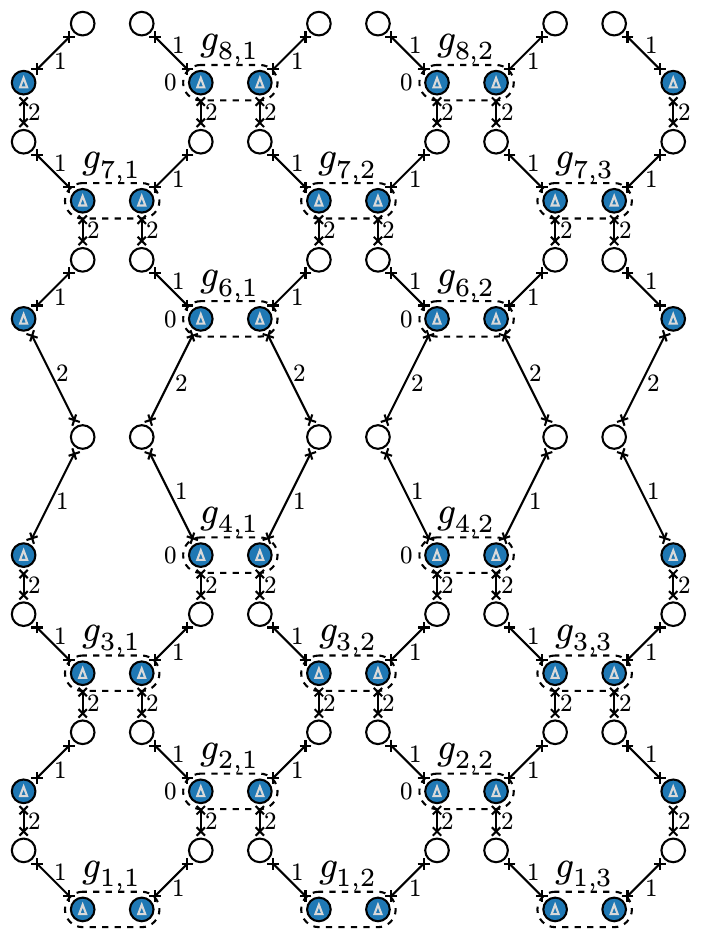}
	\caption{\label{fig:adv-tile}
		(Right) An implementation of an \(\gateAdv\) tile by ``skipping'' a layer.
		The length \(\sqrt{5}\approx 2.2\) \(\gateSWAP\) gate can be removed by ``smearing'' the jump across the entire length of the tile.
		However, the tradeoff is that qubit positions no longer globally respect a grid, and compatibility between vertically and horizontally propagating tiles is somewhat lost.
		Neither feature is required for our purposes.
	}
\end{figure}

Our tiles most naturally operate on slots that enter simultaneously.
In order to perform operations between slots separated by a time delay, it will be convenient to introduce a mechanism to accelerate the progress of slots propagating in a tile relative to other tiles.
\footnote{We can also accomplish this by a \emph{meander} which takes a longer path via side to side movement.}
Suppose that a layer of gates in a gadget is identity (as for the \(\gateI\) gadget in \zcref{assume:concatenated-codes}).
Then, we may simply skip over the corresponding station (of the tile) using a \(\gateSWAP\) of length \(\sqrt{5}\).\footnote{Recall that a similar length \(\gateSWAP\) gate was required for turning.}
See \zcref{fig:adv-tile}.

The result is that a slot exits the tile one tick earlier than it would otherwise.
We call such a tile built out of the \(\gateI\) gadget an \(\gateAdv\) tile.
The advance tile will be indicated in future figures by an identity tile with double dashed lines.

\subsubsection{Deserialization}\label{sec:frame-deserialization}

\begin{figure}
\includegraphics[width=\columnwidth]{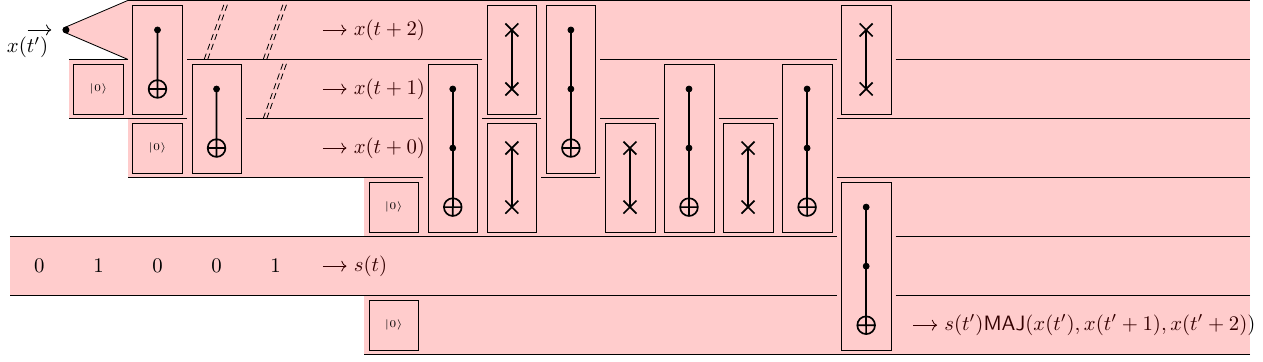}
\caption{\label{fig:deserialization-implementation}Implementation of a deserialization and computation which decodes a 3-bit repetition code and outputs \(0\) when the output is not valid.
Dashed double diagonal lines indicate \(\gateAdv\) tiles.
The metronome sequence is given as \(s\).}
\end{figure}

A basic primitive for classical processing on a classical datastream is deserialization.
Let \(x(t)\) be the encoded classical state of a slot entering the input of our tile at time \(t\).
A width \(n\) deserialization tile has \(n\) output tapes 
and tape \(i \in [n]\) outputs value \(x(t+i-\delta)\) at time \(t\) where \(\delta\) is the overall delay of the tile.
Thus, the output is a ``sliding window'' of size \(n\) along the datastream \(x(t)\).
This is accomplished using the following depth-\((n+2)\) combination of tiles:
First, initialize tapes \(1\) to \(n-1\) in \(\ket{0}\).
Then, place the following consecutive columns of tiles for each \(j \in [n-1]\): 
Copy the value from tape \(j\) to tape \(j+1\) using \(\gateCNOT\) 
controlled on tape \(j\) followed by a \(\gateAdv\) tile for all tapes \(i \le j\).
At the output, tape \(i' = n-1-i\) will have passed through \(i'\) \(\gateAdv\) tiles, 
so at time \(t\), it has the value \(x(t+i'-\delta) \) where \(\delta\) is the overall delay.
The clocked circuit is illustrated in \zcref{fig:deserialization-implementation}.

\subsubsection{Frame processing and reserialization}\label{sec:frame-processing}

Later, we will process datastreams in non-overlapping ``frames'' which are contiguous windows of size \(n\), e.g., \((x(t_0), \dots x(t_0 +n - 1))\) and \((x(t_0 + n), \dots x(t_0 + 2n - 1))\).
We would like to reassemble the processed data into a serial datastream.
In order to accomplish this, we will need the assistance of a second ``metronome'' datastream \(m(t)\) that gives the signal \(10\dots 0 = 10^{(n-1)}\) along with some arbitrary datastream \(x(t)\).
The metronome datastream will indicate when the ``sliding window'' of the deserialization of \(x\) is aligned with a frame.

Let \(P\) be a tile performing some data processing we would like to run on the data stream.
It is some combination of tiles that act on the \(n\) tapes, potentially with ancilla qubits, that implements a map \(f\colon \{0,1\}^n \to \{0,1\}^n\).
Let \(\delta\) be the delay of the data processing tile \(P\), and let \(z_i(t) = f(x(t), x(t+1), \dots, x(t+n-1))|_i\) be the output of the data processing tile at tape \(i \in [n]\) and time \(t + \delta\).
To reassemble the data, we modify the output to be \(z_i(t)' = z_i(t)m(t-\delta)\).
That is, the output is \(0\) unless the data enters the processing tile at the same time that the metronome stream is \(1\).
Then, tape \(i\) of the output \(z_i(t)'\) is advanced \(i\) times, and all tapes are summed using \(\gateCNOT\).
Let \(\delta'\) be the total delay of all previous steps.
Tape \(i\) will have the value \(z_i(t+i)'\), so value output from the sum at time \(t+\delta'\) will be \(s(t) = z_0(t)' + \dots + z_{n-1}(t+n-1)'\).
Now consider some time interval \([\tau,\tau+n]\) for which \(m(\tau-\delta') = 1\).
This must be the unique time in the interval \([\tau, \tau + n]\) for which \(m\) is non-zero, so for \(t \in [\tau, \tau + n]\) the only non-zero term in the sum is \(s(t) = z_{t-\tau}(\tau)\), the output of tape \(t-\tau\) from processing a frame beginning at time \(\tau\).

\subsubsection{Interface}\label{sec:interface}
We now need a tile capable of interfacing to the outside world.
Recall that we can only control a constant number of qubits.
Furthermore, from \zcref{assume:concatenated-codes}, we can encode and unencode a code block with error rate proporational to the gate error rate using \(\gateInject\) and \(\gateEject\).
The interface tile is the only tile where controlled qubits appear in the construction.
\begin{construction}[Interface tile]
	The interface tile has one input and one output.
	It executes \(\gateEject\) to unencode an encoded qubit and then \(\gateInject\) to encode it again.
	The single qubit at the output station of \(\gateEject\) corresponding to the output of the gadget is an \emph{controlled qubit}.
	We may perform a projective measurement and \(\ket{0}\) and \(\ket{1}\) preparations according to the control sequence.
\end{construction}
When the interface tile is used only for communication to the processor, we may omit the \(\gateEject\) tile and replace it by a physical qubit preparation (input interface).
Likewise, when we use the interface tile only for communication out of the processor, we may omit the \(\gateInject\) tile and measure the output qubit (output interface).

\subsubsection{Metronome}\label{sec:metronome}

We are now ready to produce the metronome stream needed in \zcref{sec:interface-tile} for the frame alignment.
Our frames will be of size \(\Nprog\).
Place an interface tile followed by a deserialization~(\zcref{sec:frame-deserialization}) of size \(n=\Nprog^2\).
The control sequence sends the (noise-frozen) metronome stream via the interface tile.
After the deserialization, place a circuit constructed out of primitive tiles to compute the majority of elements at positions \(t, t+\Nprog, t+2N_{\mathsf{prog}}, \dots, t+(\Nprog-1)\Nprog\) of the datastream (e.g. by a sorting network) and output it.

\subsubsection{Control tile}\label{sec:program-code-decoder}
We are now ready to construct the control tile.
It outputs a stream of noise-corrected control bits 
that may then be used for controlling the logical operations.
Recall that the program code comes with an efficient decoder \(\mathcal{D}_{\mathsf{prog}}\).
Since the program code has rate \(1/4\), 
we will need to have 4 parallel streams in order to achieve an output stream with rate \(1\).

\begin{construction}[Control tile]
	Place a metronome tile and four interface tiles which will receive codewords of the program code such that the overall delay from all interface tiles is identical.
	Use the metronome data stream from the metronome tile to execute (via primitive tiles) the program code decoder \(\mathcal{D}_{\mathsf{prog}}\) on each of the four input program code streams using the frame processing~\zcref{sec:frame-processing}.
	To assemble the four independent streams, stream \(i \in \{0,1,2,3\}\) is given a relative delay \(iN_{\mathsf{prog}}/4\) and then all streams are summed to obtain the control stream, the output.
\end{construction}

\subsubsection{Readout tile}\label{sec:readout-tile}
We will also need a means of retrieving the answer from our processor.
This can be accomplished by essentially reversing the construction of \zcref{sec:program-code-decoder}.
Consider a \emph{readout code}, which, for convenience will be taken to be the program code.

Given a stream of data, we would like to encode it into frames of the readout code which we then output via an interface tile.
\begin{construction}[Readout tile]
	Our tile receives a stream of readout data, and the metronome signal from a metronome tile.
	Use the metronome signal to encode the readout data stream (via primitive tiles) into four output streams where the frames are aligned with the metronome signal.
	Send each output stream to an interface tile.
\end{construction}
This tile has size \(\poly(\Nprog)\) primitive tiles, since the program code was assumed to have an efficient encoder.

\subsection{Operations}\label{sec:operations}
Our operations will act on the memory tape constructed in \zcref{thm:memory}.
We introduce the following three operations:
\begin{enumerate}
	\item Apply \(\gateCS\) to two consecutive qubits of the memory tape
	\item Apply \(\gateH\) to a qubit of the memory tape
	\item Measure a qubit of the memory tape and reset it to \(\ket{1}\).
\end{enumerate}
Consider a 1D-geometrically local quantum circuit\footnote{We restrict to open boundary conditions i.e. a line.} \(C\) with the Clifford + \(\gateCS\) gateset\footnote{One and two-qubit Clifford gates along with \(\gateCS\).} on \(M\) qubits and \(D\) layers.
\(C\) may be implemented with \(W=O(M)\) qubits and \(O(D)\) layers using the above operations:
Intersperse a constant density of ancilla qubits fixed to \(\ket{1}\) with qubits of \(C\).
Using the ancilla, any single-qubit Clifford operation can be implemented with \(\gateCS\) and \(\gateH\).
\(\gateCS^2\) and \(\gateH\) gives \(\gateCNOT\), which along with \(\gateCS\) itself, completes the Clifford+ \(\gateCS\) gateset.
Using \zcref{prop:long-range-sim},\footnote{We replace \(\xchg\) with \(\gateSWAP\).} a Clifford+\(\gateCS\) circuit with arbitrary connectivity can also be implemented in time \(O(WD)\) and space \(O(W)\) (e.g. \(2W/3\)).

\begin{figure}
	\centering
	\label{fig:processor}
	\includegraphics[width=5.5in]{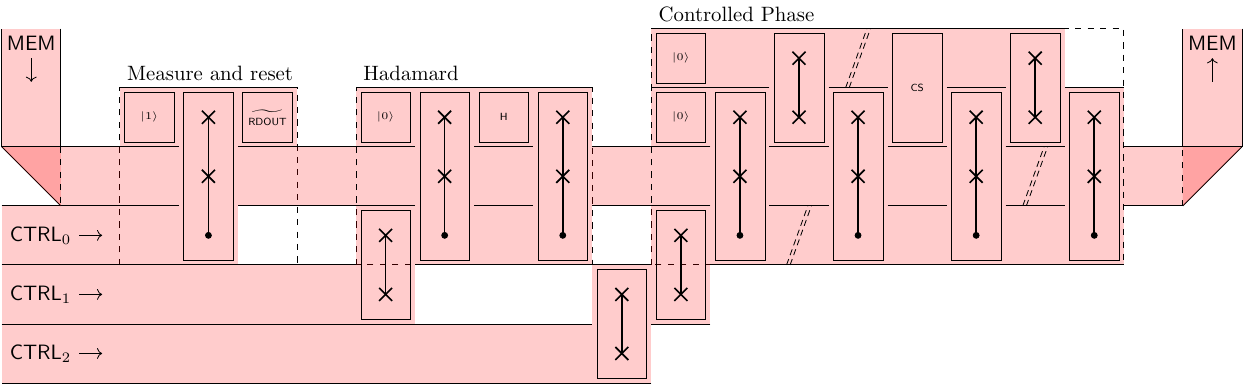}
	\caption{\label{fig:univ-processor-ops} Universal computation on a memory tape (\zcref{sec:memory}) given 3 input control streams \(\mathsf{CTRL}_0\) through \(\mathsf{CTRL}_2\). Except for the memory tape itself, all slots propagate rightward and so the layout is insensitive to the ratio of depth vs width of the gadgets.
		When a slot from \(\mathsf{CTRL}_0\) is \(\ket{1}\), the measure and reset section replaces a slot on the memory tape by \(\ket{1}\).
		The slot is (noisily) read out with an interface tile, and the slot is replaced with a \(\gateEject\) tile containing one controlled qubit.
		When a slot from \(\mathsf{CTRL}_1\) is~\(\ket{1}\), the Hadamard tile takes a slot from the memory tape, applies Hadamard, and replaces it back on the memory tape.
		When a slot from \(\mathsf{CTRL}_2\) is~\(\ket{1}\), two consecutive slots from the memory tape are routed to a controlled-\(\mathsf{S}\) tile and then returned to the memory tape.
		Note that the advance tile \(\gateAdv\) is used to access two consecutive slots.
	}
\end{figure}

By conditionally routing slots based on control streams, this gateset can be implemented on the memory tape by a simple configuration of tiles from \zcref{assume:concatenated-codes} and the advance tile. See \zcref{fig:univ-processor-ops}.
In particular, since there are only 4 possible operations, only 2 control streams are required.
The actual operation can be easily computed e.g. using a circuit in \zcref{fig:instr-decoder}.

\subsection{Processor}\label{sec:processor}\label{sec:control-sequence}
We can now assemble all the ingredients into a fault-tolerant quantum processor.
\begin{construction}[Processor]\label{constr:processor}
	Let \(\ell, L \in \mathbb{N}\) be parameters corresponding to the size of the codes from \zcref{assume:concatenated-codes} and roughly, the side length of the memory tape.
	Place a memory tape as constructed in \zcref{thm:memory} on the plane.
	The memory tape has \(W = d_\ell \left\lfloor \frac{4 w_\ell L^2}{d_\ell}\right\rfloor\) slots.
	Insert the measure and reset, Hadamard, and controlled phase operation tiles (\zcref{fig:univ-processor-ops}) into the memory tape, replacing identity primitive tiles as necessary.
	Place 2 control tiles next to the operation tiles, convert the outputs from binary to one-hot encoding, and route the outputs to the respective operation tiles.
\end{construction}

This processor is programmed as-follows: Let \(W\) be the number of slots in the memory tape.
Let \(M\) be a circuit to be simulated on \(W\) qubits and using the operations from \zcref{sec:program}.
Let \(T\) be the overall runtime.
We specify the stream exiting the control tile \(c \colon [T] \to \{0,1\}^2 \simeq \{0,1,2,3\}\) i.e. the logical level stream.
For a time \(t\) we can associate a qubit \(i\) and circuit layer \(j\) of \(C\) to it by writing \(t = i_t + W j_t\) where \(i_t \in [W]\).
\begin{align}
	c(t) = \begin{cases}
		       0 & \text{otherwise}  \\
		       1 & \text{\(\gateH(i_t)\) is at layer \(j_t\) in \(C\)} \\
			   2 & \text{Qubit \(i_t\) is reset or measured at layer \(j_t\) in \(C\)} \\
			   3 & \text{\(\gateCS(i_t, i_t+1)\) at layer \(j_t\) in \(C\)} \\
	       \end{cases}
\end{align}
The encoded stream \(\bar{c} \colon [T] \to \{0,1\}^8\) (the control sequence) can be obtained by simply computing the encoding \(\mathcal{E}_\mathsf{prog}\) for each frame.

In order to operate the processor, this encoded stream is supplied to the respective interface tiles by setting the single controlled qubit to \(\ket{0}\) or \(\ket{1}\) as part of the control sequence.
At each tick, the single controlled qubit in each interface of the readout tile is measured, and the result is written to an output tape.
The results of the output tape are written with some known latency \(\poly(N_\mathsf{prog})\) related to the time a measurement instruction is specified.
Likewise, the frames of the output code written to the output tape are aligned with respect to a delayed version of the metronome stream.
The delay is known, so the output tape may be decoded by a classical postprocessing procedure that runs the decoder of the program code.
Positions of the decoded output tape corresponding to a time at which a measurement was not made are discarded.
\begin{figure}
A figure showing compilation of a circuit into the control sequence.
\label{fig:compilation}
\end{figure}

We first will show that this processor is not too large.
\begin{proposition}
    The closed clocked circuit \(C_\bullet\) described in \zcref{constr:processor} is a 2D geometrically local circuit of size \(O(w_\ell^2 L^2 + \poly(w_\ell))\) qubits and \(O(L^2 + \poly(w_\ell))\) primitive tiles.
    It has \(13\) controlled qubits.
\end{proposition}
\begin{proof}
    First, recall that \(\Nprog = \Theta(\poly(d_\ell))\) (\zcref{sec:ft-interface}) and \(d_\ell = \poly(w_\ell)\).
    The readout, control, and metronome tiles use at most \(\poly(\Nprog)\) primitive tiles.
    We may need to insert identity primitive tiles to route between these tiles, but we will only need \(\poly(\Nprog)\) of them.
    In addition to these tiles, we also must augment the memory tape with \(O(1)\) additional tiles. (\zcref{fig:univ-processor-ops}).
    The memory tape is as in \zcref{thm:memory} and uses \(O(w_\ell^2 L^2)\) qubits.
    With the exception of the memory tape, all tiles can be laid out linearly in a single direction, so the placement can be independent of the relationship between \(d_\ell\) and \(w_\ell\).
\end{proof}

We now are ready to prove the main result.
\begin{theorem}[store={quantumprocessor},note={Quantum processor}]\label{thm:all-quantum-processor}
    Let $d_\ell$ and $w_\ell$ be the fault-tolerant gadget depth and space, respectively,
    from \zcref{assume:concatenated-codes}.
    There exists a maximum tolerable noise \(p_*^{\mathsf{comp}}>0\) such that,
    using the closed clocked circuit \(C_\bullet\) described in \zcref{constr:processor} (with \(O(1)\) controlled qubits)
    subject to \(p\)-locally stochastic noise with \(p \in [0,p_*^{\mathsf{comp}}]\),
    efficient preprocessing (to prepare the control sequence), and efficient postprocessing,
    the output of any Clifford+\(\gateCS\) quantum circuit \(M\)
    of depth \(D\) and space \(S \le 2W/3\)
    can be sampled to error \(O\left(\left[W^2SD + \poly(d_\ell)\right]e^{-2^\ell}\right)\)
    with a run time of \(O(WSD + \poly(d_\ell))\).
    The initial state of \(C_\bullet\) is arbitrary.
    If \(N\) is the number of physical qubits \(C_\bullet\) is supported on, the number of logical qubits is \(W = \Omega(N/w_\ell)\).
\end{theorem}
\begin{proof}
    The majority of the argument parallels that of \zcref{thm:memory} with the major difference occurring due to the control and readout tiles.
    Let \(M\) be a quantum circuit using the Clifford+CS gate set of depth \(D\) and width \(S \le W\) to be sampled from.
    We may compile \(M\) into operations from \zcref{sec:operations} at the cost of blowing up the depth to \(O(SD)\).
    Let \(\bar{M}\) be the compiled control sequence (as in \zcref{sec:control-sequence}) implementing these operations padded with \(\Nprog^2\) identity operations before and \(\delta+\Nprog\) identity operations after, where \(\delta = \poly(\Nprog)\) is the delay between the interface tiles of the control tile and the interface tile of the readout tile.
    \(\bar{M}\) has length \(O(WSD + \poly(\Nprog))\).
    Consider the execution of the controllable quantum circuit \(C_\bullet\) with control sequence \(\bar{M}\), \(C_{\bar{M}}\), which has depth \(3T\), excluding classical postprocessing.

    The initial padding of \(\bar{M}\) ensures that the output from the control tiles is independent of the initial state, while the final padding ensures that there is enough time to write the answer to the output tape.
    Thus, in the absence of faults, by construction, the decoded output tape is a sample from \(M\).\footnote{This is establishing the reference computation for us to compare the noisy computation against. If the noisy computational is equivalent to this reference computation then we get the right answer. Importantly this reference computation looks very close to the actual gadgets that we run.}
    
    We now must show that the noisy circuit is equivalent to this circuit with high probability.
    To do so, we will build up a set of bad fault paths and show that a locally stochastic fault does not contain any of these bad fault paths with high probability.
    Consider the execution trace of \(C_{\bar{M}}\).
    We will ignore any incomplete gadgets in \(C_{\bar{M},\mathsf{TR}}\): Any incomplete gadgets at the end of the circuit cannot affect the output tape, and any incomplete gadgets at the beginning of the circuit can be absorbed into the initial state.
    
    The next part of the argument will parallel much of \zcref{sec:11Dto20D}.
    Let \(\Omega\) be the set of tiles of \(C_\bullet\).
    For each timestep \(t\) and each tile \(g \in \Omega\), there is a corresponding gadget in the execution trace \(C_{\bullet,\mathsf{TR}}\).
    Recall that by \zcref{prop:trace-gadgets}, each gadget in \(C_{\bar{M},\mathsf{TR}}\) is fault-equivalent to a gadget of \zcref{assume:concatenated-codes}.
    For a tile \(g \in \Omega\) and time \(t\), let \(\badfaults_g^t\) be the set of bad fault paths associated to the gadget \((g,t)\) induced by the bad fault paths of the corresponding gadget of \zcref{assume:concatenated-codes}.
    Let \(\Omega_{\mathsf{comp}} \subseteq \Omega\)  be the set of tiles that are not \(\gateInject\) or \(\gateEject\).
    We will enforce that all gadget applications corresponding to these tiles are subject to a fault path that avoids the corresponding bad faults.
    We define
    \begin{align}
        \badfaults_{\mathsf{comp}} \equiv \boxplus_{t} \boxplus_{g \in \Omega_{\mathsf{comp}}} \badfaults_g^t~.
    \end{align}

    Let \(\Omega_{\mathsf{IO}} = \Omega \setminus \Omega_{\mathsf{comp}}\) be the tiles corresponding to \(\gateInject\) and \(\gateEject\).
    In \(\gateInject\) and \(\gateEject\), there are very low weight bad fault paths, so we must be willing to tolerate failures.
    Instead, we will ensure that they only fail in a way that is correctable by the following program code decoder or majority gate.
    The failure modes are as follows:
    \begin{enumerate}
        \item The output of the metronome tile \zcref{sec:metronome} may not match the true metronome stream.
        \item A frame of the program code is unsuccessfully decoded by the control tile \zcref{sec:program-code-decoder}
        \item A frame of the readout stream is unsuccessfully decoded by the post processing (\zcref{sec:readout-tile})
    \end{enumerate}
    For the first case, consider all subsets corresponding to the input of the majority in the metronome tile\footnote{Actually, we extend these subsets to the interval \([-\Nprog^2, T+\Nprog^2]\) and then restrict the final bad fault paths as needed.} \(S_t = \{t, t+N_{\mathsf{prof}}, t+2N_{\mathsf{prof}}, \dots t+(N_{\mathsf{prof}}-1) N_{\mathsf{prof}}\} \subseteq [T]\).
    For the metronome tile output to be correct, we must have less than \(\Nprog/2\) bit flips in any such input.
    Define the set of bad faults paths\footnote{For simplicity, we will enforce the constraints uniformly on all tiles of \(\Omega_{\mathsf{IO}}\).}
    \begin{align*}
        \badfaults_{\mathsf{IO},1} \equiv \bigwtsum_{g \in \Omega_{\mathsf{IO}}}\bigwtsum_{t\in [T]} \bigwtsum_{\substack{U \subseteq S_t \\ |U| = \Nprog/2}} \bigwtprod_{t' \in U} \badfaults_g^{t'}~.
    \end{align*}
    In this formula in which we count bad faults,
    $\wtsum$ should be read as ``OR'' and $\wtprod$ as ``AND.''
    For the latter two failures modes,
    we define the interval
    \(S_t' = \{t, t+1, t+2, \dots, t+\Nprog-1\} \subseteq [T]\)
    corresponding to a frame of the program or readout codes.
    For this frame to be successfully recovered,
    we must have that the number of fault bits is less than \(\alpha \Nprog \)
    where \(\alpha\) is the decoding radius from \zcref{assume:program_code}.
    Define the following set of bad fault paths which ensure that this is the case
    \begin{align*}
        \badfaults_{\mathsf{IO},2} \equiv \bigwtsum_{g \in \Omega_{\mathsf{IO}}}\bigwtsum_{t\in [T]} \bigwtsum_{\substack{U \subseteq S_t' \\ |U| = \alpha \Nprog}} \bigwtprod_{t'\in U}\badfaults_g^{t'}~.
    \end{align*}

    Consider a fault \(\fault\) of \(C_{\bar{M},\mathsf{TR}}\) that avoids \(\badfaults_{\mathsf{processor}} \equiv \badfaults_{\mathsf{comp}}\wtsum \badfaults_{\mathsf{IO},1} \wtsum \badfaults_{\mathsf{IO},2}\).
    We now show that \(C_{\bar{M},\mathsf{TR}}\) subject to \(\fault\), \(C_{\bar{M},\mathsf{TR}}[\fault]\) gives the correct output.
    Let \(R_{\mathrm{corr}} \subseteq \Omega \times [T]\) be set of gadgets of \(C_{\bar{M},\mathsf{TR}}[\fault]\) for which \(\fault\) avoids \(\badfaults_g^t\), and \(R_{\mathrm{noisy}} \subseteq \Omega \times [T]\) the set of gadgets that are followed by a gadget in \(R_{\mathrm{corr}}\).
    By construction of \(\badfaults_{\mathsf{processor}}\), \(R_{\mathrm{corr}}\) includes all gadgets that are not \(\gateInject\) or \(\gateEject\).
    We now apply the level reduction theorem~\cite[Theorem 4.11]{he2025composable} to obtain a new circuit \(\mathcal{C}_{\mathsf{logical}}^{\fault}\) where:
    \begin{itemize}
        \item All gadgets in \(R_{\mathrm{corr}}\)
        are replaced by the corresponding logical operation acting on the logical data
        and (in parallel) an auxiliary noise operation acting on an environment subsystem.
        \item All gadgets in \(R_{\mathrm{noisy}}\) are replaced by a superoperator acting jointly on the environment subsystem and the logical subsystem (corresponding to that slot).
        \item Any other gadget \(\Omega \times [T] \setminus (R_{\mathrm{noisy}} \cup R_{\mathrm{corr}})\) is at the end of the circuit and can be absorbed into the environment subsystem.\footnote{In particular, it does not interact further with the backwards lightcone of the output except via faults.}
    \end{itemize}
    In the case of \(\gateEject\) gadgets in \(R_{\mathrm{noisy}}\), the noise superoperator may only be a classical operation.\footnote{The error must preserve the classical-quantum type of the original operation.}
    In the case of \(\gateInject\) gadgets in \(R_{\mathrm{noisy}}\), the noise superoperator is a quantum operation, but it is immediately projectively measured\footnote{See \zcref{sec:interface}.} and so may be replaced by a classical operation.
    This new circuit is equivalent in the sense that it produces the same output distribution
    \begin{align}
        \circuitMap{\mathcal{C}_{\mathsf{logical}}^{\fault}} = \circuitMap{C_{\bar{M},\mathsf{TR}}[\fault]}~.
    \end{align}
    Circuit locations of \(\mathcal{C}_{\mathsf{logical}}^{\fault}\) are either in \(R_{\mathrm{noisy}}\), \(R_{\mathrm{corr}}\), or an irrelevant environment subsystem~\cite{he2025composable}.\footnote{Very strictly speaking, because of the time staggering of the gadgets, \(\mathcal{C}_{\mathsf{logical}}^{\fault}\) is a tensor network rather than a circuit because it has wires crossing layers and therefore does not possess a proper foliation. This can be fixed by inserting identity, but it doesn't matter for our purposes.}
    
    From now on we consider \(\circuitMap{\mathcal{C}_{\mathsf{logical}}^{\fault}}\).
    If \(\fault\) is \(\badfaults_{\mathsf{IO},1}\)-avoiding, then in each input to the metronome tile majority, less than \(\Nprog/2\) of the inputs originate from a \(\gateInject\) gadget for which \(\fault\) is supported on an element of \(\badfaults_g^t\).
    Likewise, if \(\fault\) is \(\badfaults_{\mathsf{IO},2}\)-avoiding, then in each frame input to the decoder, less than \(\alpha \Nprog\) of the inputs originate from a \(\gateInject\) gadget for which \(\fault\) is supported on an element of \(\badfaults_g^t\).
    In both cases, with the exception of the initial \(\Nprog^2\) ticks, this means that the input to the respective tiles match the input to the respective interface tiles up to some low weight error that the decoder/majority successfully corrects.\footnote{The circuit implementing the majority can be thought of as a gadget that sends noisy input to noiseless output and cannot tolerate any faults. Likewise with the decoder. We apply level reduction a second time to move the noise from the circuit to the environment subsystem completely.}

    Likewise, by a similar argument, less than \(\alpha \Nprog\) bits of each frame of the output code are modified by the error, so the decoder applied by the post-processing \(\mathcal{T}_{\mathsf{prog}}\) successfully recovers a bitstring output from the circuit.
    It follows that the output distribution is proportional to the originally sampled circuit.
    Note that the output is not in general normalized, because \(\fault\) was not restricted to be a CPTP superoperator
    \begin{align}
        \mathcal{T}_{\mathsf{prog}}\circ\circuitMap{\mathcal{C}_{\mathsf{logical}}^{\fault}}(\rho') \propto M~.
    \end{align}
    By an identical argument to \zcref{thm:memory}, the output distribution is then within total-variation distance \(\weightenum{\badfaults_{\mathsf{processor}}}{p}\) whenever \(\fault\) is distributed according to a \(p\)-local stochastic distribution \(\mu\).
    \begin{align}
        \frac{1}{2}\tr \left|M - \sum_{\fault} \mu(\fault) ~ \mathcal{T}_{\mathsf{prog}}\circ\circuitMap{C_{\bar{M},\mathsf{TR}}[\fault]} \right| \le \weightenum{\badfaults_{\mathsf{processor}}}{p}~.
    \end{align}
    It remains to bound \(\weightenum{\badfaults_{\mathsf{processor}}}{p}\).
    Recall that the weight enumerator adds with \(\wtsum\) and multiplies with \(\wtprod\).
    As in \zcref{thm:memory}, using the fault equivalence \zcref{prop:trace-gadgets} and the weight enumerator bound in \zcref{assume:concatenated-codes}, for \(g \in \Omega_{\mathsf{comp}}\) and \(x \in [0,p_*/3]\), we have that \(\weightenum{\badfaults_g}{x} \le \epsilon(\ell)\) where \(\epsilon(\ell) = O(e^{-2^\ell})\).
    Similarly, for \(g \in \Omega_{\mathsf{IO}}\) and \(x \in [0,p_*/3]\), we have that \(\weightenum{\badfaults_g}{x} \le 3 c_{\mathrm{inject}} x\).
    Thus, for \(x \in [0,p_*/3]\), we have the bound
    \begin{align*}
        \weightenum{\badfaults_{\mathsf{processor}}}{p} &
        \le
        \sum_{t = 0}^{T-1} \sum_{g\in \Omega_{\mathsf{comp}}} \weightenum{\badfaults_{g}}{p}
        \\&\qquad + \sum_{t = 0}^{T-1} \sum_{g\in \Omega_{\mathsf{IO}}} \sum_{\substack{U \subseteq S_t' \\ |U| = \alpha \Nprog}} \left(\weightenum{\badfaults_{g}}{p}\right)^{\alpha {\Nprog}}
        \\&\qquad + \sum_{t = 1}^T \sum_{g\in \Omega_{\mathsf{IO}}} \sum_{\substack{U \subseteq S_t' \\ |U| = \Nprog/2}} \left(\weightenum{\badfaults_{g}}{p}\right)^{\Nprog/2} \\
        &\le T |\Omega_{\mathsf{comp}}| \epsilon(\ell) + 2T |\Omega_{\mathsf{IO}}| \binom{\Nprog}{\alpha \Nprog} \left(3 c_{\mathrm{inject}} p \right)^{\alpha \Nprog} \\
        &\le T |\Omega_{\mathsf{comp}}| \epsilon(\ell) + 2T |\Omega_{\mathsf{IO}}| \left( \frac{3 e c_{\mathrm{inject}}}{\alpha} p \right)^{\alpha \Nprog}
    \end{align*}
    Recall that in \zcref{sec:ft-interface}, \(\Nprog\) was selected such that \(2^{-\alpha \Nprog} \le e^{-2^\ell}\).
    Define the constant \(p_*^{\mathsf{comp}} \equiv \min(p_*/3, \frac{\alpha}{6 e c_\mathrm{inject}})\).
    Then, for \(p \in [0,p_*^{\mathsf{comp}}]\), the contribution from the IO part of the bound is subleading (\(|\Omega_{\mathsf{IO}}| = O(1)\)) and the weight enumerator of \(\badfaults_{\mathsf{processor}}\) can be bounded as
    \begin{align*}
        \weightenum{\badfaults_{\mathsf{processor}}}{p} &\le O\left(T |\Omega_{\mathsf{comp}}| e^{-2^\ell} \right) \\
        &\le O\left((WSD + \poly(d_\ell))(W +\poly(d_\ell)) e^{-2^\ell}\right) \\
        &\le O\left((W^2SD + \poly(d_\ell)) e^{-2^\ell}\right)~.
    \end{align*}
\end{proof}

We briefly remark on the minimum number of controlled qubits possible
\begin{remark}[One controlled qubit]\label{rem:one_qubit}
    Utilizing that an interface tile can be used simultaneously for input and output, it is possible to reduce the number of interface tiles down to 1 by time multiplexing.
    The main task will be to recover which subsequence of the interleaved sequence is which.
    
    First, we will encode a control sequence \(A\) to obtain a new sequence \(A'\) such that \(0\) is encoded as \(01\) and \(1\) is encoded as \(10\) i.e. it is balanced.
    Let \(a_1a_2a_3 \dots\) be the bits of \(A'\) and let \(b_1 b_2 b_3 \dots\) be the bits of the metronome sequence.
    The multiplexed input sequence \(C\) will alternate between the sequence \(A'\) and \(B\) every tick: \(a_1 b_1 a_2 b_2 \dots\).
    
    Having received the data stream, the processor has no external way of determining which symbols are part of the \(A'\) sequence and which symbols are part of the \(B\) sequence, so it much determine this from the sequence itself.
    Fortunately, the metronome sequence is strongly biased towards \(0\) and the control sequence has been arranged to be balanced.
    By measuring the fraction of \(1\)s in a sufficiently large window, say \(\Nprog\), it is possible to identify which subset of the interleaved sequence \(C\) is which as long as the noise rate \(p\) is bounded away from \(1/2\).
    This process will be done in a memoryless fashion for each tick.
    Concretely, examine \(C\) and a copy of \(C\) advanced by 1 in sliding windows of size \(\Nprog\) as in \zcref{fig:metronomedecoder}, discard any sequence that has more than \(\Nprog (p/2 + 1/4)\) \(1\)s in the window, and output the sum.

    Having recovered \(A'\), it must be decoded, yielding \(1\) control bit per \(1/16\) ticks.
    We further multiplex the control bits to alternate between the uses in the processor.\footnote{The offset can be set by the alignment of the program code frame.}
    If we have \(c\) control tiles in the original processor, we now use every position at a multiple of \(16c\) on the memory tape.\footnote{We may also simply pick the tape length to be coprime to \(16c\).}

    Readout is done almost as-normal using the rate \(1/4\) readout code with only every 4th qubit entering the readout tile making it to the readout code.
\end{remark}

The scheme is also naturally resilient to fabrication defects.
First, we define what it means for a controllable quantum circuit to have a set of defective qubits.
Roughly, any gate of the clocked circuit supported on a defective qubit will be permitted to fail arbitrarily.
\begin{definition}[Controllable quantum circuit with defects]
    For a controllable quantum circuit \(C_\bullet\) and subset of qubits \(S_{\mathsf{defect}}\),
    we define the controlled quantum circuit  \(C_\bullet^{S_{\mathsf{defect}}}\)
    with defective qubit set \(S_{\mathsf{defect}}\)
    to be a controllable quantum circuit
    where any spacetime location with a qubit in the defective qubit set
    may be arbitrarily damaged.
    That is, a qubit in the defective qubit set may be faulty at every timestep.
\end{definition}

\begin{remark}[Fabrication defects]\label{rmk:fabrication-defects}
    One can show that \zcref{thm:all-quantum-processor} and \zcref{thm:fast-quantum-processor} hold when the controlled quantum circuit is replaced by one with an \textbf{acceptable} defective qubit set.
    There exists a maximum manufacturing defect rate \(p_*^{\mathsf{defect}}\)
    such that for a defect rate \(p^{\mathsf{defect}} \le p_*^{\mathsf{defect}}\),
    if we choose each qubit independently%
        \footnote{Actually, according to a locally-stochastic noise distribution which permits correlated failures.} 
    with probability $p^\mathsf{defect}$
    to construct a defective qubit set~\(S_{\mathsf{defect}}\),
    then $S_\mathsf{defect}$ is acceptable with constant probability, e.g.,~\(2/3\).

    \hspace{2ex} This is possible, because movement of the slots makes a defective qubit cause an error that is local in spacetime.
    For example, if a qubit at station 18 in \zcref{fig:trace} was faulty, there would always be an error at the corresponding spacetime location in each gadget execution in panel (b).
    
    \hspace{2ex} However, the circuit will not be resilient to certain low weight errors,
    such as a defective controlled qubit.
    The set of qubits that an acceptable defective qubit set must avoid is obtained by ``splitting'' each element of the bad fault set into two components.
    One component will be from the qubit defects and the other component will take the place of the original bad fault set.
    For each gadget \(g\), consider the bad fault paths \(\badfaults_g\).
    A set of defective qubits \(S\) of the corresponding tile of the clocked circuit induces a set of spacetime locations \(S_{\mathsf{loc}}\) (of each copy of the gadget in the execution trace) that will always be faulty.
    For simplicity, we will ignore the distinction between spacetime locations of the gadget itself, copies of the gadget of the execution trace, and qubits of the clocked circuit.\footnote{One can establish fault equivalence with respect to a constant degree attribution graph and use \zcref{lemma:fault-equiv} to convert between them at the cost of a small loss which can be absorbed into a slightly different choice of \(\mathcal{R}\).}
    The remaining transient part of the fault then only needs to be supported on spacetime locations outside of \(S_{\mathsf{loc}}\)
    That is, the fault must avoid the bad fault set \(\badfaults^{(S_{\mathsf{loc}})}_g = \{f \setminus S_{\mathsf{loc}} \mid f \in \badfaults_g\}\).
    
    \hspace{2ex} To ensure that the weight enumerator of \(\badfaults_g^{(S_{\mathsf{loc}})}\) is still sufficiently small, we will ensure that \(S_{\mathsf{loc}}\) avoids at least half of every element of \(\badfaults_g\).
    We can define such a set \(\mathcal{R} = \cup_{f \in \badfaults_g} \left\{g \subseteq f \mid |g| = \lceil |f|/2\rceil\right\}\) that guarantees two \(\mathcal{R}\)-avoiding faults is \(\badfaults_g\)-avoiding.
    We can easily bound its weight enumerator in terms of that of \(\badfaults_g\).
    Recall that the weight enumerator bounds the probability that a locally stochastic fault fails to avoid \(\mathcal{R}\).
    \begin{align*}
        \weightenum{\mathcal{R}}{x} &\le \sum_{f \in \badfaults_g} \binom{|f|}{\lceil f/2\rceil}x^{\lceil f/2\rceil} \le \sum_{f \in \badfaults_g} 2^{|f|} x^{\lceil |f|/2\rceil}\\
        &\le \weightenum{\badfaults_g}{2 \sqrt{x}}
    \end{align*}
    If \(S_{\mathsf{loc}}\) is \(\mathcal{R}\)-avoiding, then an \(\mathcal{R}\)-avoiding fault is also \(\badfaults_g^{(S_{\mathsf{loc}})}\)-avoiding.
    Thus, we can use \(\weightenum{\mathcal{R}}{x}\) as an estimate for \(\weightenum{\badfaults_g^{(S_{\mathsf{loc}})}}{x}\).
    In general, for the majority of the analysis in \zcref{thm:all-quantum-processor}, does not change much: One must separately ensure that the defective qubits of each primitive tile are acceptable (\(\mathcal{R}\)-avoiding) and then replace every occurrence of the bad fault set of the gadget \(\badfaults_g\) with the ``halved'' bad fault set \(\mathcal{R}\).
    The set of defective qubits is \(\mathcal{R}\)-avoiding with high probability with the exception of \(\gateInject\) and \(\gateEject\) which contain qubits that may not be defective.\footnote{This difference stems from the fact that the bad fault paths do not enter into \(\badfaults_{\mathsf{processor}}\) directly, but the restriction on the defective qubit set does.}
    However, there are only a constant number of such qubits.
\end{remark}

\section{Fast quantum processor with classical wires}\label{sec:FastProcessorWithClassicalWiresFurtherDetail}

In order to address the slowdown of the processor in \zcref{sec:quantum-only-processor}, we will introduce a modified model of computation where classical signals are permitted to propagate instantaneously through wires which must be placed at unit density.

The main usage of these signals is to create a random-access quantum memory (RAQM) which is composed of ``cells,'' each storing some number of logical qubits.
We can access qubits of the memory (that is, random-access) arbitrarily with the access time being essentially independent of the qubit position.
The classical wires are used to complete a teleportation protocol to teleport qubits from cells of the RAQM to a particular location where our universal computer will be.
That is, the processor of \zcref{sec:quantum-only-processor} is augmented by the RAQM to obtain a faster processor.

Cells of the RAQM are formed by primitive tiles, so that they will each store a superconstant number of logical qubits.
Additionally, the address of the accessed cell will also be superconstant, and so it takes some amount of time to communicate it to the processor.
These two considerations will appear when it is time to program our computer. 
We will give two compilation schemes that give different trade-offs for the runtime of quantum circuits.

\subsection{Classical-Quantum transducer}
The wires will be classical objects that are not faulty.
We model these classical degrees of freedom as assigning two bits to each lattice site.
Bits may be contained in a ``wire'' which is a contiguous subset of bits in the 2D plane.
A classical gate has input and output bits and is also noiseless.
When it writes to an output bit, all bits of the wire are set to an identical value.
If there is more than one gate writing to the same wire, then the wire takes the \(\mathsf{OR}\) of the written values.\footnote{While the ability to simultaneously take the \(\mathsf{OR}\) of many values is known to be more powerful in the bounded-depth setting, we do not use it for any computation. In our usage, the weaker model where the values `0', `1', or `Z' are output where `Z' does not modify the original value of the wire also suffices. Such values are used in hardware description languages.}

We will also need gadgets to convert an encoded qubit to a bit.
To convert a code block to a bit, we can transversally read out a code block and then execute the decoder using classical circuitry.
To convert a bit to a code block, we can simply prepare a logical \(\ket{0}\) state and then apply a transversal \(X\) controlled on the value of the classical information.
For clarity, we give an addition to \zcref{assume:concatenated-codes} describing this gadget.
\begin{ingredient}[CQ transducer]\label{assume:cq-transducer}
    There is a quantum circuit $\gadget_{\gateCtoQ,\ell}$ that has a single classical input bit and
    a nontrivial output code type $\compCodeType[\ell]$,
    and a quantum circuit~$\gadget_{\gateQtoC,\ell}$ that has a single classical output
    and a nontrivial input code type $\compCodeType[\ell]$.
    These circuits have each a set $\badfaults_{\qGate,\ell}$ of bad fault paths 
    ($\qGate = \gateCtoQ,\gateQtoC$)
    such that for all $x \in [0,p_*]$ the weight enumerator is bounded as
    \begin{align*}\label{eq:weight-enum-1dgadgets}
        \weightenum{\badfaults_{\qGate,\ell}}{x} \le \epsilon(\ell)~.
    \end{align*}
\end{ingredient}

\subsection{Memory cell}\label{sec:memory-cell}
\begin{figure}
	\centering
	\includegraphics[width=0.5\columnwidth]{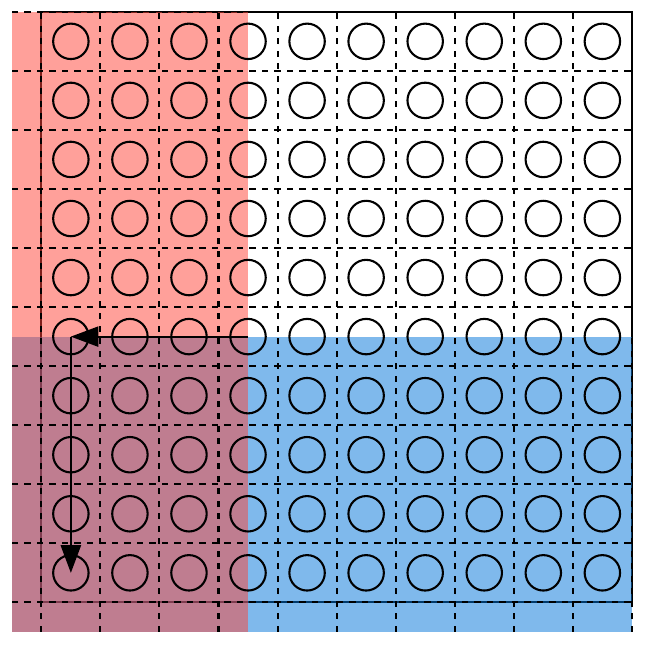}
	\caption{\label{fig:cell-address}Unary addressing scheme for an array of cells (circles). Dashed lines are row/column address lines and filled rectangles indicate the row/column value. Qubits are teleported along the rows and along the westmost column. The cell at row 4 and column 3 is being accessed.}
\end{figure}

Our teleportation-based memory will be constructed out of a primitive unit that we will call \emph{cells.}
A memory cell has two roles. 1) Similarly to the memory tapes in the previous sections, it must store slots. 2) The cell must participate in an entanglement swapping protocol to teleport qubits into and out of the memory.
The memory will have a grid of \(L \times L\) cells, and the main task of the memory will be to access a qubit stored in a cell and move it to a fixed position of the memory (e.g. the bottome left cell).
We call this cell the \textit{primary cell}.
It will play a special role. All other cells are referred to as \emph{secondary cells}.
For simplicity, let us first concentrate on the task of moving data from a particular \emph{accessed cell} to the westmost column.

The high level layout of the memory cell will contain three parts:
We will place two small memory tape arranged much like the memory tapes of the previous sections.
One memory tape will store qubits and the other will store classical bits.
The second component creates Bell pairs that are shared between tiles.
The third component is a configuration of primitive tiles that performs some processing on the memory tape which we refer to as the cell computation.
The cell computation is inserted into the memory tape, and a cell in column \(c\) and row \(r\) receives the inputs
\begin{itemize}
	\item qubit \(q_d\) and bit \(f_d\) from the memory tape
    
	\item qubits \(q_{\leftarrow, w}\), \(q_{\leftarrow, e}\), \(q_{\rightarrow, w}\), \(q_{\rightarrow, e}\),
    each forming half of a Bell pair with each neighbor:
    \(q_{\leftarrow, w}\) forms a Bell pair with~\(q_{\leftarrow, e}\)
    of the neighboring cell to the west,
    and \(q_{\rightarrow, w}\) forms a Bell pair with~\(q_{\rightarrow, e}\).
    The subscript w/e denotes which neighboring cell the qubit is in a Bell pair with. 
    The subscript \(\leftarrow\)/\(\rightarrow\) denotes the teleportation direction 
    the Bell pair will be used for.
    
	\item Column address bits \(a_c\) and \(a_{c+1}\).
    
	\item Row address bits \(b_{r}\) and \(b_{r+1}\)
	\item Fixup bits \(f_x\) and \(f_z\).
\end{itemize}
The cell computation outputs four classical bits \(m_{\leftarrow,X}\), \(m_{\leftarrow,Z}\), \(m_{\rightarrow,X}\), \(m_{\rightarrow,Z}\) which will be measurement outcomes from Bell basis measurements.
The input address bits and output measurement bits are communicated with the cell via classical wires.
Column address bits and measurements outputs are shared across columns, row address bits \(r\) are shared across rows, and fixup bits \(f_x\) and \(f_z\) are shared across the array.
Thus, there is only a constant density of classical wires in the memory.

The address will be unary encoded \(i = 1\dots 10\dots 0 = 1^i 0^{\eta-i} \) such that the cell we would like to access receives \(a_m a_{m+1} = 10\) and any cell to the west of the accessed cell receives \(a_m a_{m+1} = 11\).
Cells to the west of the accessed cell will participate in two entanglement swapping protocols:
The east-going protocol teleports a qubit from the west-most cell to the accessed cell, and the west-going protocol teleports a qubit from the accessed cell to the west-most cell.
Thus, the cell computation will perform the following process sequentially.
For convenience, we have included icons corresponding to the local state of the address lines matching that of \zcref{fig:cell-address}.
\NewDocumentCommand{\ramSUSU}{}{
\begin{tikzpicture}[scale=0.17]
    \draw[red,fill,opacity=0.5] (-1,-1) rectangle (0,1);
    \draw[blue,fill,opacity=0.5] (-1,-1) rectangle (1,0);
    \draw (0,0) circle(0.8);
\end{tikzpicture}
}

\NewDocumentCommand{\ramSSSU}{}{
\begin{tikzpicture}[scale=0.17]
    \draw[red,fill,opacity=0.5] (-1,-1) rectangle (1,1);
    \draw[blue,fill,opacity=0.5] (-1,-1) rectangle (1,0);
    \draw (0,0) circle(0.8);
\end{tikzpicture}
}

\NewDocumentCommand{\ramUUSU}{}{
\begin{tikzpicture}[scale=0.17]
    \draw[red,fill,opacity=0.0] (-1,-1) rectangle (1,1);
    \draw[blue,fill,opacity=0.5] (-1,-1) rectangle (1,0);
    \draw (0,0) circle(0.8);
\end{tikzpicture}
}

\NewDocumentCommand{\ramUUUU}{}{
\begin{tikzpicture}[scale=0.17]
    \draw[red,fill,opacity=0.0] (-1,-1) rectangle (1,1);
    \draw[blue,fill,opacity=0.0] (-1,-1) rectangle (1,0);
    \draw (0,0) circle(0.8);
\end{tikzpicture}
}

\NewDocumentCommand{\ramSSUU}{}{
\begin{tikzpicture}[scale=0.17]
    \draw[red,fill,opacity=0.5] (-1,-1) rectangle (1,1);
    \draw[blue,fill,opacity=0.0] (-1,-1) rectangle (1,1);
    \draw (0,0) circle(0.8);
\end{tikzpicture}
}

\NewDocumentCommand{\ramUUSS}{}{
\begin{tikzpicture}[scale=0.17]
    \draw[red,fill,opacity=0.0] (-1,-1) rectangle (1,1);
    \draw[blue,fill,opacity=0.5] (-1,-1) rectangle (1,1);
    \draw (0,0) circle(0.8);
\end{tikzpicture}
}

\NewDocumentCommand{\ramSUUU}{}{
\begin{tikzpicture}[scale=0.17]
    \draw[red,fill,opacity=0.5] (-1,-1) rectangle (0,1);
    \draw[blue,fill,opacity=0.0] (-1,-1) rectangle (1,0);
    \draw (0,0) circle(0.8);
\end{tikzpicture}
}

\NewDocumentCommand{\ramSSSS}{}{
\begin{tikzpicture}[scale=0.17]
    \draw[red,fill,opacity=0.5] (-1,-1) rectangle (1,1);
    \draw[blue,fill,opacity=0.5] (-1,-1) rectangle (1,1);
    \draw (0,0) circle(0.8);
\end{tikzpicture}
}\

\NewDocumentCommand{\ramSUSS}{}{
\begin{tikzpicture}[scale=0.17]
    \draw[red,fill,opacity=0.5] (-1,-1) rectangle (0,1);
    \draw[blue,fill,opacity=0.5] (-1,-1) rectangle (1,1);
    \draw (0,0) circle(0.8);
\end{tikzpicture}
}

\begin{enumerate}
\item \textbf{if} \(f_d=1\), apply the fixup \(X^{f_x} Z^{f_z}\) to \(q_d\).
\item \textbf{if} \ramSSSU \(a_c a_{c+1} b_{r}b_{r+1} = 1110\),  \textbf{then} measure \(XX\) and \(ZZ\) between \(q_{\leftarrow, w}\) and \(q_{\leftarrow, e}\),  and output the result to \(m_{\leftarrow,X}\), \(m_{\leftarrow,Z}\). Likewise, measure \(XX\) and \(ZZ\) between \(q_{\rightarrow, w}\) and \(q_{\rightarrow, e}\), and output the result to \(m_{\rightarrow,X}\), \(m_{\rightarrow,Z}\). Put \(q_d\) and \(f_d=0\) on the memory tape.
\item \textbf{if} \ramSUSU \(a_c a_{c+1}b_{r} b_{r+1} = 1010\),  \textbf{then} measure \(XX\) and \(ZZ\) between \(q_{\leftarrow, w}\) and \(q_{d}\), and output the result to \(m_{\leftarrow,X}\), \(m_{\leftarrow,Z}\). Put \(q_{\rightarrow, w}\) and \(f_d=1\) on the memory tape.
\item \textbf{if} \ramUUSU \ramUUUU \ramSUUU \ramSSUU \ramUUSS \ramSSSS \(a_c a_{c+1} b_{r}b_{r+1} \ne \ast 010\), \textbf{then} put \(q_d\) and \(f_d=0\) on the memory tape, and output \(m_{\leftarrow,X}m_{\leftarrow,Z}m_{\rightarrow,X}m_{\rightarrow,Z} = 0000\).
\end{enumerate}

\subsection{Memory operation}\label{sec:2d-mem:parity-computation}
In order to operate the memory with cells, we will need to perform some computations at the periphery of the memory.
These computations will be super-linear in \(L\) but sublinear in \(L^2\), so the majority of the memory can be devoted to the memory tapes of the cells.
Since classical signals propagate instantaneously, the precise layout of these computations is not critical as long as the density constraint on classical wires is satisfied.
For simplicity, we will perform some of the computations using methods from \zcref{sec:quantum-only-processor} instead of optimally using the classical circuitry.

To complete the entanglement swapping protocol, we must compute the overall parity of the Bell measurements.
A size \(L\) parity can be computed using a circuit of depth \(\log_2 L\).
Thus, we can compute the parity of all the columns by a classical circuit of size \(L \times \log_2 L\) in depth \(O(\log_2 L)\) at the edge of the array of cells.
The Bell measurement outcomes for the row teleportation will be sent along the columns to a tree at the north edge of the array, while the measurement outcomes for the column teleportation will be sent along the rows to a tree at the east edge of the array.

For the addressing, if we have two binary-encoded number \(i,j \in [L]\), we can compute the bit that is \(1\) when \(i \le j\) using a \(O(\log^2 \log L)\) depth sorting network on \(\log_2 L\) bits.
All row and column bits can be computed simultaneously by a classical circuit of size \(L \times O(\log(L) \log^2 \log(L))\) in depth \(O(\log_2^2 \log_2 L)\) at the edge of the array of cells.
We will use a control tile to supply the address of the accessed cell.
However, since the address has length \(2\log_2 L\) and a control tile can only provide one bit per tick, we will need to provide a complete address over multiple consecutive ticks.
We will introduce a metronome stream with periodicity \(2\log_2 L\) to mark when a complete address has been provided.

When a full address has been received, we would like to ``hold'' the value until the next time a full address is provided.
Holding the address allows different positions along the memory tape for each memory cell to be accessed.
To hold the output value, we perform the following circuit: Let \(m(t)\) be the metronome stream and let \(x(t)\) be the address stream.
First, apply a size \(2\log_2 L\) deserialization to the address stream, so that we have some new stream \(\vec{x}_1(t) = (x(t), x(t-1), \dots)\).
Then, much like the frame processing of \zcref{sec:frame-processing}, we will compute the stream \(\vec{x}_2(t) = m(t)\vec{x}_1(t)\) that is non-zero only when the metronome is \(1\).
\(\vec{x}_2(t)\) has only valid addresses or \(0\), but now we must hold the value.
To do so, we again apply a size \(2\log_2 L\) deserialization to \(\vec{x}_2(t)\), so that we have a stream which is a matrix where every row is a time delayed version of \(\vec{x}_2(t)\), \(\mathbf{X}(t) = (\vec{x}_2(t), \vec{x}_2(t-1), \dots)\).
At most one row of this matrix is non-zero at any one time, so we can output the sum of the rows of \(\mathbf{X}(t)\) to have the desired effect.
Call this sum \(\vec{x}_4(t)\).

It is also convenient to have a mechanism to avoid memory access at all.
To do so, we introduce another control tile.
Whenever the control stream from this control tile is \(0\), we output the address of the primary cell instead of \(\vec{x}_4(t)\).
To distinguish the capability of accessing a cell from actually accessing it we introduce the following terminology: If a cell may be accessed in a given tick, i.e. its address is \(\vec{x}_4(t)\), the cell is said to be \emph{open}.

\subsection{2D memory access}\label{sec:2d-mem-access}
The cells in \zcref{sec:memory-cell} are only capable of teleporting along rows.
To obtain access to arbitrary rows (i.e. teleporting along the columns), the westmost column of cells will be special.
They will essentially perform the row teleportation protocol along the columns.
This column of cells will have the additional input qubits \(q_{\downarrow, s}\), \(q_{\downarrow, n}\), \(q_{\uparrow, s}\), \(q_{\uparrow, n}\), each forming half of a Bell pair with each neighbor:  \(q_{\downarrow, s}\) (\(q_{\uparrow, s}\)) forms a Bell pair with \(q_{\downarrow, n}\) (\(q_{\uparrow, n}\)) of the neighboring cell to the south.
We will also output the four classical Bell basis measurement bits \(m_{\downarrow,X}\), \(m_{\downarrow,Z}\), \(m_{\uparrow,X}\), \(m_{\uparrow,Z}\) to classical wires running east-to-west along the array.
We now give the modified rules for the \emph{westmost} column of cells.
They are essentially identical to the row rules.
The notable exception is that rule \ref{item:ram-col-rules-hor-vert-conn} connects the horizontal and vertical teleportation protocols together.

\begin{enumerate}
\item \textbf{if} \(f_d=1\) apply the fixup \(X^{f_x} Z^{f_z}\) to \(q_d\).
\item \textbf{if} \ramSSSS \ramSUSS \(a_c a_{c+1} b_{r}b_{r+1} = 1\ast11\) \textbf{then} measure \(XX\) and \(ZZ\) between \(q_{\downarrow, s}\) and \(q_{\downarrow, n}\),  and output the result to \(m_{\downarrow,X}\), \(m_{\downarrow,Z}\). Likewise, measure \(XX\) and \(ZZ\) between \(q_{\uparrow, s}\) and \(q_{\uparrow, n}\), and output the result to \(m_{\uparrow,X}\), \(m_{\uparrow,Z}\). Put \(q_d\) and \(f_d=0\) on the memory tape.
\item \label{item:ram-col-rules-hor-vert-conn} \textbf{if} \ramSSSU \(a_c a_{c+1} b_{r}b_{r+1} = 1110\) \textbf{then} measure \(XX\) and \(ZZ\) between \(q_{\downarrow, s}\) and \(q_{\leftarrow, e}\),  and output the result to \(m_{\downarrow,X}\), \(m_{\downarrow,Z}\). Likewise, measure \(XX\) and \(ZZ\) between \(q_{\uparrow, s}\) and \(q_{\rightarrow, e}\), and output the result to \(m_{\uparrow,X}\), \(m_{\uparrow,Z}\). Put \(q_d\) and \(f_d=0\) on the memory tape.
\item \textbf{if} \ramSUSU \(a_c a_{c+1} b_{r}b_{r+1} = 1010\) \textbf{then} measure \(XX\) and \(ZZ\) between \(q_{\downarrow, s}\) and \(q_{d}\), and output the result to \(m_{\downarrow,X}\), \(m_{\downarrow,Z}\). Put \(q_{\uparrow, n}\) and \(f_d=1\) on the memory tape.
\item \textbf{if} \ramUUUU \ramSUUU \ramSSUU \(a_c a_{c+1} b_{r}b_{r+1} = \ast\ast00\) \textbf{then} put \(q_d\) and \(f_d=0\) on the memory tape, and output \(m_{\downarrow,X}m_{\downarrow,Z}m_{\uparrow,X}m_{\uparrow,Z} = 0000\).
\end{enumerate}

\subsection{2D Random-access quantum memory}\label{sec:raqm}
We now show that the 2D random access memory has a two-dimensional layout that is not too wasteful.
We will arrange such that the data required for the computation rules enters from the south side of the cell, and that the cell fits inside of a square of dimensions \(a \times a\).
Each cell will be placed on a grid with separation \(a' = O(a)\).
The space in-between cells will be used to route classical wires and produce the Bell pairs that will be sent between neighboring cells.
This space will need to be large enough to accommodate a constant number of wires and tiles for producing and transporting Bell pairs.
We will need to take care that Bell pairs produced in this in-between region arrive simultaneously at the inputs for each cell.
Accordingly, it may sometimes be necessary to turn the tape\footnote{This is referred to as a \textit{meander} in classical circuitry.} back-and-forth in order to fit the necessary number of stations.

We now must lay out the internals of the cell.
The cell computation can be accomplished using a constant number of primitive tiles.
Place this configuration of primitive tiles at the eastern-most side propagating upward.
The memory tapes will occupy the area west of the cell computation and will ``snake around'' like the memory tape in \zcref{sec:memory}.
The total number of primitive tiles in the memory tape will be \(\eta\).
This includes the cell computation tiles such that a slot on the memory tape revisits the same station every \(\eta d_\ell\) ticks.

We now must place the classical addressing circuitry at the periphery of the memory.
In general, the comparison circuit in \zcref{sec:2d-mem:parity-computation} may require a width larger than \(a'\).
To remedy this issue, we utilize the sparsity of the column wires and ``stagger'' the comparison circuitry for each column such that column addressing wires are routed through other comparison circuits as necessary.

\begin{figure}
	\centering
	\includegraphics[width=1.0\columnwidth]{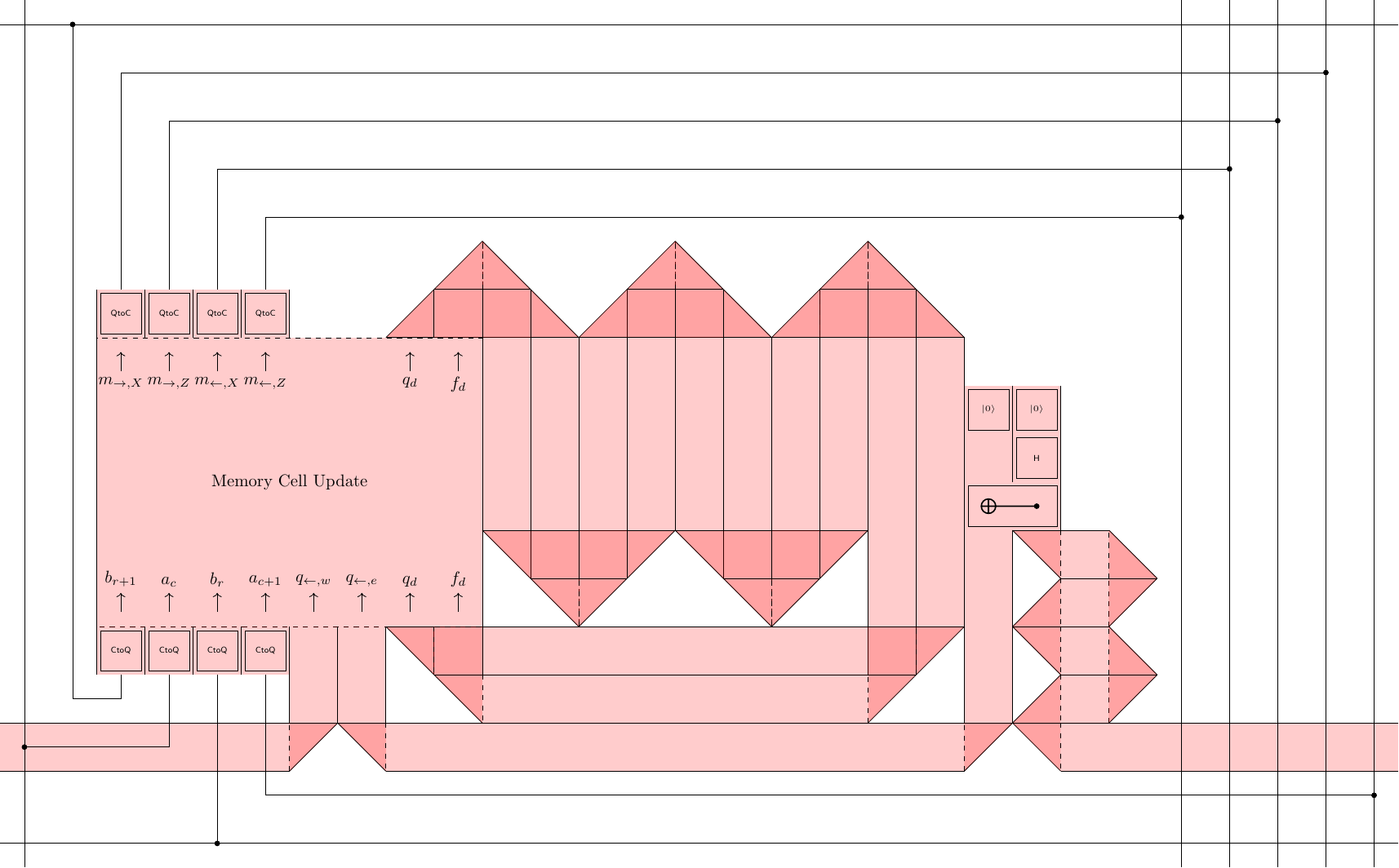}
	\caption{\label{fig:memory-cell}Layout of horizontal memory cell with \(q_{\rightarrow,w}\) and \(q_{\rightarrow,e}\) omitted for space.
    In order to make the memory cell square, the memory tape may be extended upward or further right as needed.
    The constraints for a valid placement in the plane to exist are essentially identical to the memory tape of \zcref{sec:memory}.
    Bell pairs are generated in the interstitial region between cells.
    In general, it may be necessary to place \emph{meanders} such that the delay to each cell is identical.}
\end{figure}

The parity computation is also at the periphery of the memory.
It has a minimum depth \(\log_2 L\).
Let \(\delta\) be the overall delay of the cell computation.
We pad the parity computation with identity gates such that the overall depth of the parity computation matches the time taken for the output of the cell computation to return to the input \(\eta d_\ell - \delta\).
This ensures that the necessary parity fixup is made available to the cells at the appropriate time (one cycle of the tape storing \(f_d\)).
In order for this to always be possible, we restrict \(\eta \ge \frac{ \log_2 L + \delta}{d_\ell}\).
Since the cell computation has a constant depth in terms of the number of tiles, \(\delta = \Theta(d_\ell)\).
There is a choice of \(a = \Theta\left(\sqrt{w_\ell \log L} + d_\ell\right)\) and \(\eta = \Theta\left(\frac{a^2}{d_\ell w_\ell}\right)\) such that this constraint is satisfied and there exists a placement of the tiles (similarly to \zcref{sec:memory}) inside a region of size \(a \times a\).
We show a layout of the memory cell with some lines omitted in \zcref{fig:memory-cell}.

We summarize the construction below.
\begin{construction}[Random-access quantum memory (RAQM)]
    Fix compatible parameters \(\ell\), \(L\).
    Place the classical parity calculation circuitry and addressing tiles described in \zcref{sec:2d-mem:parity-computation} at the periphery of the memory.
    The addressing tiles require two control tiles.
    Place a \(L \times L\) grid of cells where the bottom left cell is the primary cell, the leftmost column of secondary cells also have the additional column connections and rule (\zcref{sec:2d-mem-access}), and the remainder are standard secondary cells (\zcref{sec:memory-cell}).
    Prepare Bell pairs between cells and connect cells to vertical / horizontal classical wires as in \zcref{sec:2d-mem-access}.
\end{construction}

\subsection{Fast quantum processor}
We are now ready to construct the fast quantum processor.
As in previous sections, we will essentially augment the previous constructions.
The all-quantum processor from \zcref{sec:quantum-only-processor} will be a single cell of the random-access quantum memory.
In particular, we will place the control tiles to the side of the random-access quantum memory and place the computation tiles such that they are part of the cell computation of the cell in the  left corner of the random-access quantum memory.
\begin{construction}[Fast quantum processor]\label{constr:fast-processor}
	Let \(\ell, L \in \mathbb{N}\) be parameters corresponding to the size of the codes from \zcref{assume:concatenated-codes} and the number of cells along one direction of the random-access quantum memory.
	Place a random-access quantum memory with parameters from \zcref{sec:raqm}.
	Place the control tiles and operations tiles from the quantum-only processor~\zcref{constr:processor} next to the primary cell of the RAQM such that the operations tiles are part of the cell computation for the primary cell.\footnote{This increases the length of the cell computation slightly, but does not change it asymptotically. It can therefore be absorbed into the original cell computation by taking the maximum depth and padding by identity otherwise.}
\end{construction}

\subsubsection{Programming}\label{sec:fast-processor:programming}
The most native model in which to specify operation of the processor is as follows:
Label each (logical) qubit by a coordinate \((i,j) \in [L^2] \times [d_\ell \eta] \equiv [M] \times [R]\).
Rows correspond to different cells and columns correspond to different positions along the tape.
This ``matrix'' of logical qubits should be thought of as have many rows and few columns, and the total number of logical qubits in the processor is \(W = RM\).
\zcref{sec:quantum-only-processor} specifies how to perform universal logic on the first row of qubits (qubits in the primary cell) by time-encoding the accessed column.
Thus, we have universal control over this \emph{working memory}.

When a cell is open, the RAQM allows us to swap an arbitrary subset of logical qubits from the open cell to the primary cell.
However, we can only switch which cell is open every \(R\) ticks.
Concretely, if cell \(i\) is open, we can swap column-wise any subset of row \(i\) with row \(0\) and apply any depth-1 layer of gates that are nearest-neighbor along row \(0\) in time \(O(R)\).
An optimized compilation respecting these constraints would open a new cell every \(\Theta(R)\) ticks and perform some 1D circuit acting between the open cell and the primary cell.

We will instead describe two naive compilation schemes, the latter being optimal when the circuit is dense.
Consider an arbitrary circuit layer with \(v\) gates acting on logical qubits that are assigned arbitrary coordinates.
For simplicity, let us suppose that the circuit layer is composed of only two-qubit gates.
The execution of one-qubit gates is similar.

\paragraph{Compilation A} The first scheme will execute each gate one-at-a-time in time \(O(v R^2)\):
First, sequentially open the two cells containing the targeted logical qubits and transfer them to the primary cell in time \(O(R)\).
Then use the 1D routing and universal control as in \zcref{sec:quantum-only-processor} to execute the gate in time \(O(R^2)\).

\paragraph{Compilation B} The second scheme is more optimal when there are many gates per timestep \(v\approx W\):
We will pre-route logical qubits so that gates act in the same column.
This will allow us to amortize the routing cost that is present in the previous compilation scheme.
Consider the multigraph \(G\) where vertices correspond to cells and edges correspond to gates.
Compute an edge coloring of \(G\) using at most \(2R\) colors using a generalization of Vizing's theorem~\cite{berge1991short}.
For the first \(R\) colors, make an assignment of colors to columns \([R]\).
Then, sequentially sweep through the cells, opening and routing them in time \(O(R^2)\) such that a qubit supported in a gate of color \(j\) is in column \(j\).
This has a total time \(O(MR^2) = O(WR)\).
Make a second sequential sweep through the cells, executing all gates supported on a cell before moving to the next cell. 
To execute a single gate, we must open the other cell the gate is supported on and perform a constant number of operations.
This takes time \(O(R)\), and there are at most \(W\) gates.
We repeat the process for the latter \(R\) colors.
Thus, when there are \(W\) gates per layer of quantum circuit we would like to execute, the overall time per executed circuit layer is \(O(WR)\).

\subsubsection{Properties}
We now prove that \zcref{constr:fast-processor} gives a fast quantum processor.

\begin{theorem}[store={fastprocessor},note={Fast quantum processor}]\label{thm:fast-quantum-processor}
    Let $d_\ell$ and $w_\ell$ be the fault-tolerant gadget depth and space, respectively,
    from \zcref{assume:concatenated-codes}.
    There exists a maximum tolerable noise \(p_*^{\mathsf{comp}}>0\) such that,
    using the closed clocked circuit \(C_\bullet\) described
    in~\zcref{constr:fast-processor} (with \(O(1)\) controlled qubits)
    subject to \(p\)-locally stochastic noise with \(p \in [0,p_*^{\mathsf{comp}}]\),
    efficient preprocessing (to prepare the control sequence), and efficient postprocessing,
    the output of any Clifford+\(\gateCS\) quantum circuit
    of depth \(D\), space \(S \le W\), and gate count \(V\)
    can be sampled to error \(O\left(\poly(V,W) e^{-2^\ell}\right)\) using the compilation schemes from \zcref{sec:fast-processor:programming}
    with a run time $T$ with 
    \begin{align*} T \le O\left(\min\left(VR^2, D(SR + R^2)\right)\right),\end{align*}
    where \(R \le O(\log L + d_\ell^2)\) is asymptotically at most the log of the number of cells \(L^2\)
    and the square of~\(d_\ell\).
    The initial state of \(C_\bullet\) is arbitrary.
    If \(N\) is the number of physical qubits \(C_\bullet\) is supported on, the number of logical qubits is \(W = \Omega(N/w_\ell)\).
\end{theorem}
\begin{proof}
~%
\paragraph{Size and runtime}
The space of the processor is dominated by the RAQM.
Each memory cell has \(\Theta(\eta) = O\left(\frac{a^2}{d_\ell w_\ell}\right)\) primitive tiles in a footprint of size \(a^2\) for a total of \(N=O(L^2 a^2)\) physical qubits and \(W = d_\ell \eta L^2\) logical qubits.
The addressing and parity computation have size \(O(L \polylog L) = o(L^2)\) and sit at the periphery of the RAQM, so the processor has \(O(a^2 L^2) \le O\left(w_\ell L^2 \log L + L^2 d_\ell^2\right)\) qubits and bits fitting in a square of side length \(O(aL)\).

The runtime can be bounded using \(R \equiv \eta d_\ell \le O(\log L + d_\ell^2)\).
For conciseness, we have described the compilation for the case where all logical qubits are utilized. 
When the circuit uses some number \(S \le W\) qubits, we can use either compilation A or compilation B, assigning the \(S\) logical qubits to a subset of the memory cells.
The same runtime bounds hold with \(W\) replaced by \(R \lceil\frac{S}{R}\rceil \le S + R\).
That is, compilation A takes time \(O(v R^2)\) to execute a single logical circuit layer with \(v\) gates, and compilation B takes time \(O(SR + R^2)\) to execute a single logical circuit layer.

\paragraph{Fault tolerance}
The fault-tolerance argument is essentially identical to that of \zcref{thm:all-quantum-processor}.
As in the proof of \zcref{thm:all-quantum-processor}, we will focus only on the differences.
Recall that the classical components are assumed to be noiseless.
Thus, the only new component is the classical-to-quantum interface which has the same weight enumerator bound as the gadgets in \zcref{assume:concatenated-codes}.
Additionally, the total number of non-IO primitive tiles in the processor is \(O(\eta L^2)\).
This is bounded by the total number of logical qubits in the processor \(W\).

Consider execution of the processor for time \(T\) subject to a fault \(\fault\) distributed according to a \(p\)-locally stochastic noise model.
By a construction paralleling that of \zcref{thm:all-quantum-processor}, we construct a family of bad fault paths \(\badfaults_{\mathsf{fastprocessor}}\) such that when the fault \(\fault\) is \(\badfaults_{\mathsf{fastprocessor}}\)-avoiding, the output is a sample from the executed logical quantum circuit. 
The only difference is that the 
For \(p \in [0,p_*^{\mathsf{comp}}]\), the weight enumerator of \(\badfaults_{\mathsf{fastprocessor}}\) is bounded as
\begin{align*}
    \weightenum{\badfaults_{\mathsf{fastprocessor}}}{p} &\le O(T W \epsilon(\ell)) \le O\left(TW e^{-2^\ell}\right).
\end{align*}

\end{proof}

\begin{remark}[More control]\label{rem:more-control}
    As we mention in \zcref{sec:discussion}, it is possible to essentially interpolate between the standard quantum circuit model (\zcref{example:standard-model}) and a constant number of controlled qubits (\zcref{example:autonomuous-model}) such that we can recover the speed of the standard quantum circuit model as more control is provided to us.
    The first way is to directly transmit full cell addresses to avoid the \(\log L\) slowdown in \zcref{sec:fast-processor:programming}.
    The second, more extreme method, resembles classical multicore processors: place many small copies of the fast processor device from \zcref{thm:fast-quantum-processor} and augment the processors with an additional teleportation-based \(\gateSWAP\) instruction to other cores e.g. in a nearest neighbor connectivity or one of the many supercomputing network topologies.
    If the initial quantum algorithm to be run is specified in the form of a dense quantum circuit, then the workload is inherently parallel and the speedup can be maintained for a very large number of cores.\footnote{At some point the work per core will be too small to amortize the operation latency of the cores and memory (\(\sim \poly( \log L, d_\ell)\)). Likewise, if the input circuit is not initially 2D, then there will be an inherent routing overhead associated with 2D execution. If one imagines each core as living on separate chips, then this overhead can be reduced in practice via non-local networking e.g. fiber optics.}
\end{remark}

\printbibliography
 
\end{document}